\documentclass[twocolumn]{aastex61}
\usepackage[utf8]{inputenc} % para acentos

\accepted{November 9, 2017}
\submitjournal{ApJS}
\shorttitle{AzTICA}

\begin{document}

\title{Multiple component decomposition from millimeter single-channel data}

\correspondingauthor{ Iván Rodríguez-Montoya }

\author{Iván Rodríguez-Montoya }
\affiliation{Consejo Nacional de Ciencia y Tecnología.
			Av. Insurgentes Sur 1582, 03940, Ciudad de México}
\affiliation{Instituto Nacional de Astrofísica, Óptica y Electrónica.
			Apartado Postal 51 y 216, 72000. Puebla Pue., México }

\author{David Sánchez-Argüelles }
\affiliation{Instituto Nacional de Astrofísica, Óptica y Electrónica. 
            Apartado Postal 51 y 216, 72000. Puebla Pue., México }

\author{Itziar Aretxaga}
\affiliation{Instituto Nacional de Astrofísica, Óptica y Electrónica. 
            Apartado Postal 51 y 216, 72000. Puebla Pue., México }

\author{Emanuele Bertone}
\affiliation{Instituto Nacional de Astrofísica, Óptica y Electrónica. 
            Apartado Postal 51 y 216, 72000. Puebla Pue., México }

\author{Miguel Chávez-Dagostino}
\affiliation{Instituto Nacional de Astrofísica, Óptica y Electrónica. 
            Apartado Postal 51 y 216, 72000. Puebla Pue., México }

\author{David H. Hughes}
\affiliation{Instituto Nacional de Astrofísica, Óptica y Electrónica. 
            Apartado Postal 51 y 216, 72000. Puebla Pue., México }

\author{Alfredo Montaña}
\affiliation{Consejo Nacional de Ciencia y Tecnología.
			Av. Insurgentes Sur 1582, 03940, Ciudad de México}
\affiliation{Instituto Nacional de Astrofísica, Óptica y Electrónica. 
            Apartado Postal 51 y 216, 72000. Puebla Pue., México }

\author{Grant Wilson}
\affiliation{Department of Astronomy, University of Massachusetts,
			 Amherst, MA 01003, USA}
             
\author{Milagros Zeballos}
\affiliation{Instituto Nacional de Astrofísica, Óptica y Electrónica. 
            Apartado Postal 51 y 216, 72000. Puebla Pue., México }

\email{ irodriguez@inaoep.mx }
\email{domars@inaoep.mx}
\email{itziar@inaoep.mx}

%% Mark off the abstract in the ``abstract'' environment. 
\begin{abstract}
We present an implementation of a
blind source separation algorithm
to remove foregrounds off millimeter surveys
made by single-channel instruments.
In order to make possible such a decomposition over
single-wavelength data:
we generate levels of artificial redundancy,
then perform a blind decomposition,
calibrate the resulting maps,
and lastly measure physical information.
We simulate the reduction pipeline using mock data:
atmospheric fluctuations, extended astrophysical foregrounds,
and point-like sources, but we apply the same methodology to
the AzTEC/ASTE survey of the
Great Observatories Origins Deep Survey-South (GOODS-S).
In both applications, our technique robustly decomposes
redundant maps into their underlying components,
reducing flux bias,
improving signal-to-noise, and minimizing information loss.
In particular, the GOODS-S survey is decomposed into
four independent physical components,
one of them is the already known map of point sources,
two are atmospheric and systematic foregrounds,
and the fourth component is an extended emission
that can be interpreted as
the confusion background of faint sources.
\end{abstract}

%http://journals.aas.org/authors/keywords2013.html#Resolved_and_unresolved_sources_as_a_function_of_wavelength

\section{Introduction}
\label{sec_intro}

Astronomical observations in millimeter (mm) wavelengths
provide crucial information to comprehend the formation
and evolution of structures in the Universe at all scales,
from galaxy clustering \citep{Carlstrom2002_SZ}
to circumstellar debris disks \citep{Chavez2016EE}.
This observational window also led to the discovery of
a whole new population of bright dust-obscured sub-mm galaxies (SMGs),
mostly unresolved by single-dish telescopes, but whose detection
is available within a wide range of high redshifts \citep{Casey14}.
Moreover, cold dusty sources are brighter in mm-wavelength,
allowing a relatively easy detectability from ground based observatories.

Even though mm-astronomy already span a few decades,
there are still some relevant challenges for ground-based observations,
concerning especially foreground removal and calibration of data.
First,
water vapor and oxygen emit lines at different microwave lengths,
making the Earth atmosphere partially opaque to millimeter emissions;
alongside the difficulty that atmospheric fluctuations
are non-stationary and often abrupt.
Second,
even astrophysical foregrounds may hinder 
the inference of some physical quantities.
For example, bright patches of an extended emission
could be confused with SMGs or other compact sources.
Conversely, point-like objects stand as a contamination for an extended source.
Third,
for single-channel instruments multi-wavelength separation is impeded,
making foreground removals quite challenging.
But even with multi-channel instruments,
in order to maximize their profit,
the challenge dwells in developing advanced decomposition algorithms.

Thus, for any ground-based mm-wavelength experiment,
it is crucial to explore new strategies to improve data cleaning,
astrophysical component separation, and enhance sensitivity.
In this spirit,
here we present a new implementation of two well known methodologies:
Principal Component Analysis (PCA) and Independent Component Analysis (ICA).
Our main goal is to propose and test a new technique able
to perform multi-component separation in defiance of the single-channel limitation.
Second, we want to propose and test strategies to calibrate the separated components.
Finally, we want to apply our ideas to real data in order to
probe our ability to recover previous measurements, and
get an insight on the potential benefits to use our PCA-ICA technique.

In previous studies, atmospheric cleaning has been attempted by
removing common modes along the detector-array \citep{Sayers2010}.
On the other hand, ICA was used in space-based multi-channel experiments
to clean the Cosmic Microwave Background from its astrophysical foregrounds
\citep[for a review see][]{IchikiReview2014}.
Similar algorithms have evolved and successfully applied to a
variety of astrophysical observations,
from exoplanetary light curves \citep{Morello2015Exoplanet}
to forecasts of interferometric 21 cm cosmological signals \citep{Zhang2016HiemICA}.
In context with literature, we are reporting the first
multi-component analysis of single mm-wavelength data,
and for a ground-based telescope.
The core of our proposal relies on a technique to increase data redundancy,
whose closest discussion was made in \citet{Waldmann2014Exoplanets}.
Although we focus on AzTEC,
a 144 bolometer camera currently operating in a single (1.1mm) channel \citep{Wilson08}
and coupled to the Large Millimeter Telescope \citep{LMT2010},
our approach could be extended to other single- or even multi-channel experiments.

This paper is organized as follows:
%2
In section \ref{sec_algorithms},
we motivate our methodology,
introducing the theoretical basis
of the PCA and ICA algorithms.
%3
In section \ref{sec_pipeline},
we describe the AzTEC instrument,
the observational data, and the
numerical code used to process the time-domain data
into an astrophysical map.
In subsection \ref{subsec_ica_pipeline},
we introduce our proposals of using
PCA in time-domain followed by ICA in map-domain,
as an extension to the standard AzTEC pipeline.
%4
Section \ref{sec_mock} is devoted to
implement and test our techniques with simulations.
We describe the mock data employed,
the simulated reduction process,
the decomposition parameters, calibration strategies,
and steps to extract astrophysical information.
%5
In section \ref{sec_real}
we apply the same tools to
the Great Observatories Origins Deep Survey-South (GOODS-S),
observed with AzTEC when it was installed on
the 10 m Atacama Submillimeter Telescope Experiment (ASTE)
\citep[hereafter $KS10$]{Scott2010},
in order to recover previous measurements
and discuss them in connection with our simulation results.
%6
Our conclusions are summarized in section \ref{sec_conclusions}.

\section{PCA and ICA algorithms}
\label{sec_algorithms}

\begin{figure*}
 \noindent\makebox[\textwidth]
{ \centering
   \includegraphics[width=0.98\textwidth]{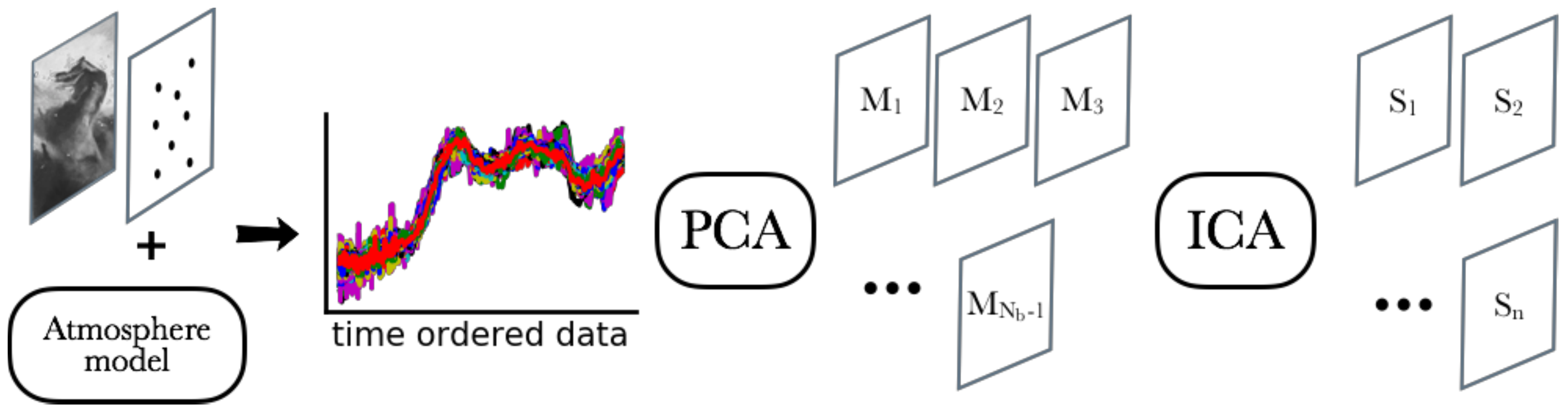}  }
 \caption{ Schematic description of the data processing in this paper.
   For simulations, the time-streams are built from
   an atmospheric model and mock (extended and point-like)
   astrophysical data.
   PCA is used in time-domain to generate a series of redundant maps
   $\left\lbrace M_i \right\rbrace$, and
   ICA is applied in map-domain for multi-component decomposition.
   Later on, the independent signals $\left\lbrace S_j \right\rbrace$
   are calibrated and astrophysical information is extracted.}
    \label{fig_diagram}
\end{figure*}

Why PCA and ICA?
Our concern is that atmospheric and astrophysical
emissions are mixed along some range of scales.
It is appealing that both PCA \& ICA
are blind (non-parametric) separation algorithms,
so we do not need to rely on physical models,
but just on the statistical properties of data.
PCA computes uncorrelated projections of data,
while ICA demands the stronger condition of statistical independence.
Before going into formal details of each algorithm
\citep[for an introductory tutorial see][]{Stone2004},
let us intuitively discuss their respective roles in our implementation
(see also figure \ref{fig_diagram}).

PCA is used in time-domain as follows:
Due to their intense brightness and angular scale,
atmospheric fluctuations induce
large correlations along the bolometer array.
PCA computes a vector basis where
the fluctuation modes are uncorrelated,
the first few are attributed to atmospheric contamination and removed,
leaving modes dominated by astrophysical signal and noise.
PCA is very efficient, especially for point source recovery;
but some motivations to explore more advanced algorithms include
that PCA uses only second-order statistical moments,
\textit{i.e.}, non-Gaussian information is unexploited.
More importantly, at least part of the astrophysical information
is lost when the subset of (bad) principal components is discarded.

Formally, ICA is an extension of PCA,
but the cases of interest and criteria
to apply each of these algorithms may be significantly different.
Provided that the underlying emissions are non-Gaussian,
ICA employs high-statistical moments to find a basis of 
(not only uncorrelated but) statistically independent components.
Using ICA, it should be possible to reduce information loss because
every component is in principle isolated, containing no information about others.
In general practice,
ICA is fed with $m \geq 2$ mixed signals to be decomposed
into $n\leq m$ independent components.
Unfortunately, ICA bears an inherent incompatibility
with single-channel instruments:
a single-channel instrument yields a single map
at a given wavelength band ($m=1$),
but ICA needs at least two input maps in order to decompose them
into at least two independent components
(see \S\ref{subsec_theoryICA} and \S\ref{subsec_ica_pipeline} for details).
Still, we can turnaround this limitation by producing
hierarchical levels of (artificial) redundancy,
mimicking the maps from a multi-channel survey,
which can be used as input for ICA.
Specifically, in this paper we propose to apply ICA in map-domain
over a series of redundant maps produced with PCA in time-domain.

In the rest of this section we briefly overview
the basics of both algorithms, so that
the reader familiar with the theoretical aspects
may jump directly to \S\ref{sec_pipeline}.

\subsection{Principal Component Analysis (PCA) }
\label{subsec_theoryPCA}

Let us assume that we are working with an astronomical survey
whose raw data $\lbrace x_i \rbrace _{i=1}^{N_b}$
is a collection of $N_b$ timestreams (say \textit{e.g.} the number of bolometers),
each of them with sampling size $T$.
Thus, our time-ordered data can be represented by a vector
$\mathbf{x}$, such that $\dim{(\mathbf{x})} = N_b \times T$.
Assume also that the timestreams were centred in a pre-processing step,
$x_i \leftarrow x_i - \left\langle x_i \right\rangle $,
where $\left\langle \cdot \right\rangle$ denotes the expectation value.

The covariance matrix is defined by
$ \mathbf{C_x} = \left\langle \mathbf{x} \cdot \mathbf{x}^T \right\rangle$,
$ \dim{ (\mathbf{C_x}) } = N_b \times N_b$.
If it would be the case that every timestream $x_i$ were Gaussianly distributed,
then $\mathbf{C_x}$ would contain all the information available,
and higher-order statistical moments would be either zero 
or trivially rewritten in terms of $\mathbf{C_x}$.

PCA is formally an eigenvalue problem
for the covariance matrix $\mathbf{C_x}$,
the goal is to find a vector basis
$\lbrace \mathbf{e}_i \rbrace _{i=1}^{N_b}$
such that they project the raw-data $\mathbf{x}$ into a new set 
of uncorrelated components $\lbrace y_i \rbrace _{i=1}^{N_b}$,
\begin{eqnarray}
  \mathbf{y} &=& \mathbf{E}^{T} \cdot \mathbf{x}, \\
  \left\langle y_i \cdot y_j \right\rangle &=& \lambda_i \, \delta_{ij}, \nonumber
\end{eqnarray}  
where the columns of $\mathbf{E}^T$ are the unit eigenvectors $ \mathbf{e}_i$,
and $\lbrace \lambda_i \rbrace _{i=1}^{N_b}$ are the eigenvalues of $\mathbf{C_x}$.
It turns out that the new covariance matrix is diagonal
$\mathbf{C_y} = \text{diag}( \left\lbrace \lambda_i \right\rbrace )$,
whose elements are the variances of the projected time-streams,
$\lambda_i = \text{var}(y_i)$.
The PCA distinctive step is to order the projected timestreams
according to an eigenvalue hierarchy
$\lambda_1 \geq \lambda_2 \geq \cdots \geq \lambda_{N_b} $,
then, $y_1$ is called the first principal component,
$y_2$ the second principal component, and so on.
When projected back to the raw data $\mathbf{x}$,
the principal component $y_1$ is evidently the major source of correlation,
contrary to $y_{N_b}$, which is the source of least correlation.

The cleaning step is simply to discard $N_{\text{atm}}$
principal components that are attributed
to large scale foregrounds (typically the atmosphere).
Then, the remaining components
$\left\lbrace y_i \right\rbrace _{i=N_{\text{atm}}+1}^{N_b}$
are projected back into the original space.
Notice that the cleaned data vector $\mathbf{x}'$
also has $\dim{(\mathbf{x}')} = N_b \times T$.

\subsection{Independent Component Analysis (ICA)}
\label{subsec_theoryICA}

In signal processing analysis,
the \textit{cocktail party} problem is referred to as
the linear mixing of $n$ true signals into the recordings of $m$ sensors,
under the condition $n\leq m$
\citep[for a comprehensive review see][]{Hyvarinen2004book,Comon2010book}.
In this subsection the notion of a sensor is meant quite generic.
For instance, a sensor may be an astronomical survey
measured at a given wavelength along with
the process to make an observation map; in this scheme,
the mixed and true signals live in pixel-domain.
Keeping this broader notion in mind, let us denote $x_i$ as the mixed signal
of the $i$-th sensor, for $i=1,2,\cdots ,m$.
The $j$-th true signal is then denoted as $s_j$, for $j=1,2,\cdots ,n$.
The raw data is then modelled by ICA as an instantaneous mixing
of the true signals in terms of a linear combination,
\begin{equation}
    x_i = a_{i1} \; s_1 + a_{i2} \; s_2 + \cdots + a_{in} \; s_n.
    \label{eq_ica_aij}
\end{equation}
The mixing coefficients $\left\lbrace a_{ij} \right\rbrace$
are real numbers that may be interpreted as the transmission/extinction
information of the true signals through the sensing process.
The model can also be written in matrix notation,
\begin{eqnarray}
   \mathbf{x} &=& \mathbf{A} \cdot \mathbf{s}, \\
   \mathbf{s} &=& \mathbf{W} \cdot \mathbf{x},
   \label{eq_ica_matrix}
\end{eqnarray}
where $\mathbf{x}$ is an array whose column-vectors are the $m$ mixed signals,
$\mathbf{A}$ is called the mixing matrix ($m \times n$).
The goal is to estimate the unmixing matrix
$\mathbf{W} = \mathbf{A}^{-1}$,
retrieving the true signals $\mathbf{s}$.
Since both $\mathbf{W}$ and $\mathbf{s}$ are simultaneously estimated,
the unmixing problem becomes too complicated for classical methods.

ICA relies on the assumption that the true signals are
statistically independent and non-Gaussian.
Two random variables $y_1$ and $y_2$ are
statistically independent if and only if
their joint probability distribution is equal to
the product of their marginal probability distributions,
\begin{equation}
  	p(y_1 ,y_2) = p(y_1) \cdot p(y_2). 
 \label{eq_independence}
\end{equation}
Statistical independence implies uncorrelatedness,
though uncorrelatedness does not necessarily imply independence.
ICA then appeals to the central limit theorem, which says that
the sum of two non-Gaussian distributions
is more Gaussian than the initial distributions;
conversely, independent signals are maximally non-Gaussian.
Thus, to solve the unmixing problem in equation (\ref{eq_ica_matrix}),
ICA estimates those coefficients $w_{ji}$ that maximize
the non-Gaussianities of $\left\lbrace s_{j} \right\rbrace _{j=1}^{n}$.

Non-Gaussianities are measured by high-order statistical moments.
For example, the skewness measures the symmetry of the distribution,
the kurtosis measures how spiky (or flat) the distribution is.
For Gaussian distributions both the skewness and kurtosis are zero
because only the first two moments are relevant,
namely the mean and variance.
Although the simplest approach would seem
to maximize skewness or kurtosis,
they are easily biased by outliers, thence
more robust non-Gaussianity estimators should be used instead
\citep[for a concise review see][]{Choi2012}.
Here, we focus on the concept of \textit{negentropy}.

Entropy is the basic concept in Information Theory and
it is particularly interesting for measuring non-Gaussianities
\citep{Hyvarinen2000}.
Defined as,
\begin{equation}
   H(y) = - \int dy \, p(y) \log{p(y)},
\end{equation}
entropy is related to the degree of information contained in the random variable $y$.
As the variable is more unstructured and unpredictable, its entropy is larger as well.
Indeed, the Gaussian distribution possesses the maximum entropy,
\textit{i.e.}, it is the most random, least structured, and least informative distribution.
In the same vein,
negentropy is defined as the deviation from the maximum entropy,
\begin{equation}
   J(y) =  H(y_G) - H(y).
   \label{eq_negentropy_general}
\end{equation}
Here, $y$ represents the signal of interest,
$y_G$ is a Gaussian distribution with the same mean and variance as $y$.
Negentropy is always nonnegative $J(y) \geq 0$ and
equals zero if the signal $y$ is Gaussianly distributed $J(y_G) = 0$.
Clearly, the larger the negentropy
the more informative is the distribution,
and the more independent is the signal.
Hence, negentropy is the optimal estimator
of non-Gaussianity and statistical independence.

As defined in equation \ref{eq_negentropy_general},
measuring negentropy is difficult
and some approximated estimators, like higher-order cumulants,
are frequently used,
\begin{equation}
  J(y) = \left[ \left\langle G(y) \right\rangle - \left\langle G(y_G) \right\rangle \right]^2,
  \label{eq_negentropy}
\end{equation}
where $G$ is a non-quadratic function that may be conveniently chosen.
In principle any power-function higher than quadratic is a valid choice for $G$,
but in practice, a wise choice may boost the speed of the algorithm.
For instance, $G(y) = y^4$ is the kurtosis-based approximation,
useful when the independent signals are flat-like distributed,
but otherwise non-robust against outliers.
Some commonly used functions are
\begin{eqnarray}
  G_1(y) &=& 1/c_1 \log\left(\cosh{ (c_1 y) }\right), \label{eq_G1} \\
  G_2(y) &=& -\exp{ \left( -y^2/2 \right)}, \label{eq_G2}
\end{eqnarray}
where $1 \leq c_1 \leq 2$ is a constant often equal to one.
With this negentropy approximation,
an optimizing algorithm can find
the numerical values of the unmixing coefficients
such that maximize the negentropy of the independent components expressed in equation (\ref{eq_ica_matrix}).
To this end, in this paper we use FastICA \citep{Hyvarinen1997,Hyvarinen2000},
which is a very efficient fixed-point algorithm
that has been widely used and tested,
as it is the most standard ICA-algorithm.

There remain, though, two ambiguities inherent to ICA.
These are natural consequences of the fact that
we are dealing with a system with
less equations than unknown variables.
Henceforth, as a post-decomposition step,
the independent components ought to be calibrated,
before any physical information may be inferred.

The \textit{permutation ambiguity} means that the order
of the independent components is basically random;
this is because equation (\ref{eq_ica_aij}) is invariant 
under permutations of the mixing matrix.
The seriousness of this ambiguity depends on the number of
independent components and how distinctive they are.
If a given problem required to control
the order of many independent components,
the permutation ambiguity could become too prohibitive
for an ICA application.

The \textit{scaling ambiguity}, is the inability
to determine the variance of each independent component.
Notice that equation (\ref{eq_ica_aij}) remains invariant
under the transformation
$s_j \leftarrow a_j s_j$ and $a_{ij} \leftarrow a_{ij}/a_j$,
where $a_j$ is a real scale factor.
Then, one may choose arbitrary scales right after the decomposition.
As a convention, we will set every $s_j$ to unit-std,
but scale-calibrations ought to be pursued afterwards.

\section{The AzTEC instrument and pipeline}
\label{sec_pipeline}

\subsection{The instrument}
\label{subsec_aztec_instrument}

AzTEC (Aztronomical Thermal Emission Camera, \textcite{Wilson08})
is a continuum millimeter wavelength receiver containing
144 Si$_3$Ni$_4$ spiderweb mesh bolometers.
The receiver is configured to operate in the 1.1mm atmospheric window.
The bolometers are arranged in an hexagonal array divided in
six slices or hextants, distributed in a closed packed configuration.
The footprint of the bolometer array covers a roughly circular area
of $\sim$8 arcmin diameter on the sky.
These time-ordered data, or \textit{timestreams},
are later processed along with the telescope pointing information
to construct an image of the sky surface brightness.
For a ground-based (sub)millimeter camera,
a single detector timestream $d$ can be described as
\begin{equation}
  d = \mathcal{P}s+\mathcal{A}+N \,,
\end{equation}
where $s$ is the surface brightness distribution
of astronomical objects,
$\mathcal{P}$ is the pointing matrix,
$\mathcal{A}$ is the atmosphere emission,
and $N$ is the instrumental noise.
It is important to note that the atmosphere fluctuations are
between 1 and 4 orders of magnitude
larger than the astronomical emission.
Therefore, it is necessary to calculate and remove
an estimation of the atmospheric component, in order to
retrieve an image of the brightness distribution of faint sources.
This process is critical for ground based observations,
where both the telescope scanning pattern and
the map projection code are designed to decouple
the astronomical emission from the atmospheric foregrounds.
In particular, for the data described in the sections below,
AzTEC observations were carried out with a modified
\textit{Lissajous} pattern;
this is a parametric curve constructed from
two sinusoidal waves in orthogonal directions.
The projection of the scan track over the sky,
relative to the map center, can be described by:
\begin{eqnarray}
 \delta A(t)&=&5.5\arcmin\sin{9t}+2.0\arcmin\sin{9t/30} , \\
 \delta E(t)&=&5.5\arcmin\sin{8t}+2.0\arcmin\sin{8t/30},
\end{eqnarray}
where $t$ is the observation elapsed time in seconds,
$\delta A$ and $\delta E$ are the track offsets
from the map center in azimuth and elevation respectively.
In the following subsection we briefly overview
the standard map projection code, namely, the reduction pipeline.

\subsection{The reduction pipeline}
\label{subsec_macana}

A deep millimeter survey is a number of observations of a sky-patch,
stored in raw data files containing the recorded timestreams,
along with telescope parameters for calibrations.
For surveys made with the AzTEC camera,
each observation contains $N_b$ timestreams,
corresponding to the effective number of bolometers
(\textit{i.e.} the bolometers with
the best electronic responsivity).
The AzTEC reduction pipeline uses PCA
to remove the atmospherical signal
(hereafter the \textit{cleaning process})
and projects the cleaned timestreams into a bi-dimensional grid,
delivering an astronomical image as a result.
(For a detailed description of the AzTEC pipeline,
we refer the reader to \citet{Wilson08,Scott08}.)

\textit{Time sampling.}
Typically, a single observation lasts 20 min,
but the timestreams are sampled in
\textit{time-chunks} of 10 or 29 s.
We denote with $\mathbf{x}$ the chunk made of
$N_b$-timestreams with length $T$.
Every time-chunk is worked out sequentially for each observation,
but the reduction code runs in parallel for multiple observations.

\textit{Signal conditioning}.
All timestreams are corrected for instrumental glitches 
and large spikes induced by cosmic rays.
A low pass filter is applied in order to 
minimize the contamination of high frequencies.

\textit{Atmospheric removal}.
As explained in \S\ref{subsec_theoryPCA},
the principal component is the major source of correlation 
and is blamed for atmospheric contamination.
But also the second and third principal components
are often contaminated.
The question is how many principal components shall be discarded,
regarding a compromise between contamination removal
and information loss.
Because PCA is applied to every time-chunk,
the amount of information contained in each principal component
depends on the chunk-length $T$,
the larger $T$ the more components shall be discarded.

\textit{The} \textsc{pca}2.5$\sigma$ \textit{procedure}.
The AzTEC pipeline has a semi-automated process
optimized for point sources \citep{Wilson08}. 
Choosing a small time-chunk length ($T$=10 s customarily),
the number of discarded components is estimated
from the eigenvalue distribution:
the 2.5 std outliers are iteratively rejected,
and their corresponding eigenvectors discarded.
Using this procedure, typically 12 principal components are discarded per time-chunk.

\textit{Map making}.
The sky positions per bolometer are continuously recorded,
according to the telescope pointing calibration,
the bolometer-array geometry, and the scanning strategy.
This information is contained in an object
called the pointing matrix $\mathcal{P}$.
Using $\mathcal{P}$, the cleaned timestreams are projected into
a single grid called the coadded map.

\textit{Noise estimation}.
Around 100 jackknife simulations of the cleaned timestreams
are projected into a single \textit{weight map} $W(p)$
that stores the inverse noise variance per pixel $p$.
The noise of every coadded map depends on:
1) the effective sensitivity
$\sigma_{\text{eff}}$ or the rms noise,
which is nearly uniform along the map, and
2) the sample number per pixel,
namely the \textit{hitmap} $H(p)$.
Hence, the weight map $W(p)$ can be approximated by
$W^{1/2}(p) \approx  w_{\text{eff}} \, H(p)$.
Both the effective weight $w_{\text{eff}}=1/\sigma_{\text{eff}}$
and $H(p)$ can be normalized such that $0 < H(p) < 1$.
This approximation is accurate to better than 0.1\% for AzTEC maps.
Then, the signal-to-noise (S/N) map can be directly obtained from
the signal and weight maps.

\textit{Filtering.}
The AzTEC pipeline is equipped with low- and high-pass filters.
The low-pass filter is a Gaussian one with 
a FWHM of the size of the telescope beam, and it is necessary
for removing spurious high-frequency fluctuations.
A Wiener filter with a point-like kernel can be used
to boost the detection of point sources
\citep{WienerFilter, Downes2012};
it enhances compact sources in detriment of extended signals.

The \textit{detection of bright sources}
is performed inside the uniformly covered area.
The first step is to locate the highest S/N pixel
and enclose the bright source within a beam-radius area.
Then, pixels closer than twice the beam-size are discarded
as candidates for the next brightest source.
The search continues down to a specified limit
\textit{e.g.} S/N$>$4.

\subsection{The ICA extension to the pipeline}
\label{subsec_ica_pipeline}

Despite the fact that ICA is formally an extension of PCA,
ICA cannot be directly applied to clean the bolometer timestreams.
The main obstacle is the permutation ambiguity, explained in \S\ref{subsec_theoryICA}.
As mentioned in the preceding subsection,
keeping control of the timestream order is crucial
because it contains the bolometer positions on the sky,
without which the map-making step could not be accomplished.

For surveys made with multi-channel instruments,
it is possible to generate a map per channel,
containing redundant information at different wavelengths.
ICA is often used to gain leverage from these
multi-wavelength signals in map-domain
\citep[for instance][]{IchikiReview2014}.
Since typically one would have only a few channels
(\textit{e.g.} $m=3$),
the permutation ambiguity is not an issue in map-domain.
Unfortunately, the number of intrinsic signals within the maps
could be typically larger than the number of available channels,
that is, $n>m$, then preventing a proper decomposition.
In single-channel instruments like AzTEC
the limitation is obviously worse,
lacking any leverage of multi-wavelength redundancy.

With the aim to overcome the single-channel limitation,
we are proposing the generation of \textit{artificial} redundancy.
Specifically, we generate redundant maps
by applying different thresholds in the PCA technique:
map $M_1$ is made by discarding the first principal component,
for $M_2$ the first and second principal components, and so on,
up to $M_{N_b-1}$, where $N_b$ is the effective number of bolometers. Our main assumption is that the components present in the map are coupled by different mixing coefficients in the redundant maps.
We choose a relatively large time-chunk (120 s)
in order to generate a smoother transition in
the degree of redundant information.
Notice that PCA in time-domain is basically employed as a filter
to generate hierarchical levels of redundancy.
Consequently, the redundant maps are strongly correlated
at several angular scales, and now, ICA can be applied on them.

Following \S\ref{subsec_theoryICA},
we model the set of redundant maps
$\left\lbrace M_i \right\rbrace _{i=1}^{N_b-1}$
as a mixture of $n$ independent components
$\left\lbrace S_j \right\rbrace _{j=1}^{n}$.
The $i$-th redundant map is explicitly,
\begin{equation}
    M_i = a_{i1} \; S_1 + a_{i2} \; S_2 + \cdots + a_{in} \; S_n ,
    \label{eq_ica_maps}
\end{equation}
or in matrix notation,
\begin{eqnarray}
  \mathbf{M} &=& \mathbf{A} \cdot \mathbf{S}, \\
  \mathbf{S} &=& \mathbf{W} \cdot \mathbf{M}.
\end{eqnarray}
Here $\mathbf{M}$ is an array containing the $N_b-1$ redundant maps,
$\mathbf{A}$ is the $(N_b-1) \times n$ mixing matrix,
$\mathbf{W} \approx \mathbf{A}^T$ is the unmixing matrix, and
$\mathbf{S}$ is an array containing the $n$ independent maps.
In this paper we use FastICA \citep{Hyvarinen2000}
to estimate $\mathbf{W}$ and $\mathbf{S}$
by maximizing the negentropy of the independent maps.

In FastICA we simply use the \textsf{logcosh}
function as in equation (\ref{eq_G1}).
For visualization purposes,
we also use the following expression for negentropy,
\begin{equation}
  J(y) = 20 \frac{\left|\left\langle G(y_N) \right\rangle - \left\langle G(y) \right\rangle\right|}%
                 { \left\langle G(y_N) \right\rangle },
  \label{eq_negentropy_mine}
\end{equation}
which is equivalent to equation (\ref{eq_negentropy})
but differs by a couple of normalization factors
(the constant $20$ and $\left\langle G(y_N)\right\rangle$)
included to increase the contrast between negentropy values.
Here, $y$ and $y_N$ are unit-std distributions centered at zero,
$y$ represents the pixel-values of a map,
$y_N$ is the normal standard distribution,
$G(y)=\log{(\cosh{y})}$.
Notice that this expression satisfies the negentropy properties
$J(y) \geq 0$ and $J(y_N) = 0$.

It is wise to restrict the pixel-data to
the better sampled region on the sky.
We adopt the convention to feed ICA with the map-area where
the (outmost) coverage is at least $\sim$30\% of the maximum,
but to extract physical information only from
the 50\% uniformly covered region.
After a successful decomposition,
in order to tackle the permutation and scaling ambiguities,
we must calibrate the independent maps.
To this end, we implement some calibration alternatives in \S\ref{subsec_calibration}.

To end this section,
a few alternatives to generate redundancy
can be mentioned from the literature.
An interesting approach could be
a decomposition in a convenient wavelet space,
with the ancillary beneficial ability
to calibrate the independent components
\citep{Waldmann2014Exoplanets}.
Another arguably possible alternative would be
to use observations taken at different time-intervals
as the input for ICA
\citep[see \textit{e.g.}][]{Funaro2003,Waldmann2012}.
We do not follow this approach because 
the amount of Gaussian noise in every individual observation
is much larger than the co-added observation, thus,
making quite difficult the decomposition for ICA.
Besides, given that this set of time-ordered maps 
were not observed simultaneously,
they would break the basic assumption of instantaneous mixing.

\begin{figure}[b]
  \centering
    \includegraphics[width=0.47\textwidth]{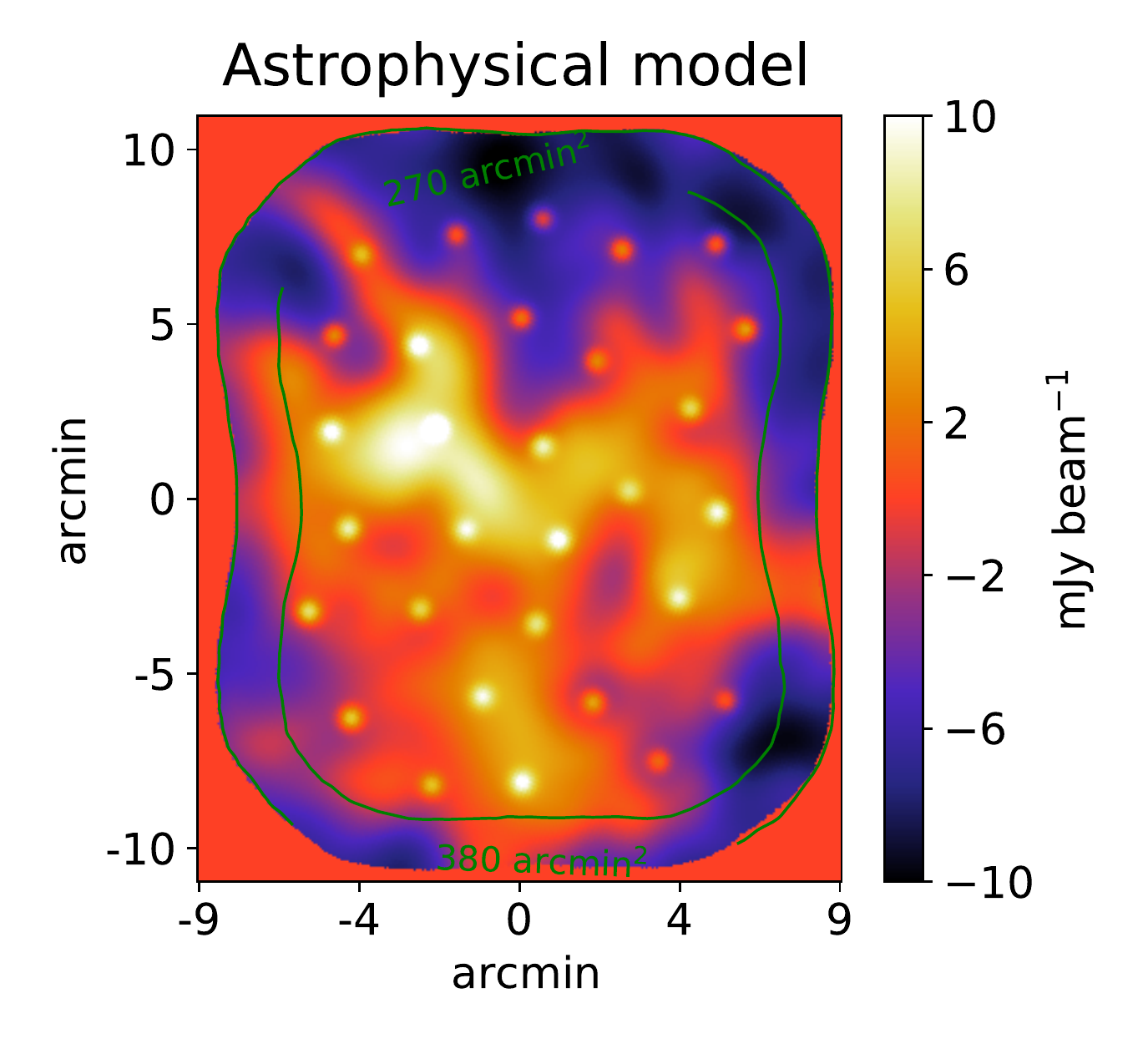} 
  \caption{ Astrophysical mock data used in our simulations:
	1) a set of 30 beam-sized sources compose the point model $P$,
	and 2) a smoothly ($3\times$beam) varying source
    constitutes the extended model $E$.
	The inner 270 arcmin$^2$ contour represents 
	the area where the telescope coverage is
	at least 50\% of the maximum.}
  \label{fig_model}
\end{figure}

\begin{figure*}
 \noindent\makebox[\textwidth]
{
\centering
 \begin{tabular}{ccc}
   \includegraphics[width=0.33\textwidth]{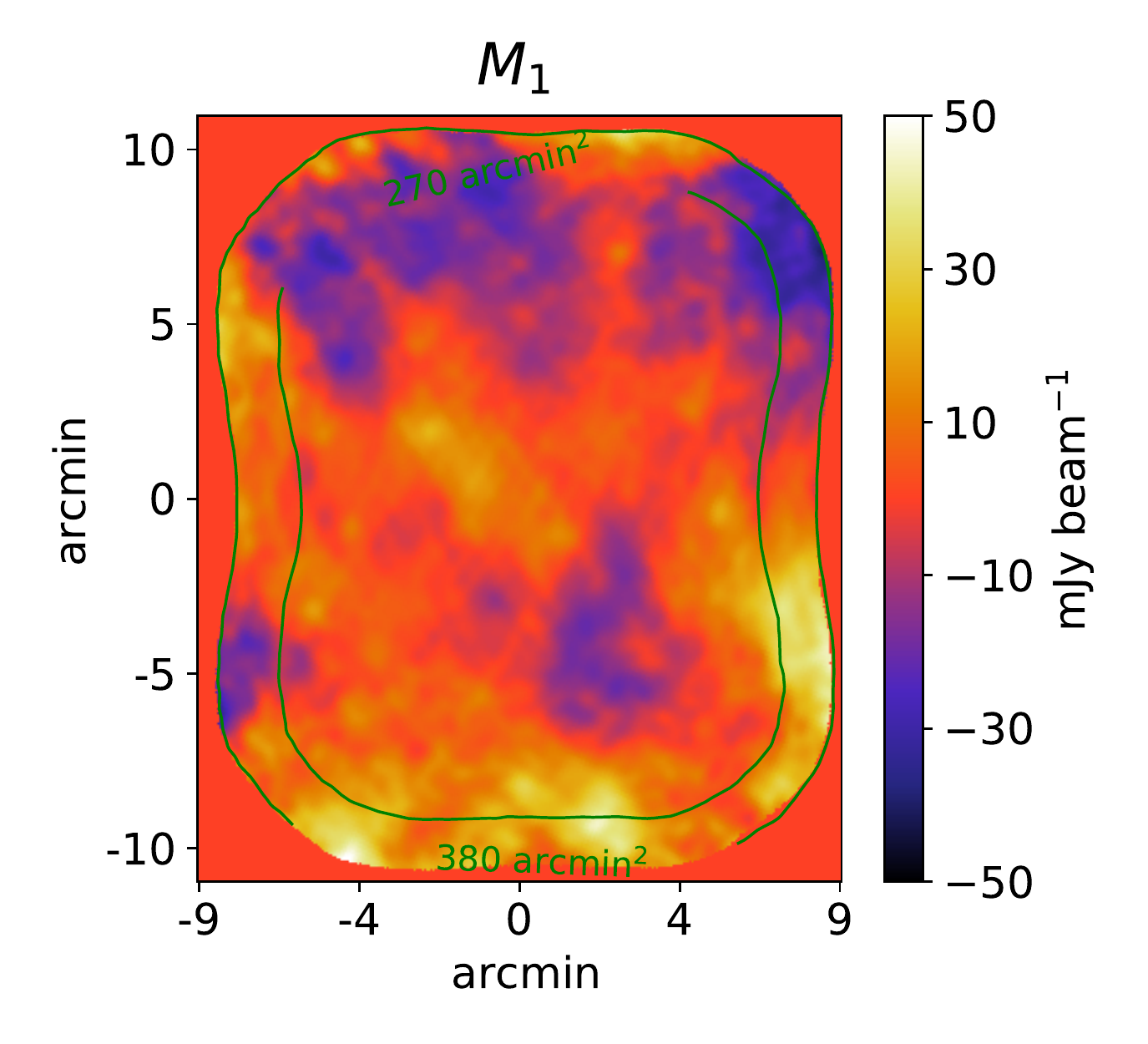} &
   \includegraphics[width=0.33\textwidth]{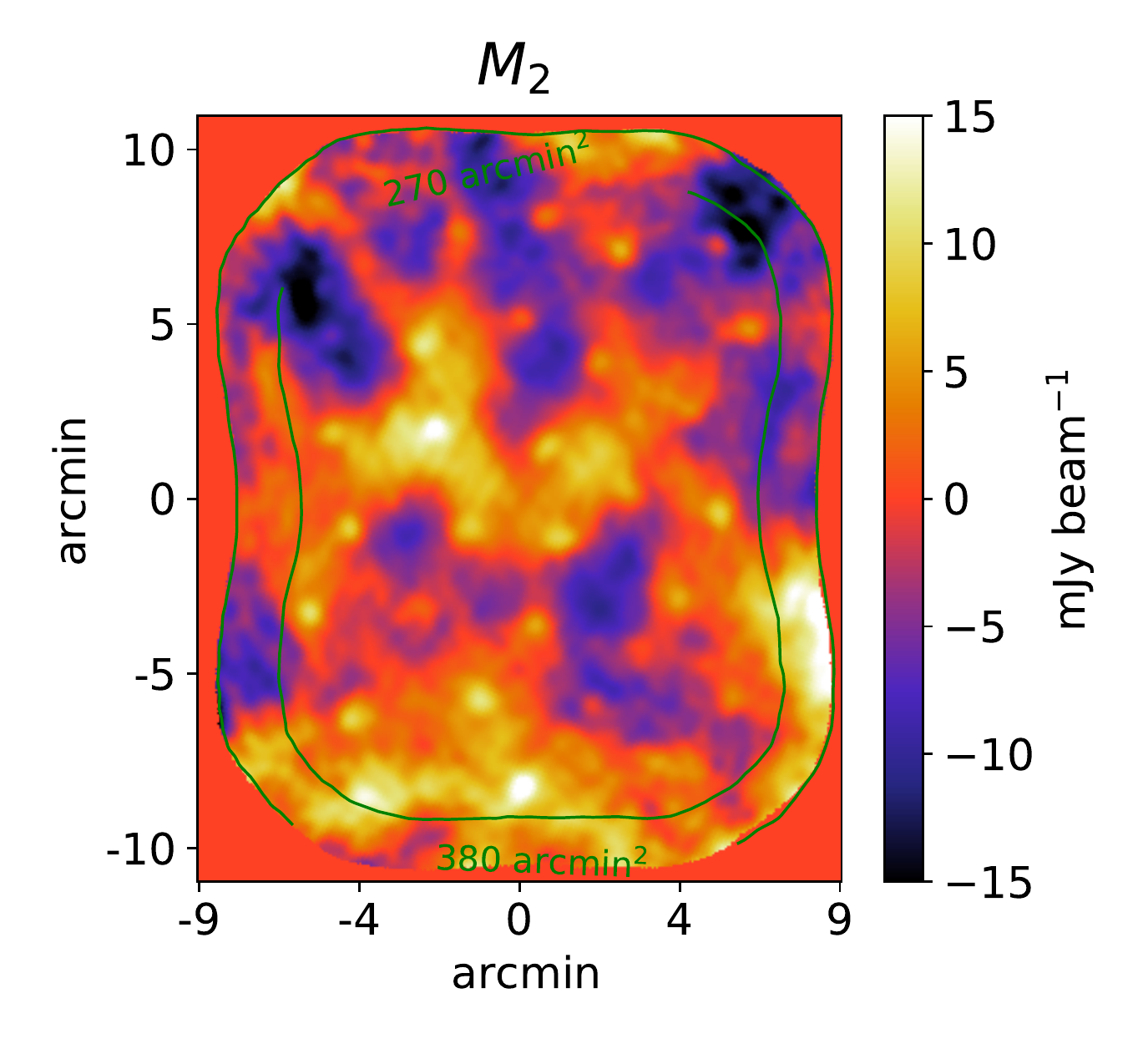} &
   \includegraphics[width=0.33\textwidth]{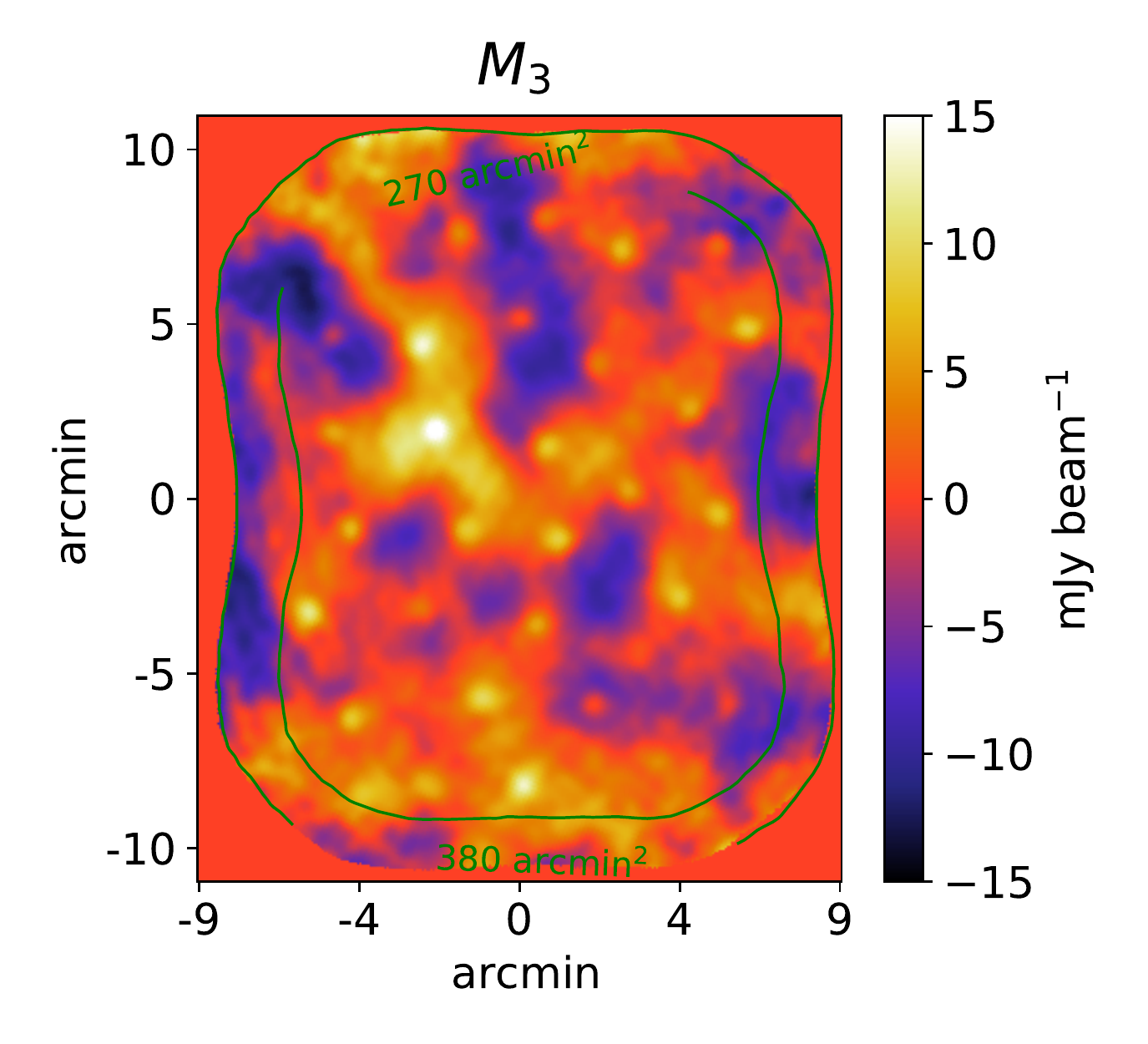} \\
	%    \vspace{1cm} \\
   %
   \includegraphics[width=0.33\textwidth]{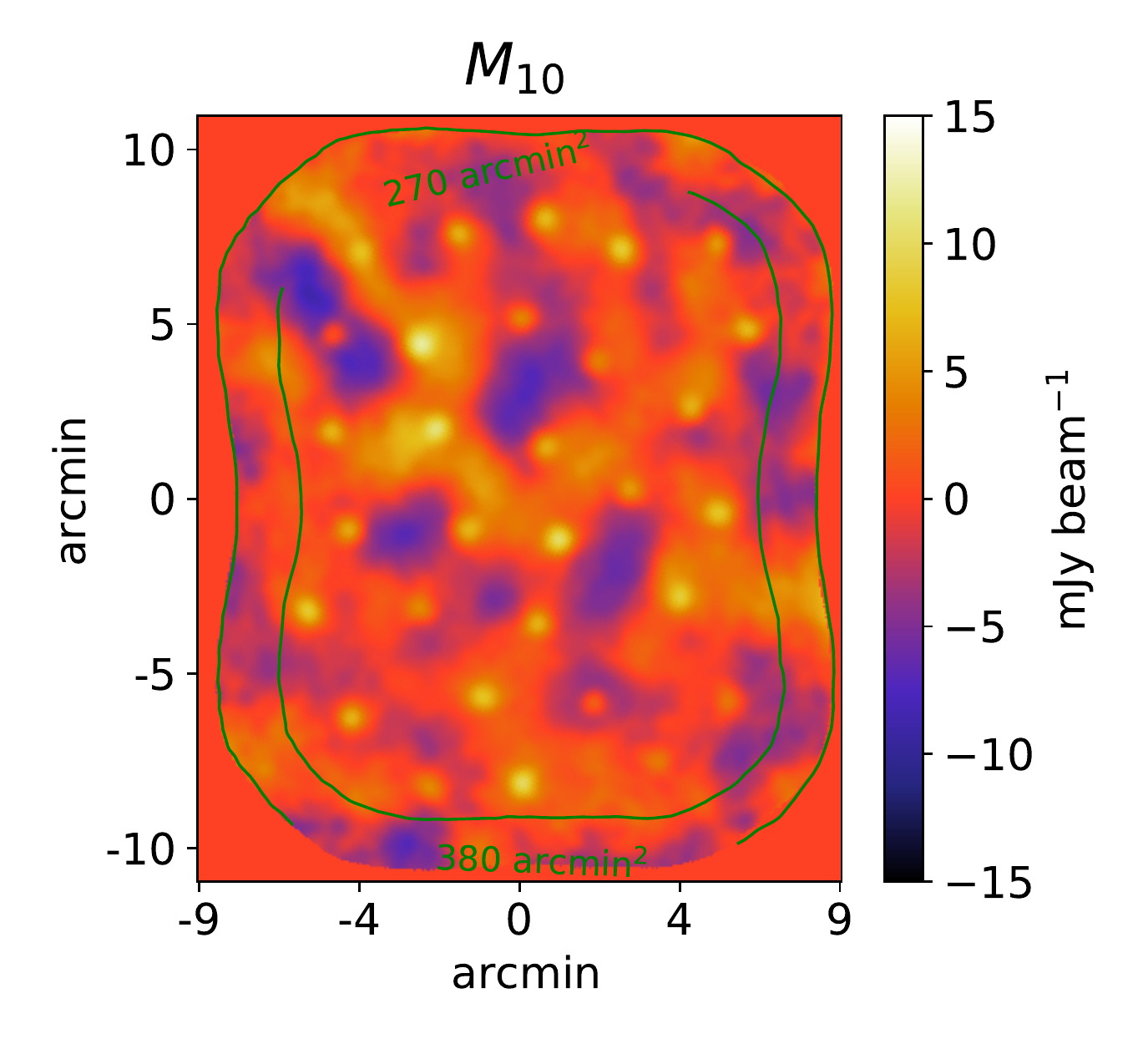} &
   \includegraphics[width=0.33\textwidth]{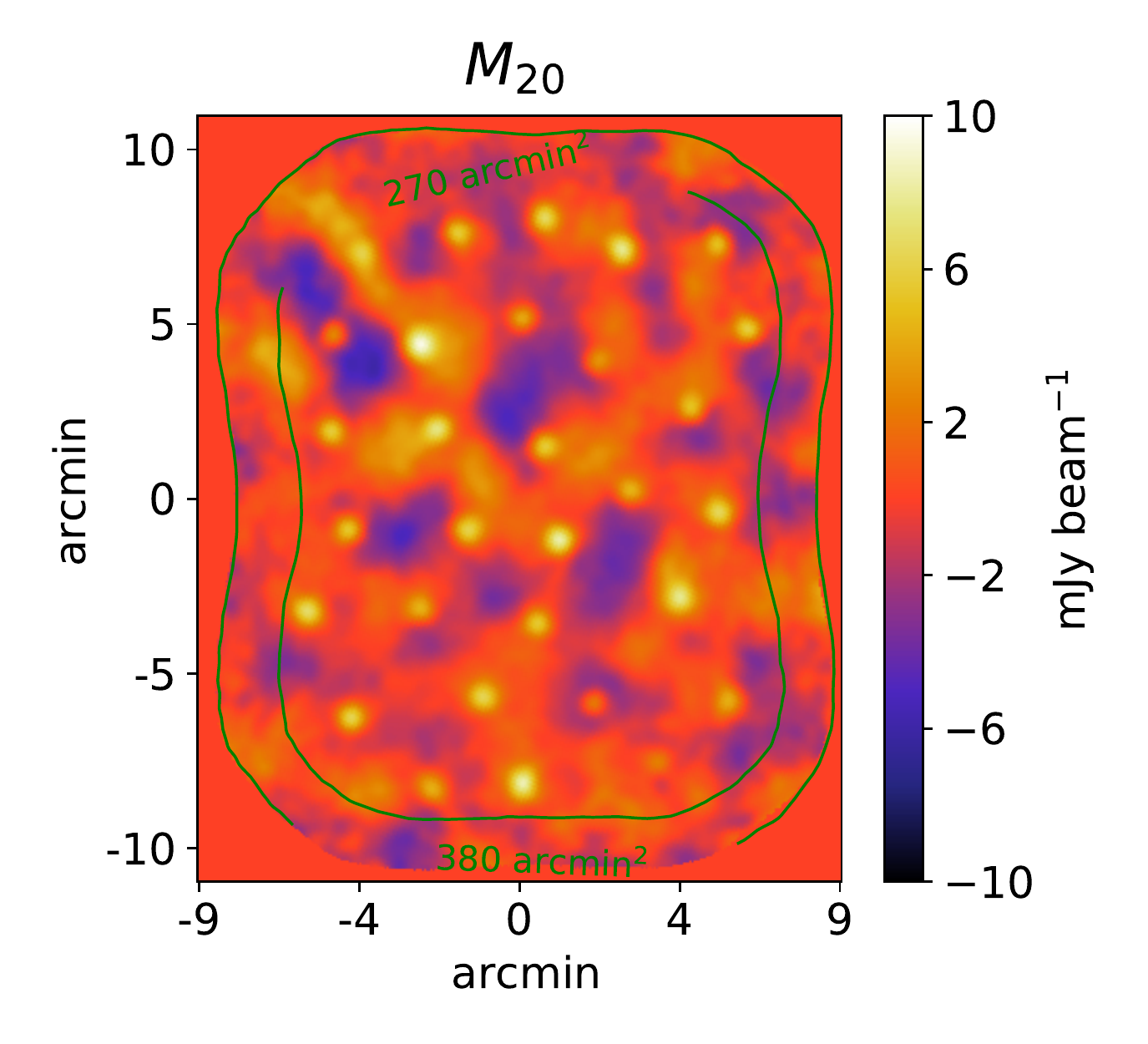} &
   \includegraphics[width=0.33\textwidth]{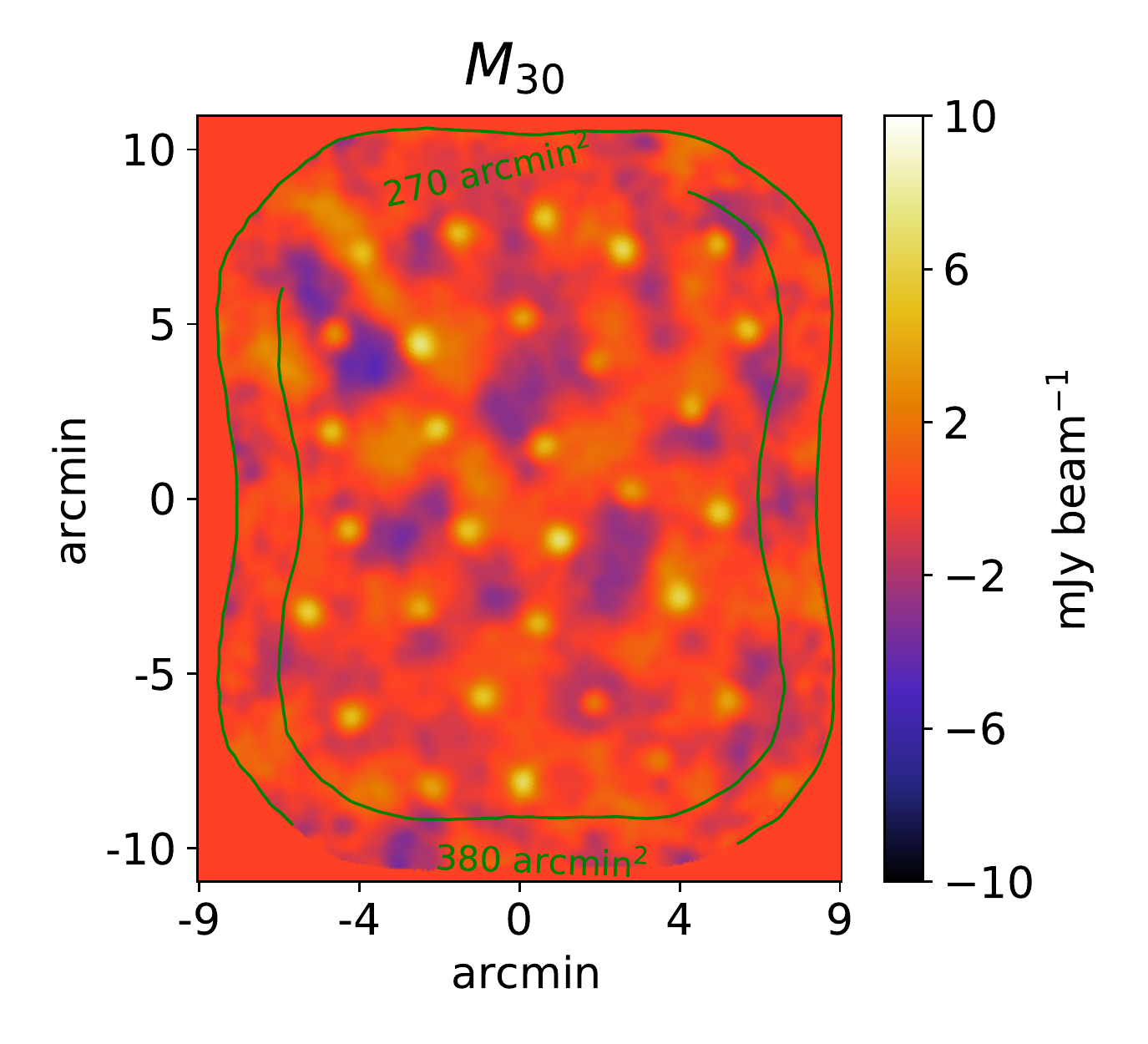} 
 \end{tabular}     }
 \caption{ Redundant maps $M_i$ made with the AzTEC pipeline,
	removing $i$-principal components, respectively.
	Notice that atmospheric and extended foregrounds
    are progressively removed.
	Point sources persist in all redundant maps
    with decreasing brightness;
	the last (not shown) redundant maps contain
    nearly Gaussian noise.
	Contours as in figure \ref{fig_model}.}
    \label{fig_sim_redundant_maps}
\end{figure*}

\section{Decomposition of mock data}
\label{sec_mock}

In this section, we fabricate a set of atmospheric and
astrophysical signals mixed in timestreams,
simulate the pipeline process to perform a decomposition,
propose strategies for calibration, and
finally measure (mock) astrophysical information,
which is useful to probe our technique.

As a benchmark, we use the AzTEC GOODS-S survey,
in which a Lissajous scan was performed,
the telescope beam FWHM was approximately 30 arcsec,
and the pixel size was chosen to 3 arcsec.

\subsection{Building simulations}
\label{subsec_build}

The atmospheric signals are created in time-domain,
where an inverse $f^{-\alpha}$ filter is used to generate 
atmosphere realizations statistically similar to observations.
The $i$-th detector signal is
\begin{equation}
 \mathcal{S}_i = \mathcal{F}^{-1}\lbrace (\mathcal{P}_i)^{1/2} \mathcal{G} \rbrace ,
 \label{eq_atm_sim}
\end{equation}
where $\mathcal{P}_i$ is the $i$-th power spectrum,
$\mathcal{G}$ is the Fourier transform of a random Gaussian sequence
with the same length of timestreams, and
$\mathcal{F}^{-1}$ is the inverse Fourier transform operator.
To preserve the statistics of the correlation matrix,
we use a single $\mathcal{G}$-realization for all the detectors.
We include the effect of an elevation gradient and
a differential air-mass change in the line-of-sight,
\begin{equation}
 F_i = F_0 e^{\tau/\tau_0} \sec\left({\pi/2-\varepsilon_i}\right) ,
 \label{eq_atm_gradient}
\end{equation}
where $\varepsilon_i$ is the elevation track,
$\tau$ is the opacity,
$F_0\simeq 80$ mJy beam$^{-1}$
\footnote{1Jy $=10^{-26}$W m$^{-2}$ Hz$^{-1}$.}
and $\tau_0\simeq 0.06$
are typical flux and opacity normalization factors at 1.1 mm.
The noise levels resulting from our simulations
are usually smaller than real data by a factor
$\delta n_{\scalebox{0.5}{S}} \simeq 1.25-1.75$
(possibly because real atmospherical data
may contain additional patterns,
though we expect them to be less dominant).
Thus, we propagate this factor as
$W_{\scalebox{0.5}{S}} \leftarrow\delta n^{\scalebox{0.7}{-2}}_{\scalebox{0.5}{S}} W_{\scalebox{0.5}{S}}$,
allowing us to make a proper comparison
between simulation and real S/N.

The mock astrophysical data, as shown in figure \ref{fig_model},
is a group of 30 point sources labeled as $P$,
embedded in an extended source labeled as $E$.
$P$ resembles for example a population of SMGs,
while $E$ represents an extragalactic extended emission.
The point sources are randomly located Gaussian distributions,
spaced-out at least 5 times the beam size,
with fluxes between 4.5 and 8 mJy beam$^{-1}$.
For $E$ we use a facsimile of 30 Doradus in
the Large Magellanic Cloud \citep{Herschel14},
adapted for our interests:
smoothed with a 90 arcsec Gaussian kernel,
centered at the mean, and maximum flux of 10 mJy beam$^{-1}$.

Finally,
we map the mock astrophysical data back to time-domain,
using the GOODS-S scanning strategy and pointing information.
To reduce computation time,
we take advantage of the Lissajous scan continuity,
applying a custom bi-linear interpolation algorithm
to efficiently convert pixel information into mock timestreams.

\subsection{Reduction and decomposition of mock data}
\label{subsec_reduction_mock}

\begin{figure}
  \centering
    \includegraphics[width=0.47\textwidth]{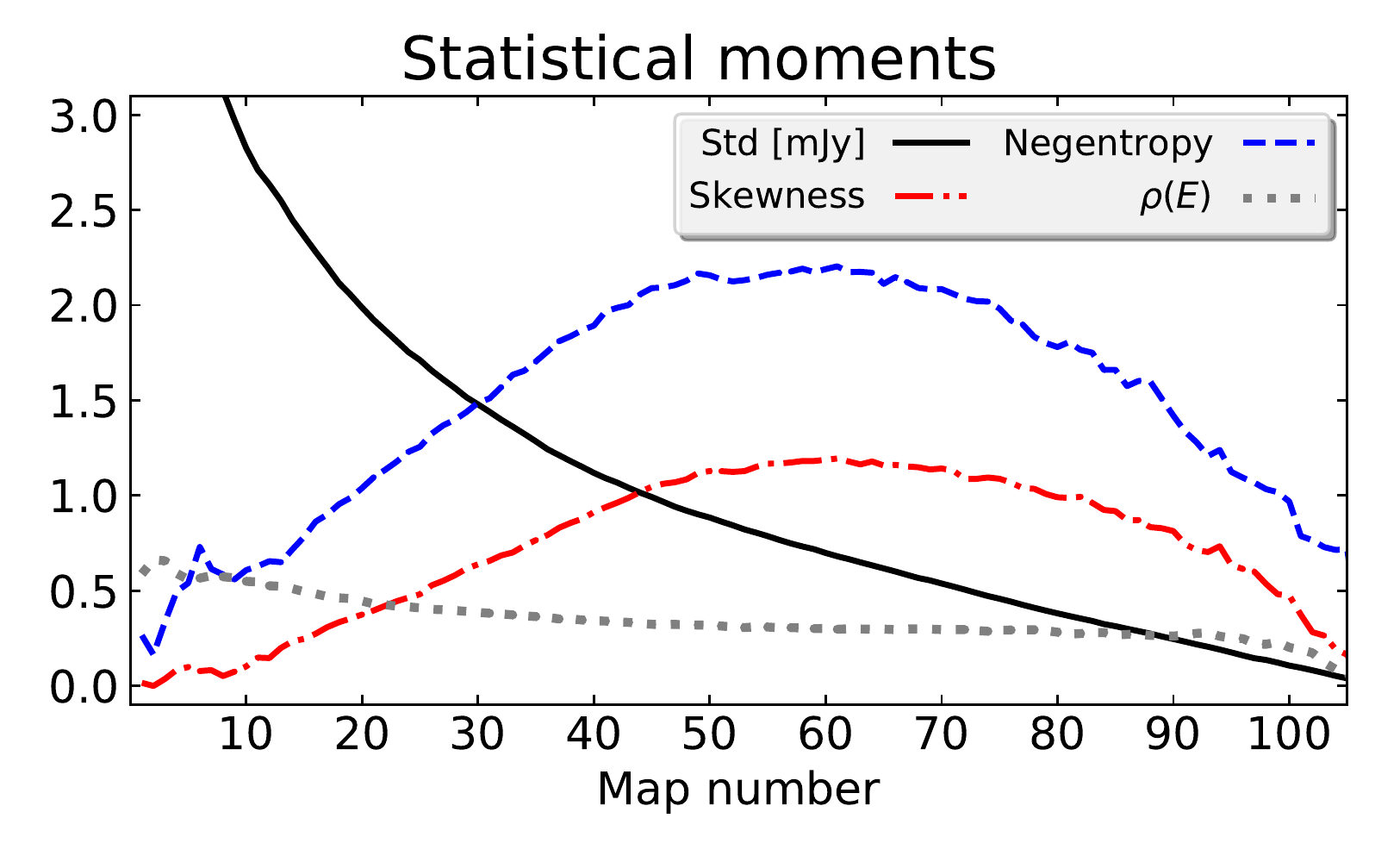} 
  \caption{ Statistical moments of
    (inner 270 arcmin$^2$ area) redundant maps.
    Negentropy is computed with equation (\ref{eq_negentropy_mine}),
    and $\rho(E)$ stands for pixel-correlation
    with the extended model $E$.}
  \label{fig_sim_stats}
\end{figure}

A total of $N_b =106$ redundant maps are produced
with the AzTEC pipeline,
and a few of them are shown in figure \ref{fig_sim_redundant_maps};
it is interesting to see their statistical moments
because they reflect their mixing degree.
We adopt the 270 arcmin$^2$ (50\% of uniform coverage)
map area to measure information and to perform our analyses.
In figure \ref{fig_sim_stats} we plot 
the standard-deviation (std), skewness,
and our approximation to negentropy.
Due to the dominant brightness of atmospheric emission, 
the std is high for the first map and falls off
to zero at the last map.
The most mixed map is $M_1$, correspondingly,
its negentropy and skewness are close to zero.
We also computed a set of maps with the atmospheric model only,
which helped us to confirm $M_{1 \leq i \leq 5}$
as the most atmospheric-contaminated maps.
However, atmospheric and extended emissions
are progressively removed from redundant maps,
increasing negentropy until a maximum around $M_{60}$;
we assert that $M_{60}$ is the least mixed map.
Afterwards, only point-like sources are left
but become gradually fainter,
until the last maps are dominated by nearly Gaussian noise;
likewise, negentropy and skewness fall off
close to zero at $M_{106}$.

\begin{figure}
  \centering
    \includegraphics[width=0.47\textwidth]{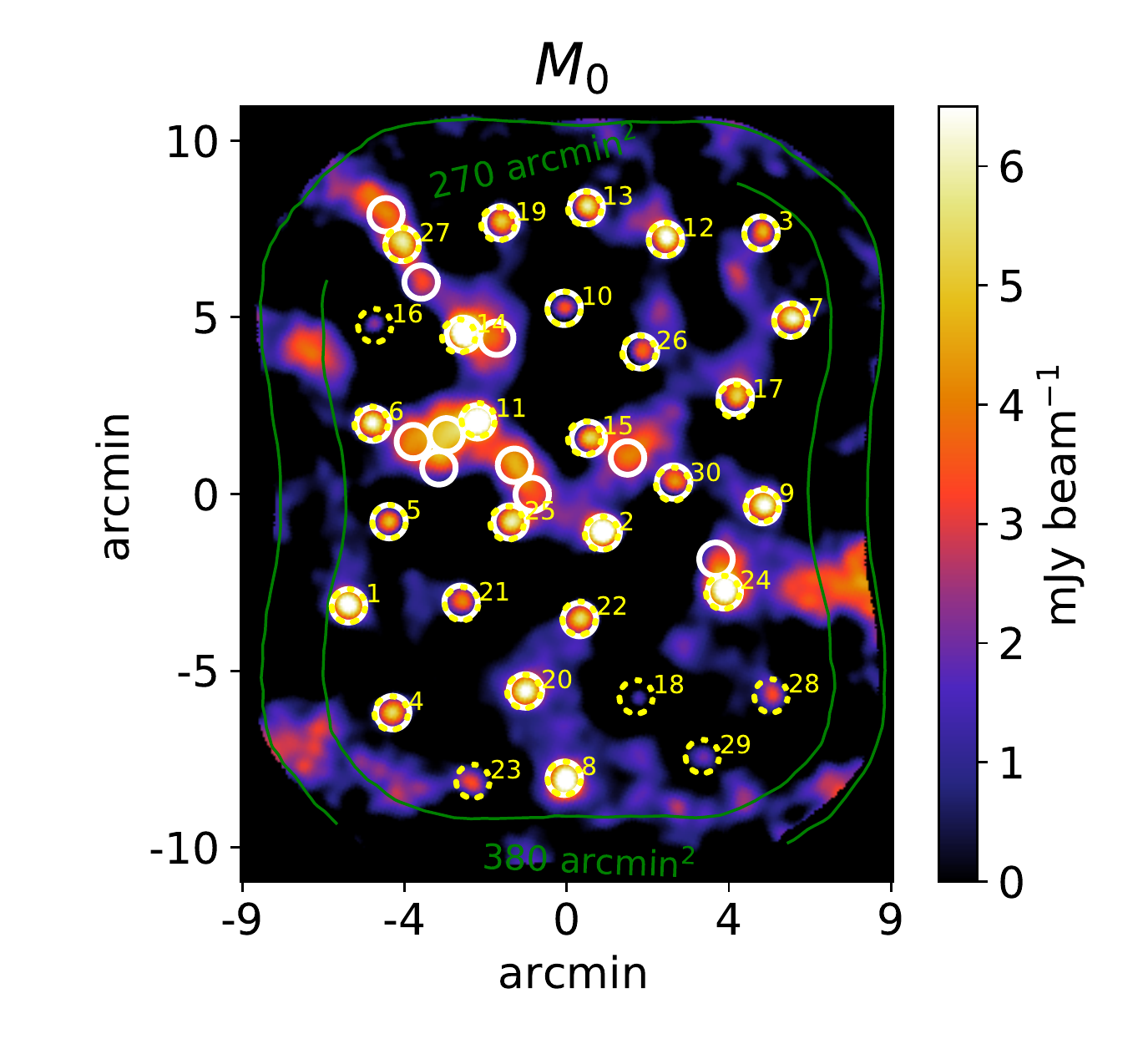} 
  \caption{ Reference $M_0$ map,
     computed with the \textsc{pca}2.5$\sigma$ procedure,
     described in \S\ref{subsec_macana}.
     Sources with S/N$>$4 are accounted as detections
     and circled with solid-white lines.
     The 30 initial mock point sources are numbered and
     circled with dotted-yellow.
     Notice that not all the initial/detected
     point sources are coincident,
     some detections correspond to bright patches
     of the extended model $E$.}
  \label{fig_sim_M0}
\end{figure}

We also perform the customary \textsc{pca}2.5$\sigma$ procedure
for point sources, as explained in \S\ref{subsec_macana}.
The resulting map $M_0$ can be used as a reference
to appreciate the leverage of our PCA-ICA technique,
compared to the simplest PCA approach in time-domain.
$M_0$ is shown in figure \ref{fig_sim_M0} and
its statistical moments are listed in table \ref{tabla_sim_comp}.
Indeed, we do not see bright atmospheric residuals in $M_0$,
like those evident in $\left\lbrace M_{1\leq i \leq 5} \right\rbrace$.
Still, we see important residuals from
the astrophysical extended model,
all mixed with the point sources.
As we mentioned, these astrophysical residuals can be harmful
because they bias measurements intended for compact sources.

We use the following set of FastICA parameters.
We choose to work with $n=4$ independent components;
this number is data-dependent,
but a good choice of $n$ must yield physically meaningful and robust solutions.
We test a large number of random initialization matrices,
and check that the solution preserves small negentropy dispersion.
We use the FastICA parallel algorithm and
the tolerance parameter tol=$10^{-12}$ \citep{Hyvarinen2000}.

\begin{figure}
  \centering
    \includegraphics[width=0.47\textwidth]{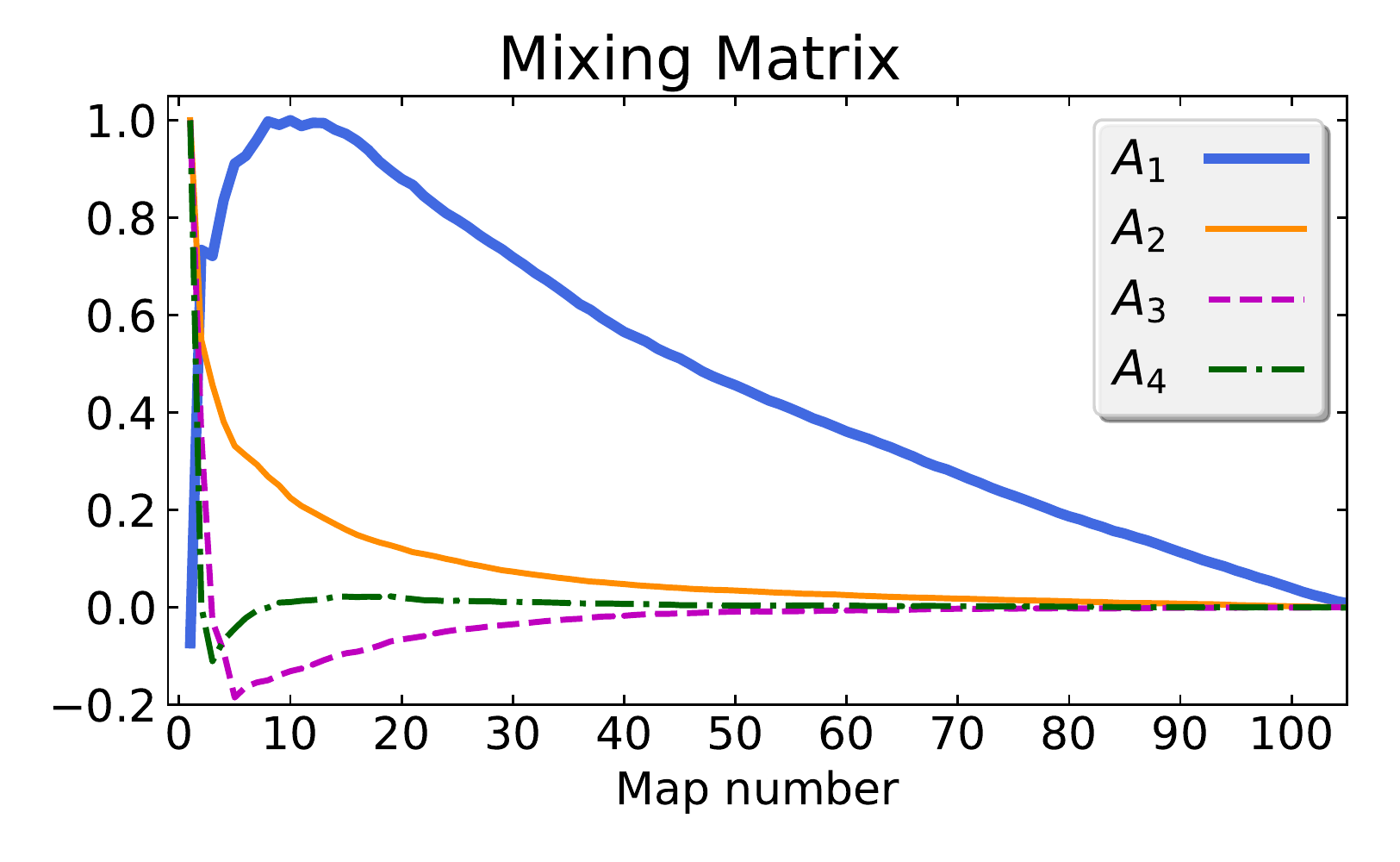} 
  \caption{ Mixing matrix coefficients from equation
   (\ref{eq_ica_maps}).
   The $j$-th component $S_j$ in figure \ref{fig_icasims}
   is coupled to the redundant maps
   $\left\lbrace M_{i} \right\rbrace$
   by the mixing coefficients
   $A_j = \left\lbrace a_{ij} \right\rbrace$.
  }
  \label{fig_sim_mix}
\end{figure}

\begin{figure*}
 \noindent\makebox[\textwidth]{
 \centering
  \begin{tabular}{cc}
    \includegraphics[width=0.47\textwidth]{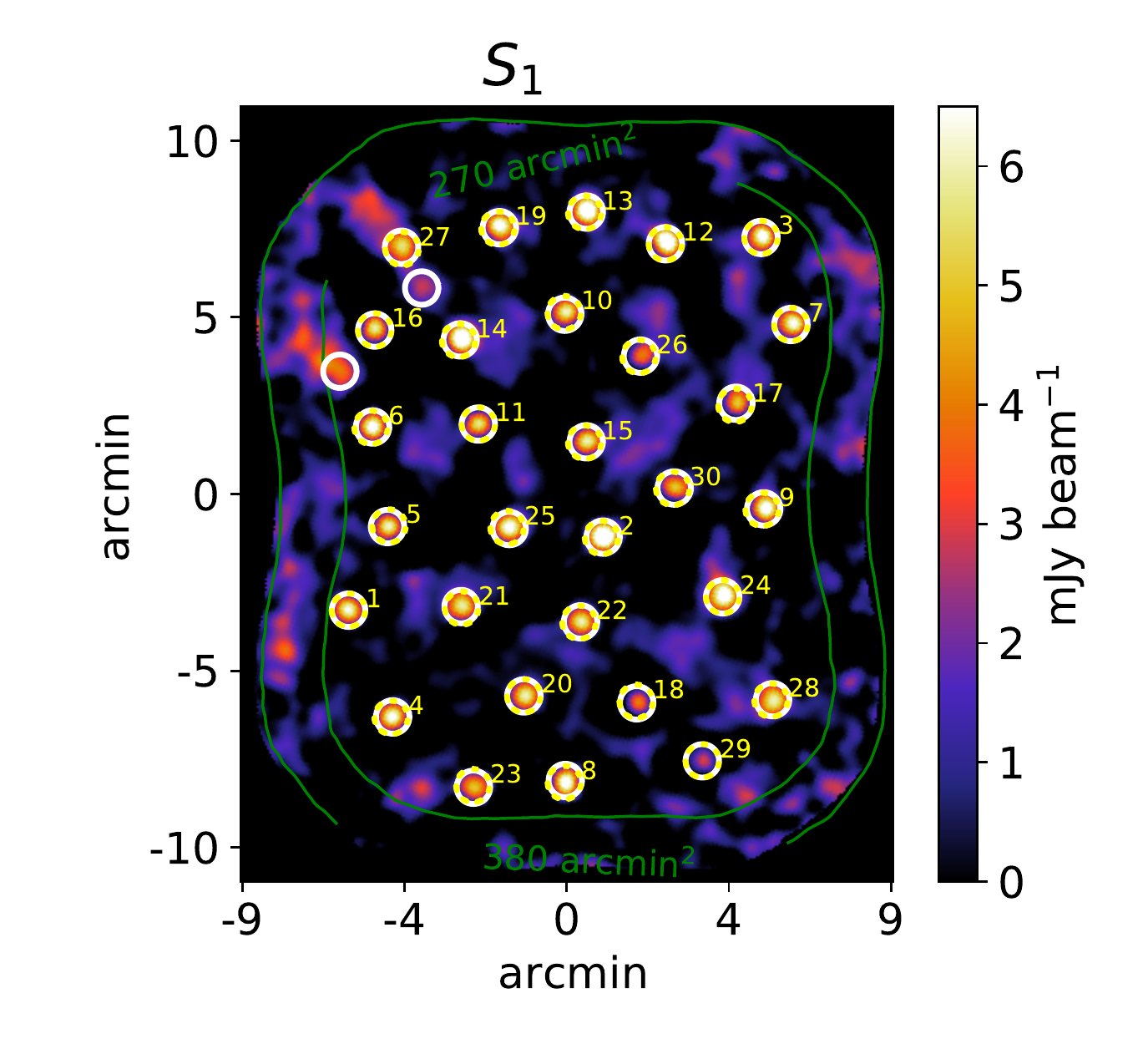} &
    \includegraphics[width=0.47\textwidth]{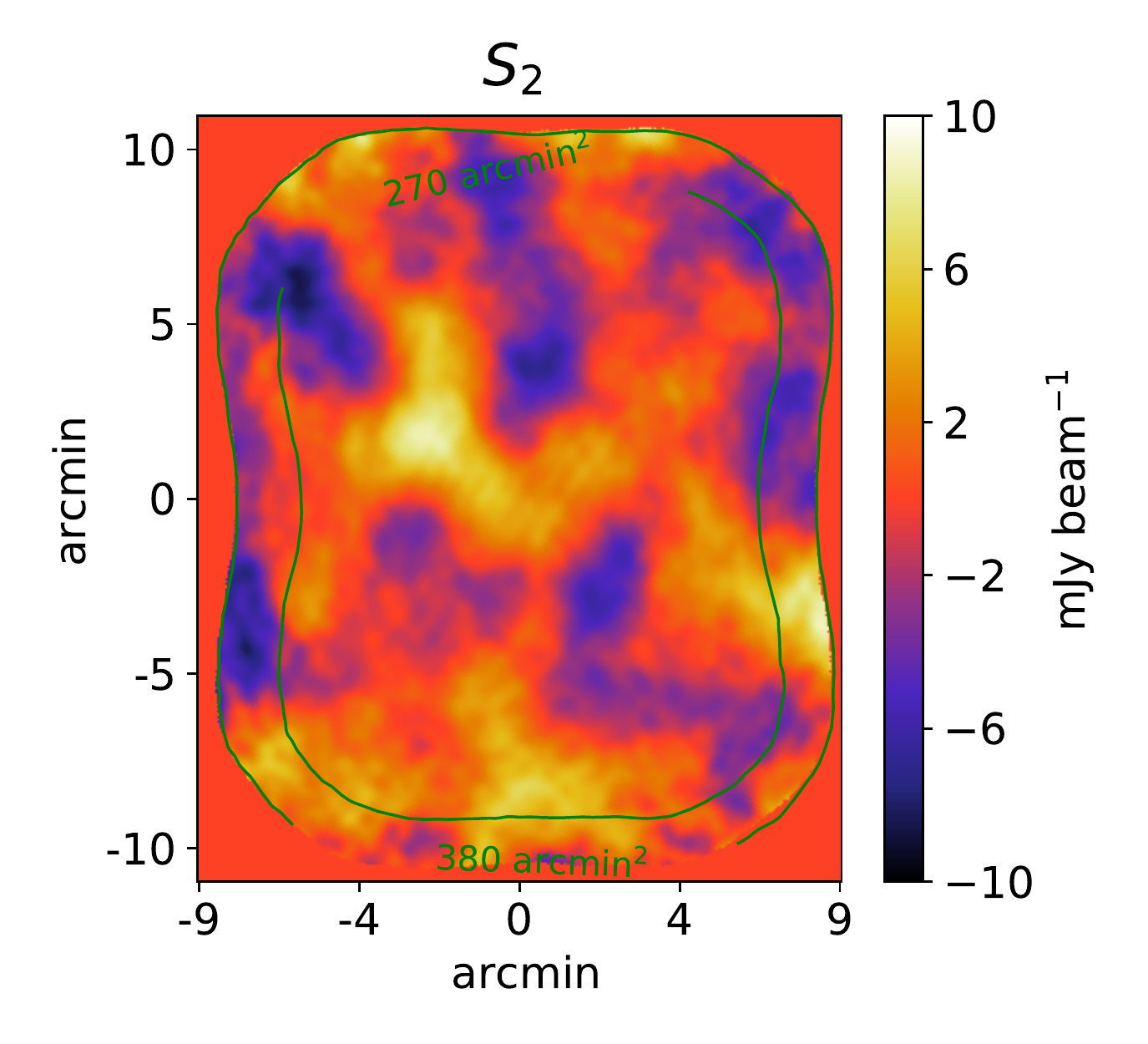} \\
    \includegraphics[width=0.47\textwidth]{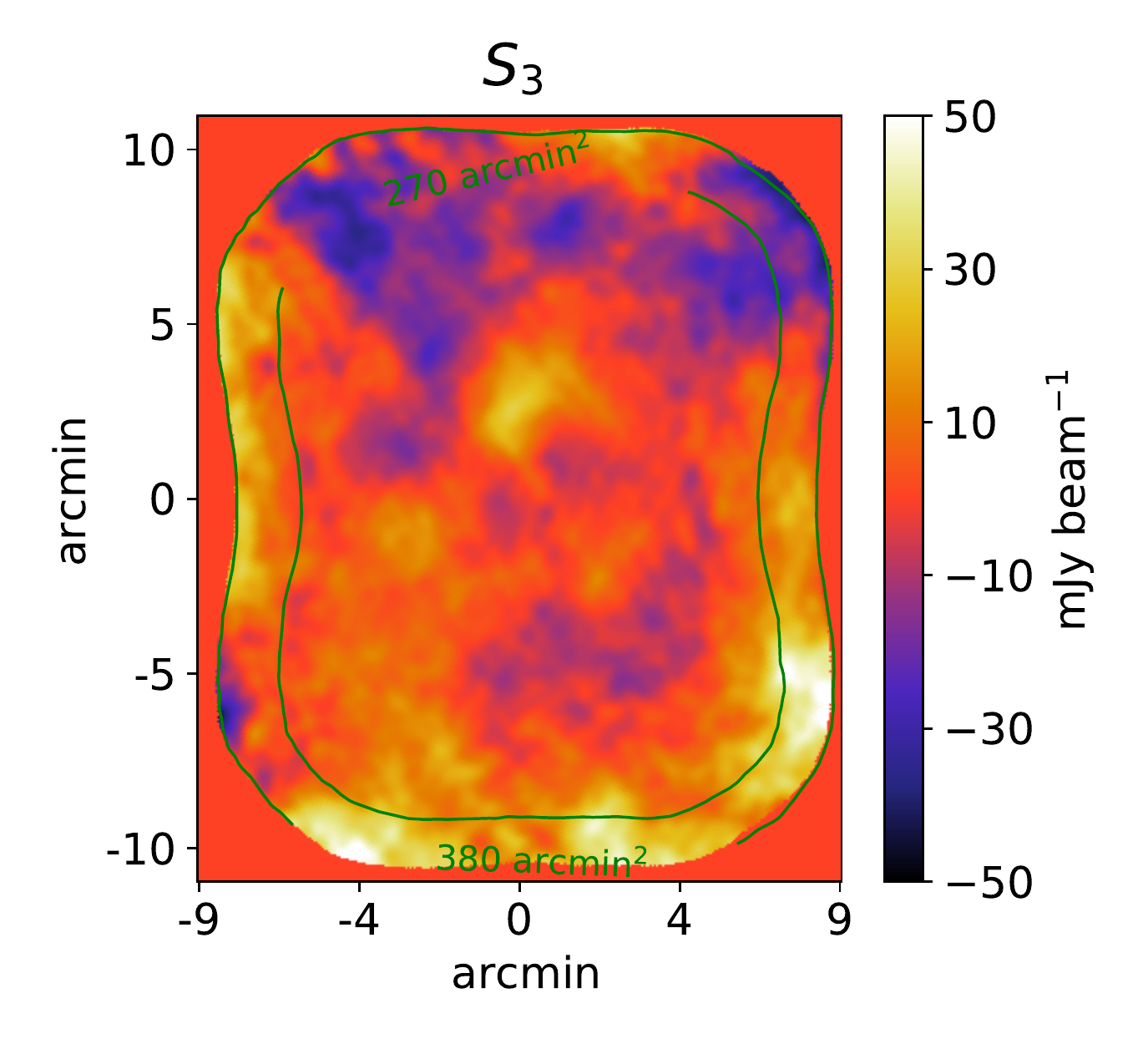} &
    \includegraphics[width=0.47\textwidth]{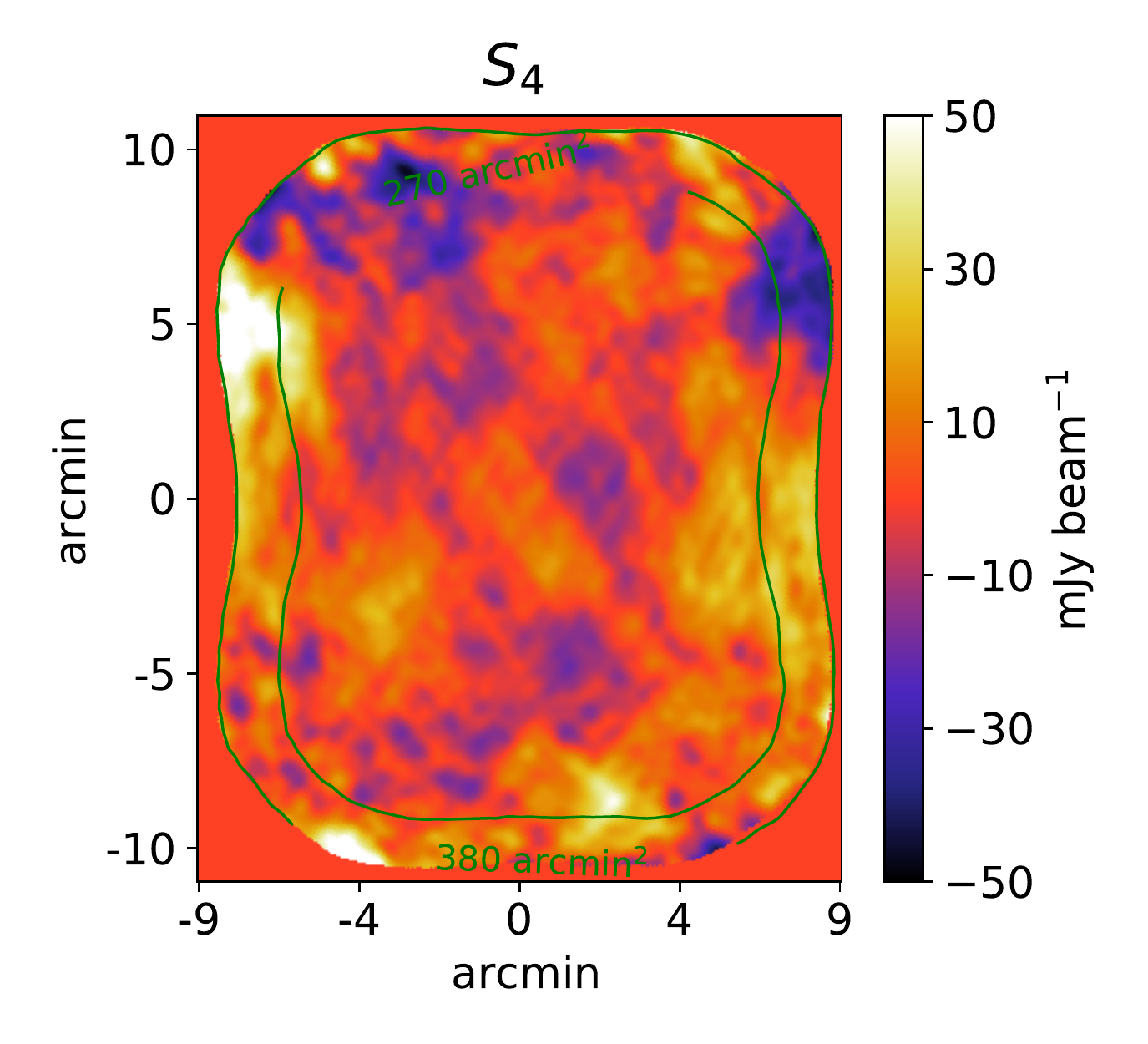} 
  \end{tabular}     }
  \caption{ Independent components decomposed
    from redundant maps as in equation (\ref{eq_ica_maps}),
    and calibrated as explained in \S\ref{subsec_calibration}.
	$S_1$ is interpreted as the point-like component of
    the astrophysical model.
	The white circles enclose S/N$>$4 detections and
    the yellow-dotted circles locate the original mock point sources;
	all but two detections are coincident with the point model.
	$S_2$ is interpreted as the extended component of
    the astrophysical model.
	The component $S_3$ is interpreted as an atmospheric foreground.
	The stripes featured in map $S_4$ are interpreted as
    effects of the Lissajous scan.
	Contours as in figure \ref{fig_model}.}
     \label{fig_icasims}
\end{figure*}

The decomposition results are both the mixing matrix
shown in figure \ref{fig_sim_mix} and the independent components
shown in figure \ref{fig_icasims}.
By visual inspection, $S_1$ can be easily identified with
the point model $P$ and $S_2$ with the extended model $E$.
$S_3$ is made of smooth bright fluctuations,
so we identify it as an atmospheric foreground.
$S_4$ looks less familiar because it contains symmetric stripes;
actually this pattern can be identified as an effect
due to the Lissajous scanning strategy.

\begin{figure}
  \centering
    \includegraphics[width=0.47\textwidth]{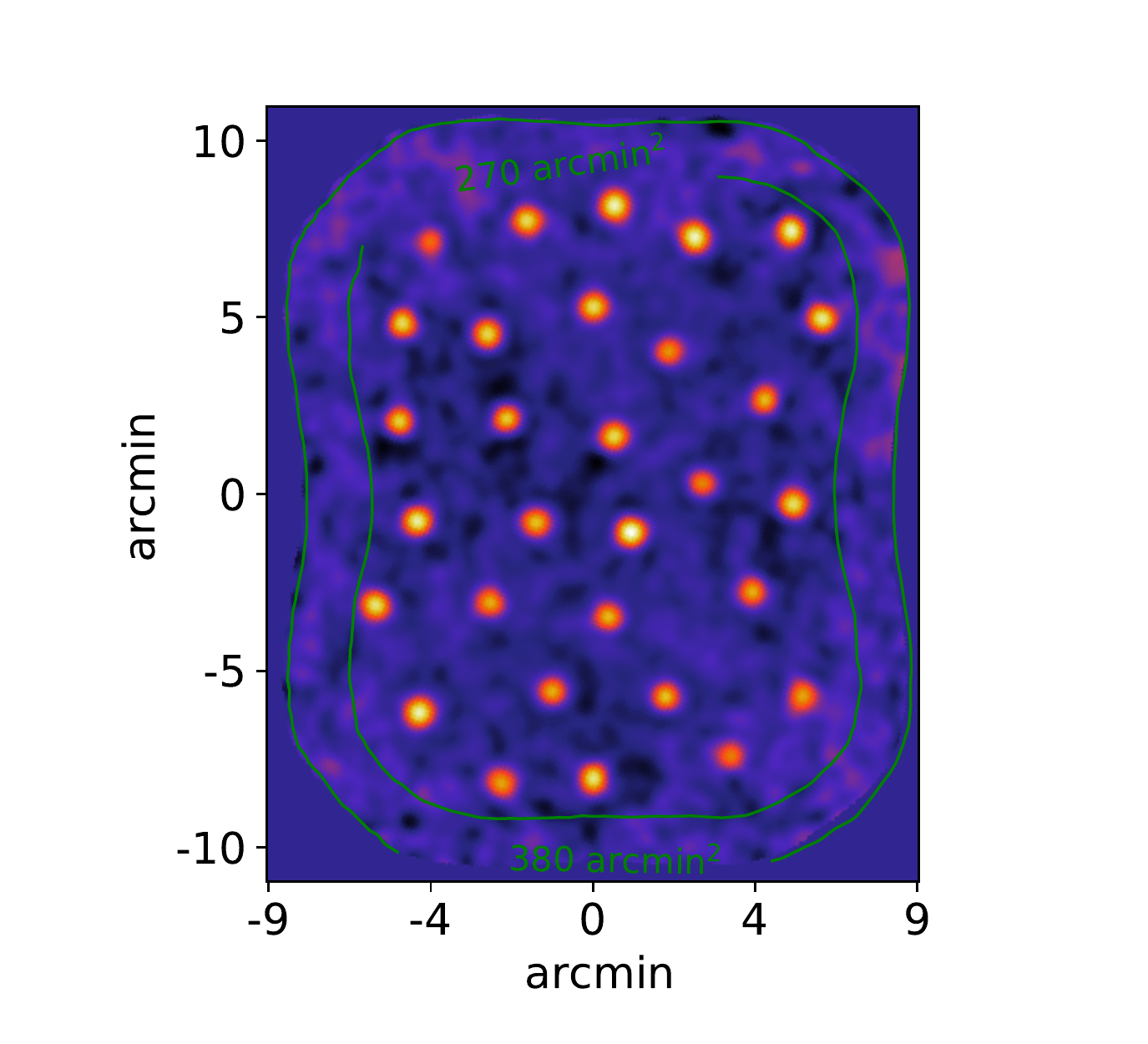} 
  \caption{ Point (uncalibrated) component
    decomposed from arbitrarily highly redundant maps
    (the sets $M_i$ and $N_i$ explained in the text).}
  \label{fig_inf_redund}
\end{figure}

In order to get an intuitive insight about
the usefulness of redundancy,
we also perform ICA over a highly redundant set of maps.
We produce another set of redundant maps denoted by
$\left\lbrace N_i \right\rbrace$;
they contain the same mock atmosphere and $E$, but not $P$.
We then perform an ICA decomposition of
the $2(N_b-1)$ redundant maps
$\left\lbrace M_i \right\rbrace$ and 
$\left\lbrace N_i \right\rbrace$,
and the results are remarkable:
as seen in figure \ref{fig_inf_redund},
the point sources are almost perfectly isolated.
The flux distribution is delta-like with a heavy positive tail
due to the point source signal;
hence, detections would have extremely high S/N.
Of course this degree of redundancy is unrealistic,
but at least from a qualitative point of view,
this idealization helps us to gain intuition
about what we may expect from a decomposition of redundant maps.

\subsection{Calibration of independent components}
\label{subsec_calibration}

Because of the ICA ambiguities,
the independent components need some calibrations
before being ready to extract physical information.
Below we discuss some calibration strategies.

\textit{The permutation ambiguity}
could be even trivially solved by eye
as we just did in the previous subsection,
but when the ICA decomposition is part of a pipeline,
we need an autonomous algorithm to identify
physical components \textit{on-the-fly}.
As can be seen in figure \ref{fig_sim_mix},
we find heuristically that the sum of
the $P$-mixing coefficients is always the largest,
followed by the $E$-mixing coefficients,
the atmospheric foregrounds, and the scanning pattern.
This effect is related to the typical angular scales
of the objects contained in each independent component.
We can use this hierarchy for blind identification of components.
Our main target is the point-sources component
because of the scale-calibration described below.
\citep[An alternative criterion to handle the permutation ambiguity
can be found in][]{Waldmann2012}.

The scaling ambiguity may be split into
\textit{sign} and 
\textit{absolute-scale} ambiguities.
To solve the sign ambiguity we demand that
$\sum_{i=1}^{N_b-1} a_{ij}>0$, for each $j=1,..,4$.
The intuitive reasoning is that the mixing coefficients $a_{ij}$
represent the degree of $S_j$ mixing into every individual $M_i$,
and the average mixing should be positive.
The absolute-scale ambiguity could be approached with the std criterion:
every independent component is scaled with 
the std of the most \textit{akin} redundant map.
With `most akin', we mean the $M_i$
whose pixel correlation with $S_j$ is maximum.
This inaccurate criterion might be useful only for visualization purposes,
with the atmospheric foregrounds for instance.

We also propose a scale calibration
using external artificial information.
We refer to a \textit{witness} as a mock source
similar to the physical signal of interest.
It is inserted into the redundant maps,
and retrieved after an ICA decomposition.
By definition, a witness must fulfill two conditions:
\textit{i}) it does not alter the statistical properties
of actual data, and
\textit{ii}) it is always recovered in the ICA map under calibration.
These conditions warrant that except for the scale,
the decomposition preserves the witness information.

For the point-source component,
a witness can be a two-dimensional Gaussian with beam-sized symmetrical widths.
First, 
we insert one witness into the redundant maps,
creating a new set of slightly perturbed
$\left\lbrace M'_i \right\rbrace$ maps.
The witness is wisely located in regions where
the flux distribution is locally uniform
and at least 3.5 beams away from any (mock) astrophysical point source.
We check that the general statistical properties of the redundant maps
are not modified by the insertion of a witness.
After $\left\lbrace M'_i \right\rbrace$ are ICA-decomposed,
the witness is always found in $S_1$.
We set to unit the scale of $S_1$ inside the 270 arcmin$^2$ area.
We fix the sign ambiguity and make a
bi-variate fit to the witness in $S_1$.
The scale factor is found by the ratio of
the actual witness flux to the resulting fitted flux.
The process is randomly repeated to generate a statistical distribution,
whose mean is the calibration-scale $a_1$,
and its std is added in quadrature as a systematic error.
Using about 10,000 witnesses, we find $a_1=1.44 \pm 0.06$
and the calibrated std is listed in table \ref{tabla_sim_comp}.
Then, the companion weight map $W_1$ can be computed
according to our discussion in \S\ref{subsec_macana}.

For the extended component $S_2$,
it is not obvious what kind of artificial signal
meets the conditions to be used as a witness.
Hence, a pixel-by-pixel fit to a subset of redundant maps $M_i$
seems a better approach for calibration.
We should consider the pixel correlation with $S_2$ and
discard the most atmospheric contaminated maps to choose that subset.
For this simulation,
we actually know the underlying (mock) astrophysical component,
and hence we can fit $S_2$ directly to $E$
in order to help us to choose the subset of maps for the fit,
according to their degree of pixel correlation with $S_2$. 
Following this approach we choose $M_{4<i<7}$,
whose correlations with $S_2$ are large ($\gtrsim 0.75$).
The fit is performed simultaneously for
the four independent components.
We use the outer 380 arcmin$^2$ area for calibration
(the same as for the ICA decomposition).
The resulting scale factor from this calibration is $a_2=2.78\pm0.46$,
and  the corresponding std is reported on table \ref{tabla_sim_comp}.

For comparison, from the same fit,
the scale factor for $S_1$ is found to be $1.48\pm 0.42$.
Hence, the calibration scales obtained from
the witness and fitting approaches
are consistent within the error bars.

\subsection{Inference of astrophysical information}
\label{subsec_astro_inference}

\begin{figure}
  \centering
    \includegraphics[width=0.47\textwidth]{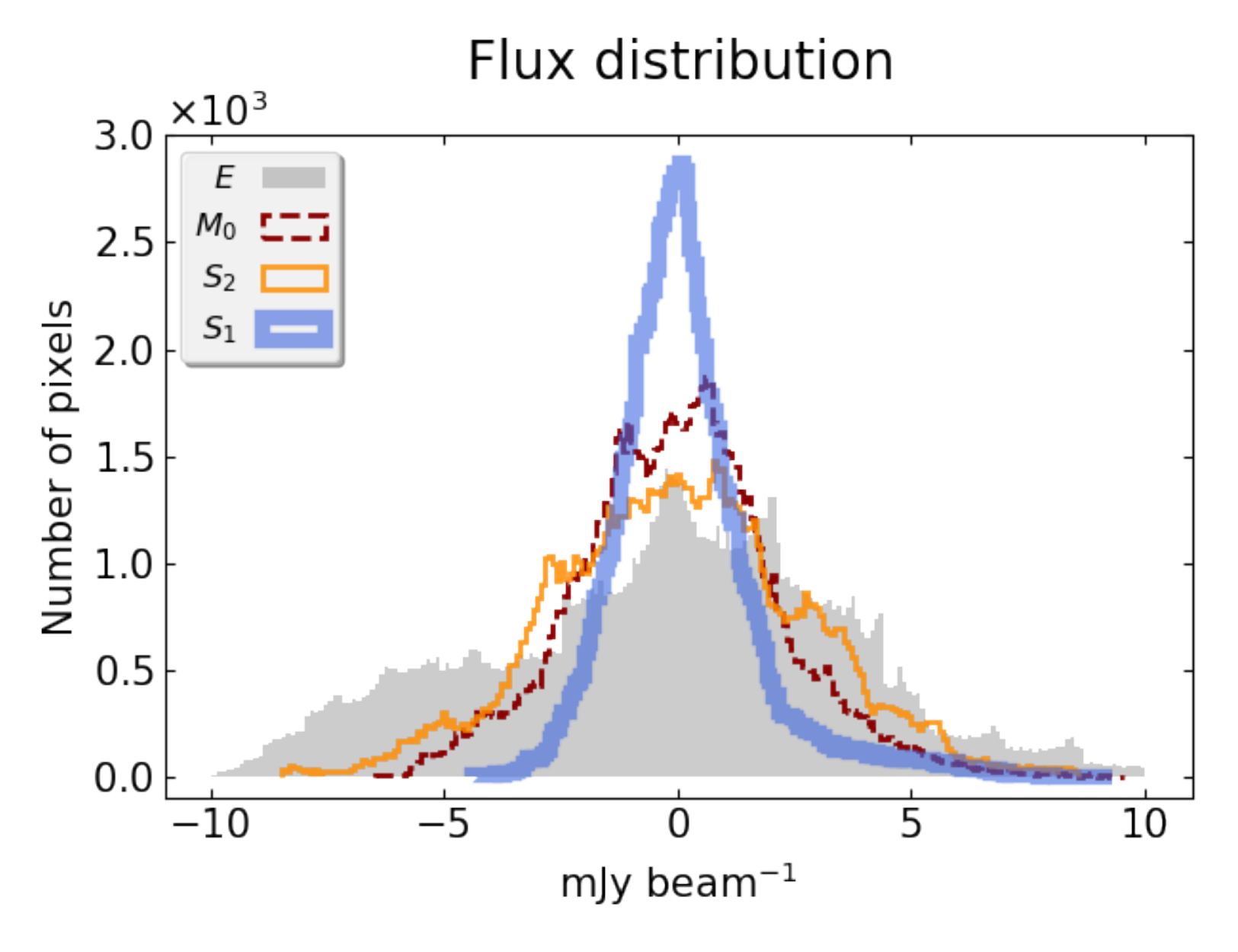} 
  \caption{Flux distribution within the inner 270 arcmin$^2$ 
  			region of the following maps:
  			The extended model $E$ in figure \ref{fig_model},
  			the reference map $M_0$ of figure \ref{fig_sim_M0},
  			and the independent components
            $S_1$ and $S_2$ in figure \ref{fig_icasims}.}
  \label{fig_sim_histos}
\end{figure}

After the redundant maps are decomposed
and the corresponding calibrations are performed,
we can analyze the $M_0$, $S_1$, and $S_2$ maps in more detail.
The statistical properties of these maps can be read
from table \ref{tabla_sim_comp} and seen on their
flux distribution in figure \ref{fig_sim_histos}.

\begin{table}[h]
\centering
\caption{Statistical properties per simulation map}
\begin{tabular}{cccccc}
	Map		&	Std [mJy]	&	Skewness	&	Negentropy	& $\rho(P)$ & $\rho(E)$ \\
	\hline 
	$M_0$	&	2.18		&	0.27		&		0.54	&	0.47	&	0.48 	\\
	$S_1$	&	1.51		&	1.26		&		2.15	&	0.70	&	0.12	\\
	$S_2$	&	2.61		&  0.05		&		0.10	&	0.08	&	0.71	\\
	\hline
\end{tabular}
\tablecomments{	Statistical moments quantified within
  270 \textsl{arcmin}$^2$ area.
  Negentropy is computed with equation
  (\ref{eq_negentropy_mine}),
  $\rho(P)$ stands for the pixel-correlation
  with the point-model, and $\rho(E)$
  for the correlation with the extended model.}
  \label{tabla_sim_comp}
\end{table}

The point-source measurements are improved after our decomposition.
In table \ref{tabla_sim_comp},
all the statistical properties of $S_1$ are improved
compared to $M_0$.
Negentropy, in particular, indicates that
$S_1$ contains more information than $M_0$; as a consequence,
we should expect a S/N boost in $S_1$ compared to $M_0$.
Furthermore, $M_0$ evidently contains extended emission residuals,
so we could expect more biased flux measurements in $M_0$.
We also read in table \ref{tabla_sim_comp} that $M_0$ is
equally correlated to the punctual $P$ and extended $E$ models,
whereas $S_1$ is mostly correlated with $P$.
From the histograms in figure \ref{fig_sim_histos}, we see that
the flux distribution of $M_0$ is relatively flat,
resembling the distribution of the extended model $E$.
Conversely, the $S_1$ is sharply peaked at its mean,
with a hard positive tail corresponding to the signal of point-sources;
as expected from our discussion in \S\ref{subsec_reduction_mock}.

The extended source is fairly isolated after decomposition.
The $S_2$ flux distribution in figure \ref{fig_sim_histos}
is much more spread out than $S_1$, actually, it resembles
the flux distribution of $E$ more closely than $M_0$.
Because $S_2$ is mostly correlated with the extended model $E$,
but negligibly correlated to the point model $P$
(table \ref{tabla_sim_comp}), we can assert that
the astrophysical information contained in $S_2$
is not contaminated by the point sources.

\begin{figure}
  \centering
    \includegraphics[width=9cm]{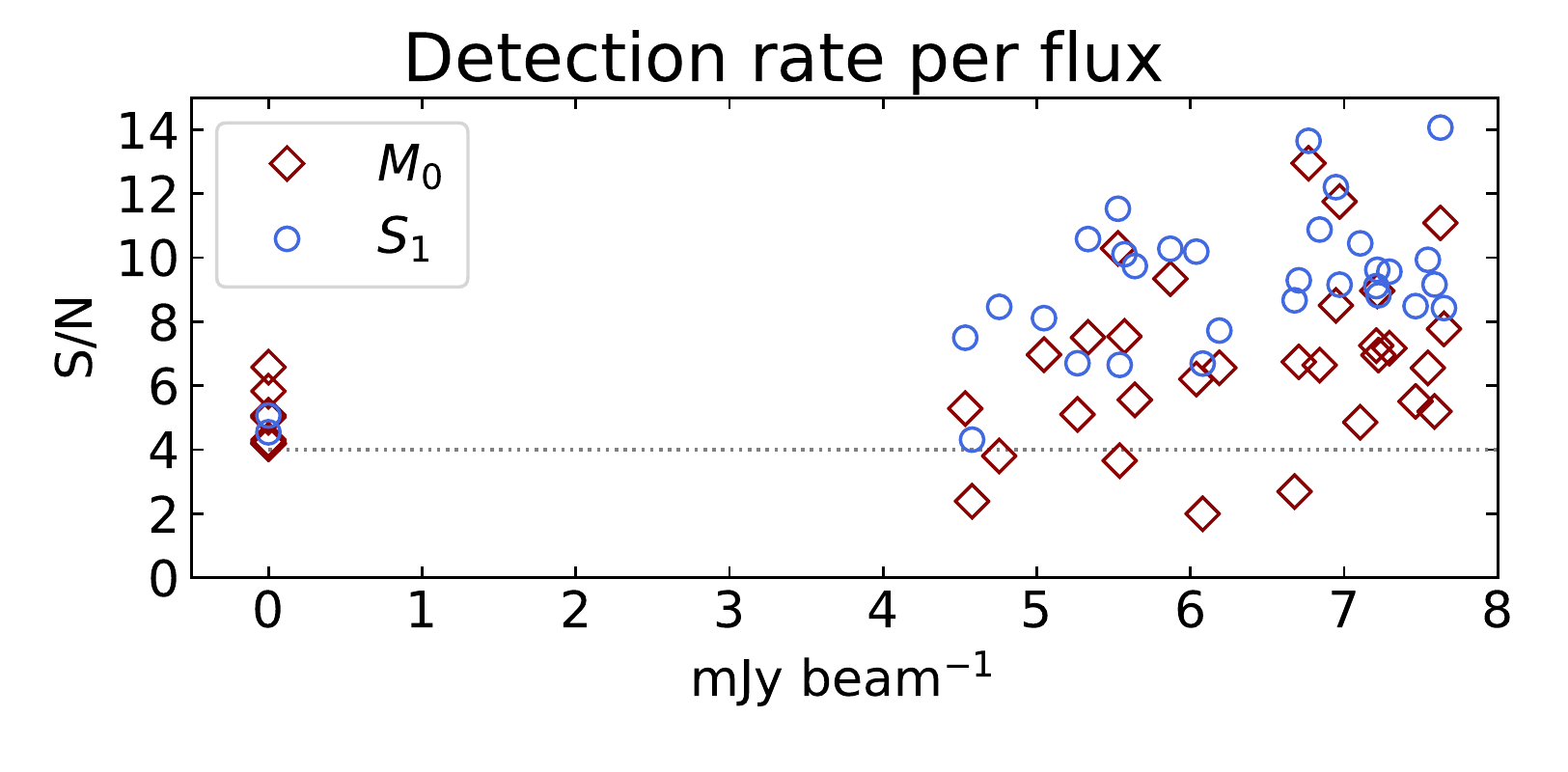} 
  \caption{ Detection rate (S/N) of point sources
    as a function of flux.
    Sources with S/N$>$4 are counted as detections:
    10 (2) detections on $M_0$ ($S_1$)
    do not match the point model.}
  \label{fig_sim_SNrate}
\end{figure}

Now we can detect point sources as described in
\S\ref{subsec_macana}.
In $M_0$, we detect 35 bright sources with S/N$>$4;
from which, 10 do not match any mock point sources,
and 5 of the 30 mock point sources are not detected in $M_0$.
On the other hand,
in $S_1$ we detect 32 bright sources with S/N$>$4, from which, 
only 2 sources (with S/N$\lesssim$5) do not match $P$.
Inside 270 arcmin$^2$,
the level of noise is not abruptly varying,
so that one would expect that the S/N depends
primarily on the flux of the point source.
In figure \ref{fig_sim_SNrate} we plot
the detection rate as a function of point source flux.
The sources detected in $S_1$ seem to follow the expected trend,
while in $M_0$ some bright sources ($>5$ mJy) are found with very low S/N.
These anomalies in $M_0$ can be attributed to the evident 
extended emission residuals in figure \ref{fig_sim_M0}.

\begin{figure}
  \centering
    \includegraphics[width=0.47\textwidth]{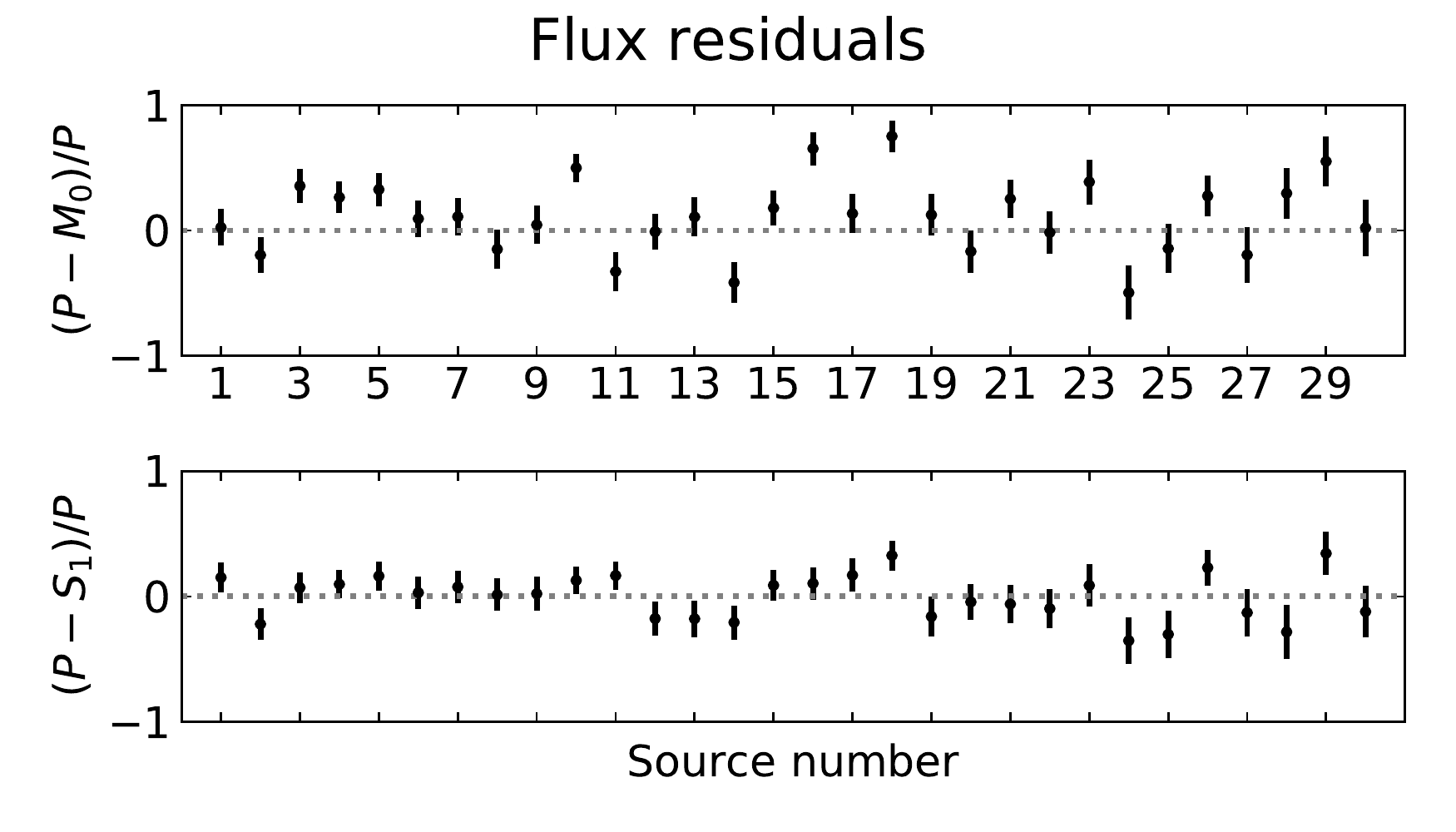} 
  \caption{ Point-source flux residuals:
	the model flux minus the measured flux
    in the $M_0$ and $S_1$ maps.}
  \label{fig_sim_flux_comp}
\end{figure}

We also check the point source flux recovery.
Indeed, flux biases are not unexpected after the reduction process;
for example, the Gaussian filter could smear fluxes in the map.
Usually, these biases are negligible for blank fields
(see next section),
but in the presence of an extended source like in our mock data,
the bias can be a problem to deal with.
In figure \ref{fig_sim_flux_comp},
we show the residuals between initial and measured fluxes
for $M_0$ and $S_1$ respectively.
The error bars are computed as usual
 (\textit{e.g.}, for the residual $R=(F_1 - F_2)/F_1$,
 the error is
 $\sigma_R = F_2/F_1 \sqrt{(\sigma_1/F_1)^2+(\sigma_2/F_2)^2}$).
 For the $P$ known point fluxes, we use the effective sensitivity
 $\sigma_{\text{eff}}\approx 0.58 \text{ mJy}$,
 taken from the weight map $W_1$ (see \S\ref{subsec_macana}).
In $M_0$ we measure a significant rms of 0.31 mJy,
again attributed to the remains of the extended emission.
Contrarily, $S_1$-residuals are much less scattered,
with an rms deviation of 0.19 mJy,
a behavior closer to the expectation from a blank field.
(Notice that these values account for statistical errors only.)
Moreover, our $S_1$-scale calibration gets reassured;
any (positive or negative) tendency would hint a fail in calibration.
Satisfyingly, the points are roughly
symmetrically distributed around zero,
hence, indicating an accurate $S_1$-scale calibration.

\begin{figure}
  \centering
    \includegraphics[width=0.47\textwidth]{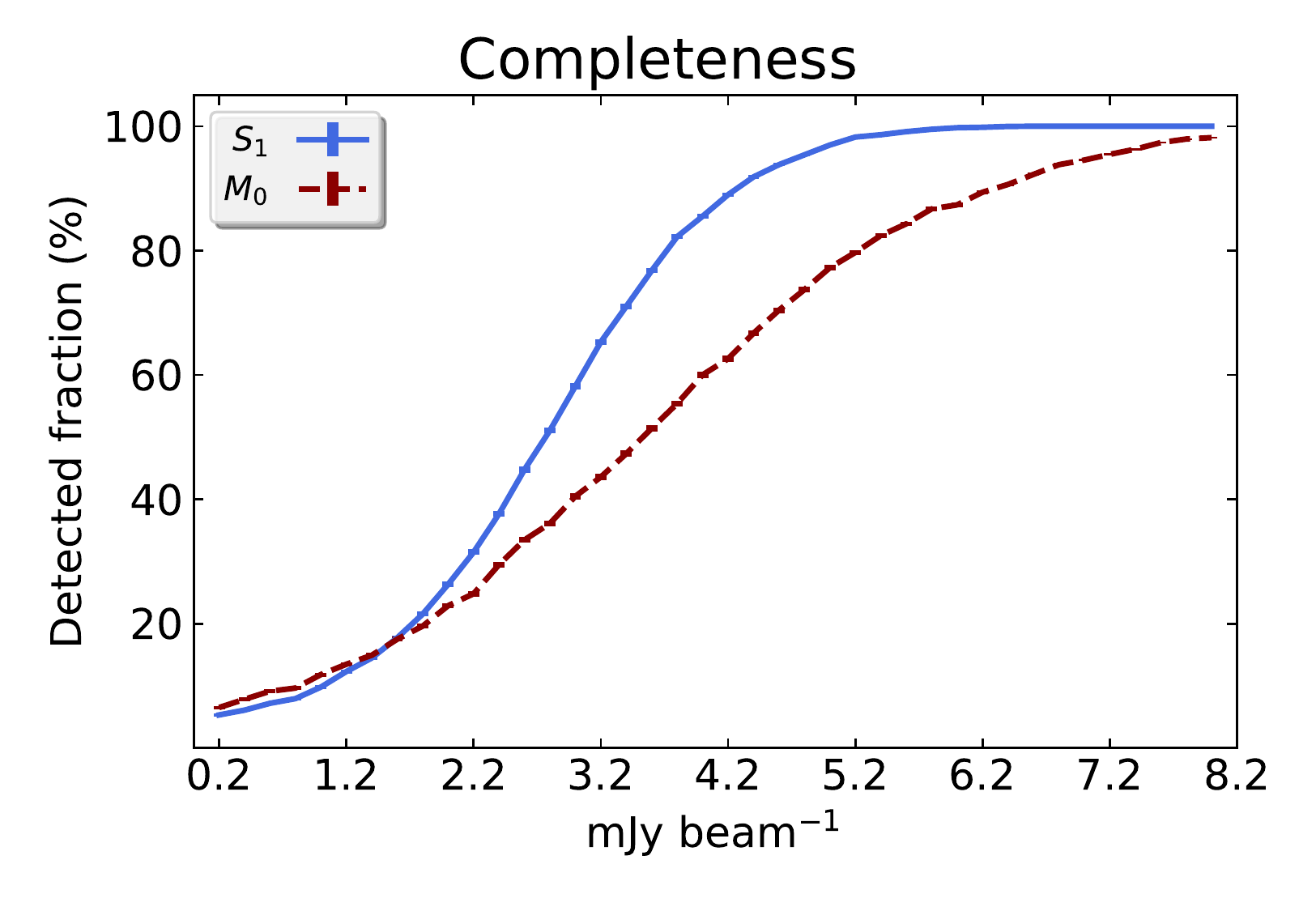} 
  \caption{ Completeness of $M_0$ and $S_1$ maps.}
  \label{fig_sim_complete}
\end{figure}

The detection rate in figure \ref{fig_sim_SNrate} is representative
only of the particular set of point sources in our mock data.
A deeper characterization of $S_1$ and $M_0$
requires a larger sample of point sources,
and is known as the \textit{completeness} of the map.
To compute it, we insert one additional point source,
excluding the locations of the 30 initial point sources
and surrounding (beam-radius) areas.
The artificial source is said to be recovered if it is found
with S/N$>$4 around a circle of half-beam radius.
For each equidistant flux step,  we insert 10,000
point sources (one at a time) at random locations.
We estimate the error bars assuming a binomial distribution.
The results for $M_0$ and $S_1$ are shown in figure \ref{fig_sim_complete}.

\begin{figure}
  \centering
    \includegraphics[width=0.47\textwidth]{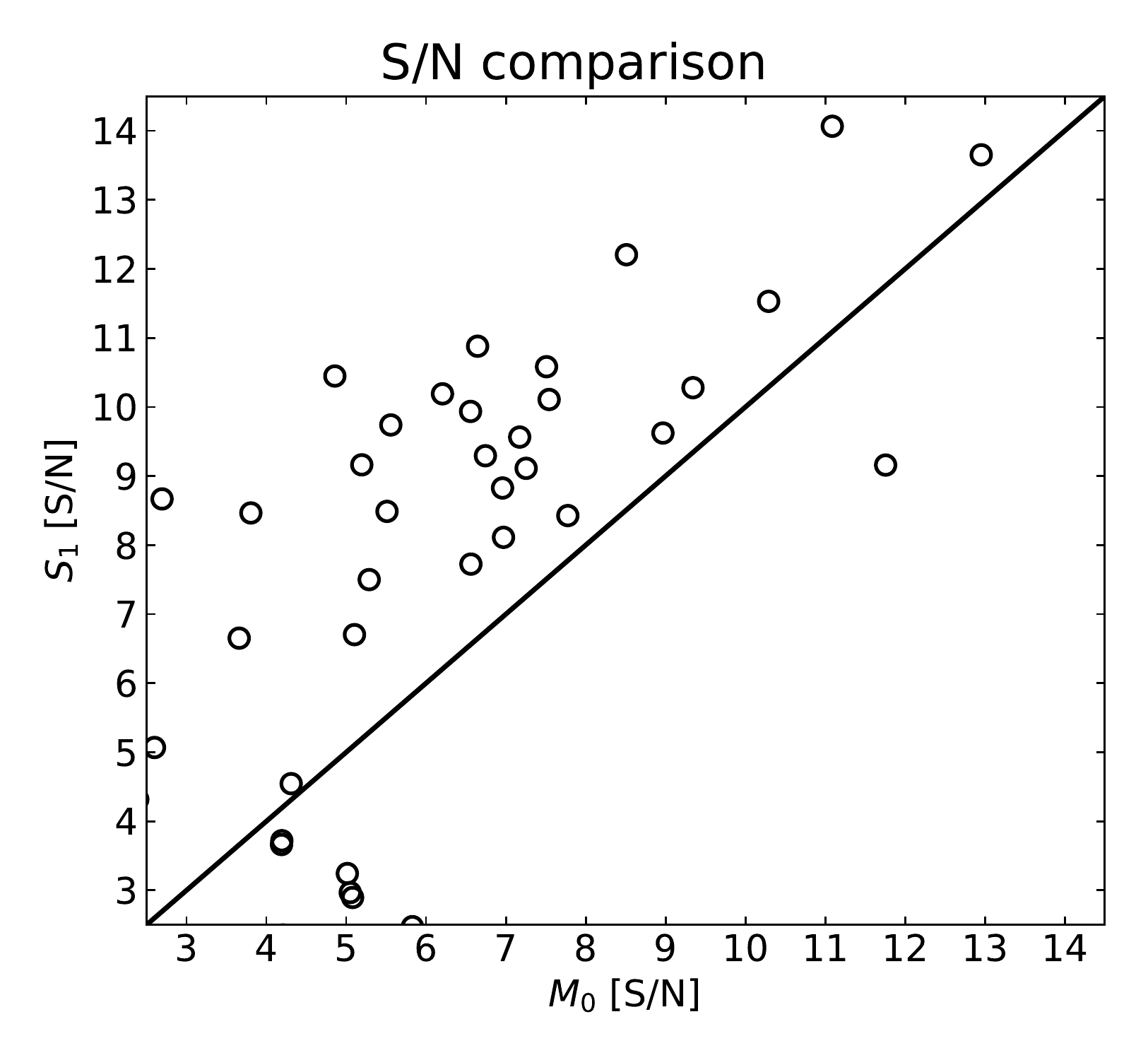} 
  \caption{ S/N comparison between point sources detected on $M_0$ or $S_1$.}
  \label{fig_sim_sncomp}
\end{figure}

Finally, as expected from table \ref{tabla_sim_comp},
the point sources in $S_1$ are detected with higher S/N than
in $M_0$, (see figure \ref{fig_sim_sncomp}).
Only false point-source detections show higher S/N in $M_0$.

\section{Decomposition of real data}
\label{sec_real}

Our next step is to apply our techniques to a set of real data
with the aim to check the recovering of previous results.
For that, we revisit the AzTEC/ASTE GOODS-S survey,
which is considered to be a blank field.
%A \textit{blank field} is a region of the sky
%representative of the average surface density of SMGs
%in the Universe, where no overdensities (or underdensities)
%on the SMG population are known \textit{a priori}.
For that reason, this survey has been extensively studied
and used as a trial data set for extensions to the AzTEC pipeline
\citep{Scott2010,Scott2012,Downes2012,Yun2012}.
Although astrophysical foregrounds are not expected,
these observations were certainly contaminated by
bright atmospheric foregrounds.
Hence, the AzTEC/ASTE GOODS-S survey is ideal for our purposes,
mainly to test our calibration strategies on point-like sources.
Besides, our previous simulations were purposely designed
very similar to the AzTEC/ASTE GOODS-S survey,
which consists of 74 observations, each one containing
$N_b$=106 effective bolometer timestreams.
Then, we can apply the same methodology as in \S\ref{sec_mock}.

The AzTEC/ASTE GOODS-S observations reported
a 1$\sigma$ depth of about $0.48-0.73$ mJy beam$^{-1}$,
which is below the estimated confusion background limit
of 2 mJy beam$^{-1}$ \citep{Scott2010}.
The confusion background is
the sea of faint unresolved sources in the sky,
creating an extended emission that can potentially bias
detections below the confusion background limit \citep{Hogg2001}.
Unfortunately, this uncertainty cannot be reduced
by increasing the observation time; however,
as the confusion background should be non-Gaussianly distributed,
then, an ICA decomposition is not precluded \textit{a priori}.

\begin{figure}
  \centering
    \includegraphics[width=0.47\textwidth]{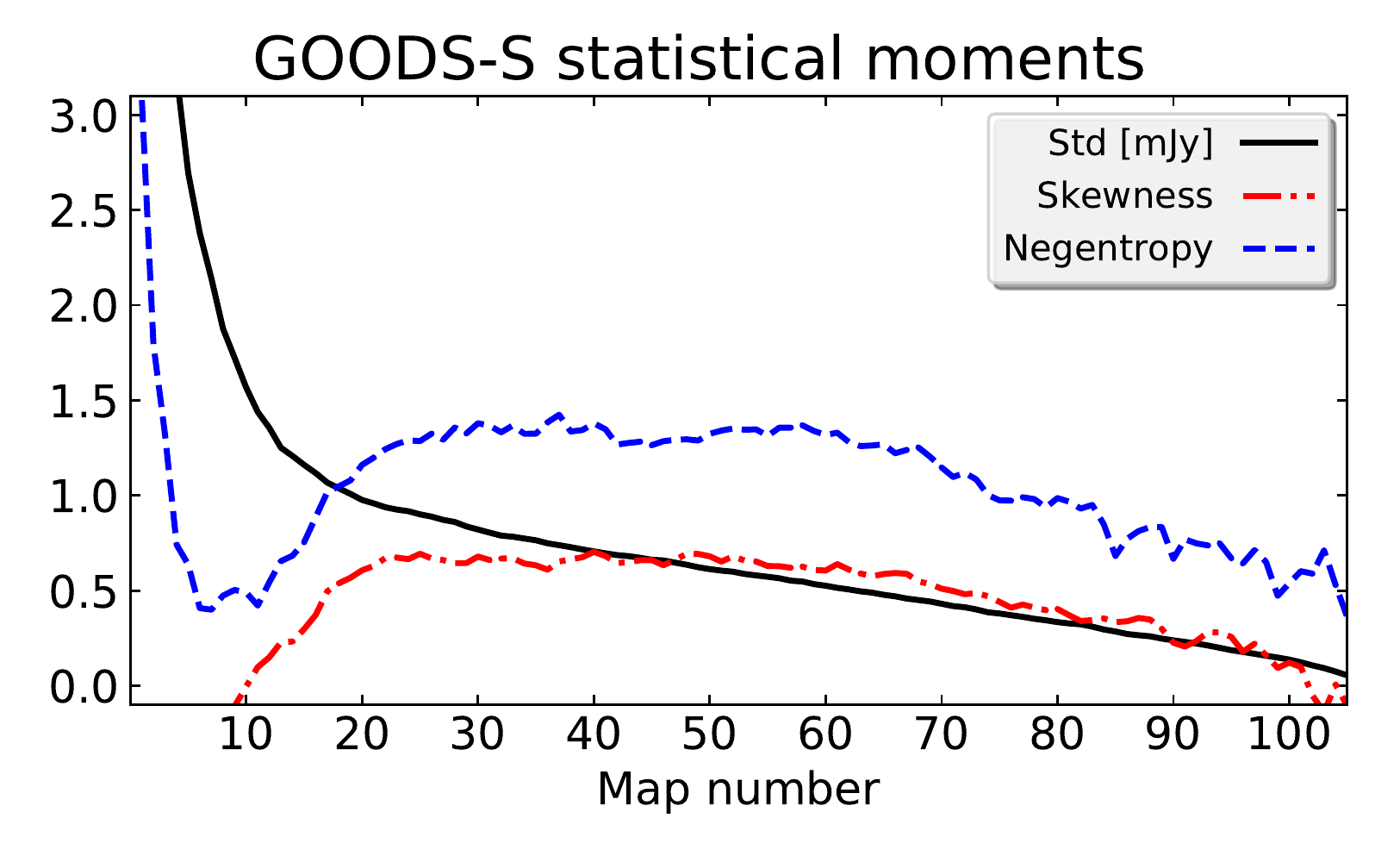} 
  \caption{Statistical moments of GOODS-S redundant maps,
  				analogous to figure \ref{fig_sim_stats}.}
  \label{fig_stats}
\end{figure}

\begin{figure}
	\centering
    \includegraphics[width=0.47\textwidth]{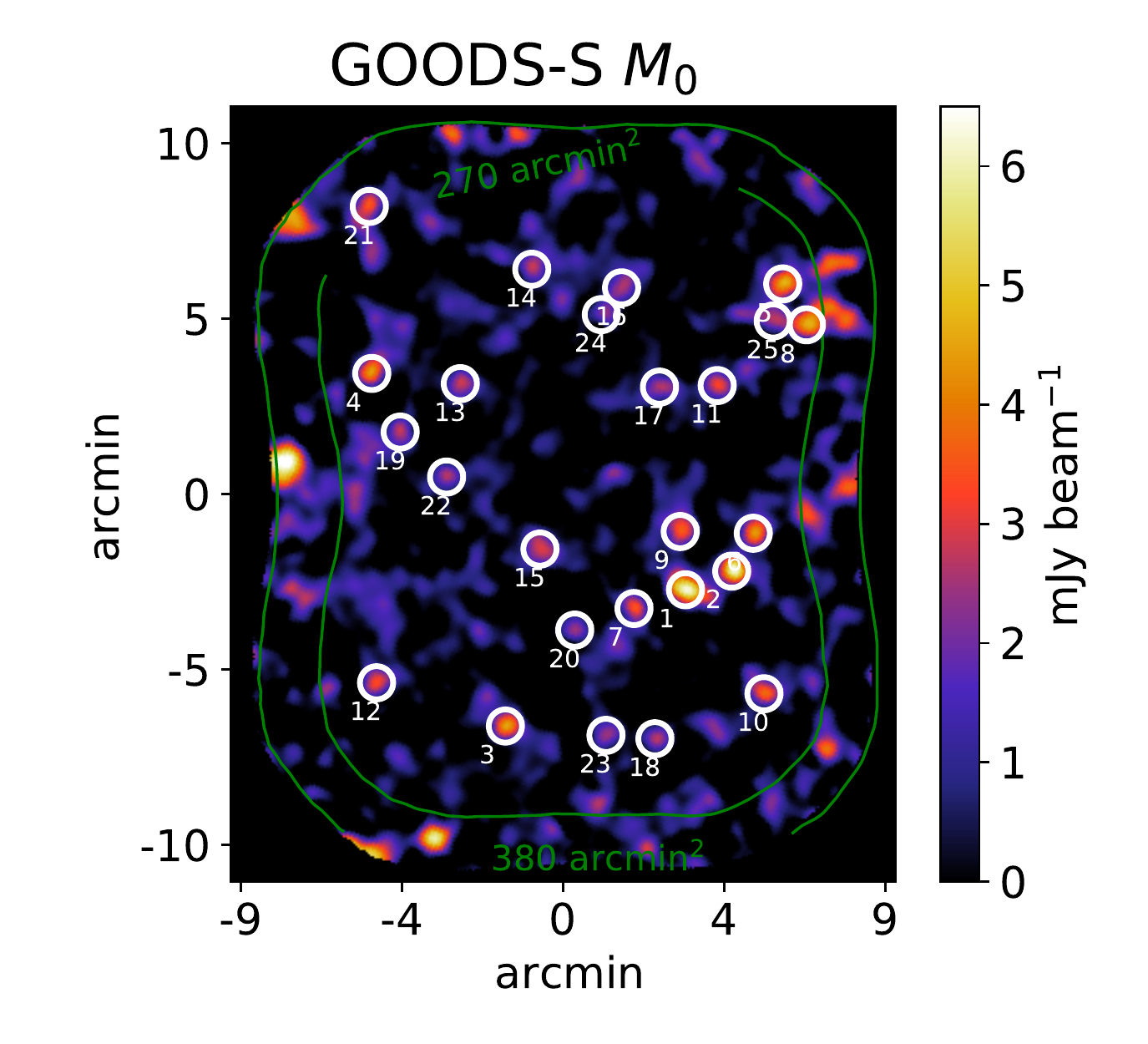} 
  \caption{ GOODS-S reference map $M_0$ computed with
    the \textsc{pca}2.5$\sigma$ procedure,
    described in \S\ref{subsec_macana}.
    Sources with S/N$>$4 are accounted as detections
    and circled with solid-white lines.
    External contours as in figure \ref{fig_model}.}
  \label{fig_m0}
\end{figure}

\begin{figure*}
 \noindent\makebox[\textwidth]{
 \centering
  \begin{tabular}{ccc}
    \includegraphics[width=0.33\textwidth]{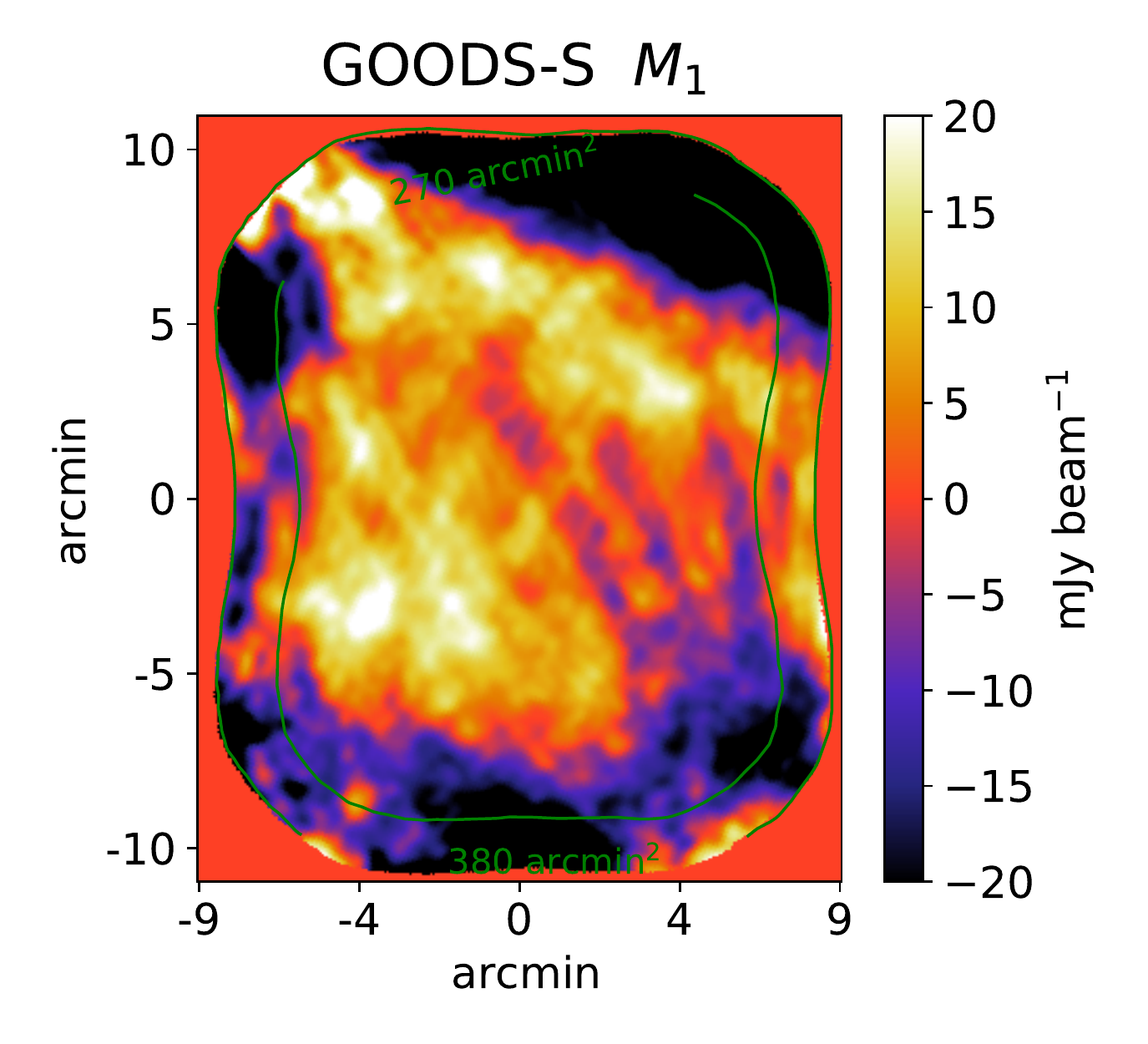} &
    \includegraphics[width=0.33\textwidth]{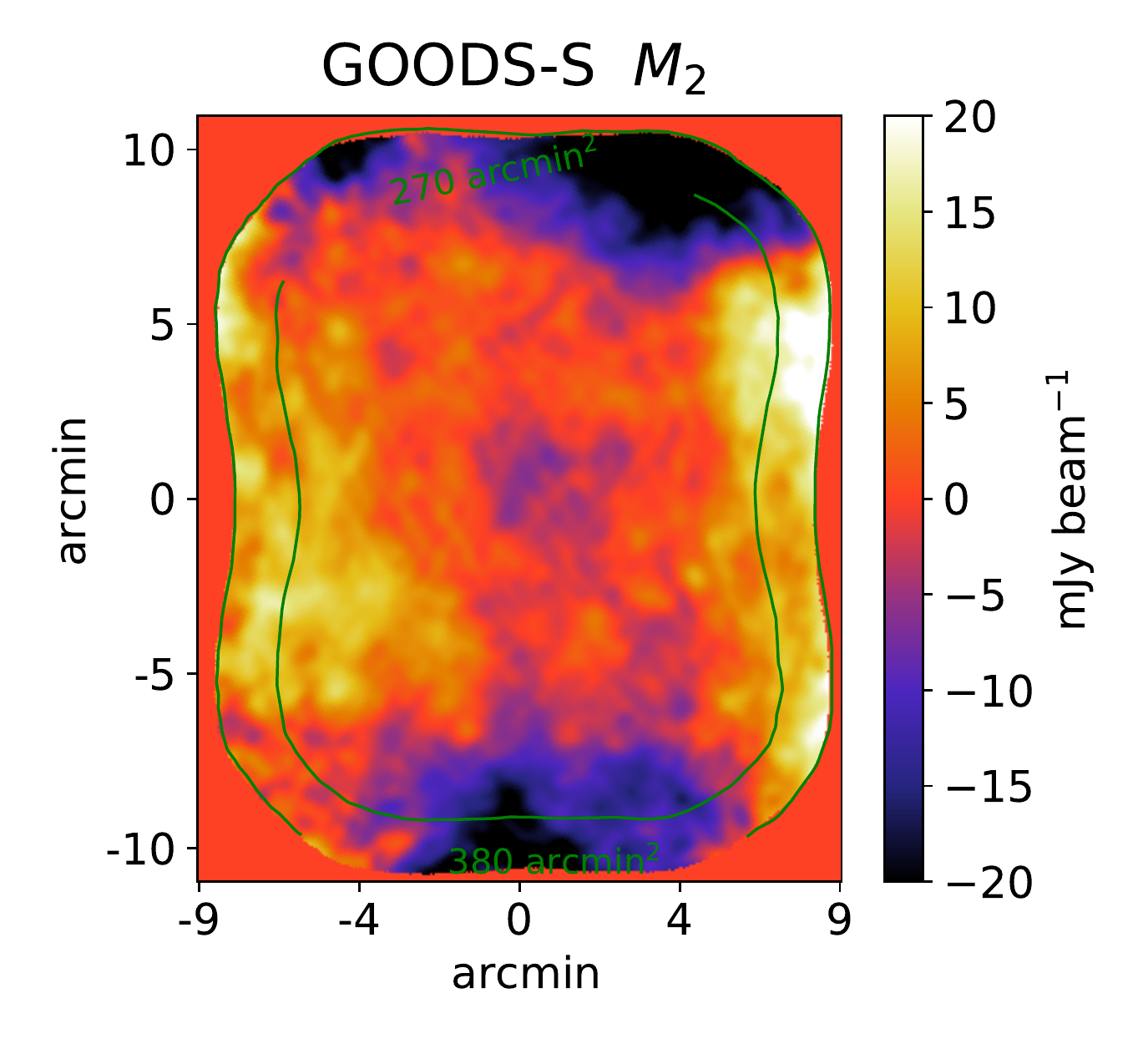} &
    \includegraphics[width=0.33\textwidth]{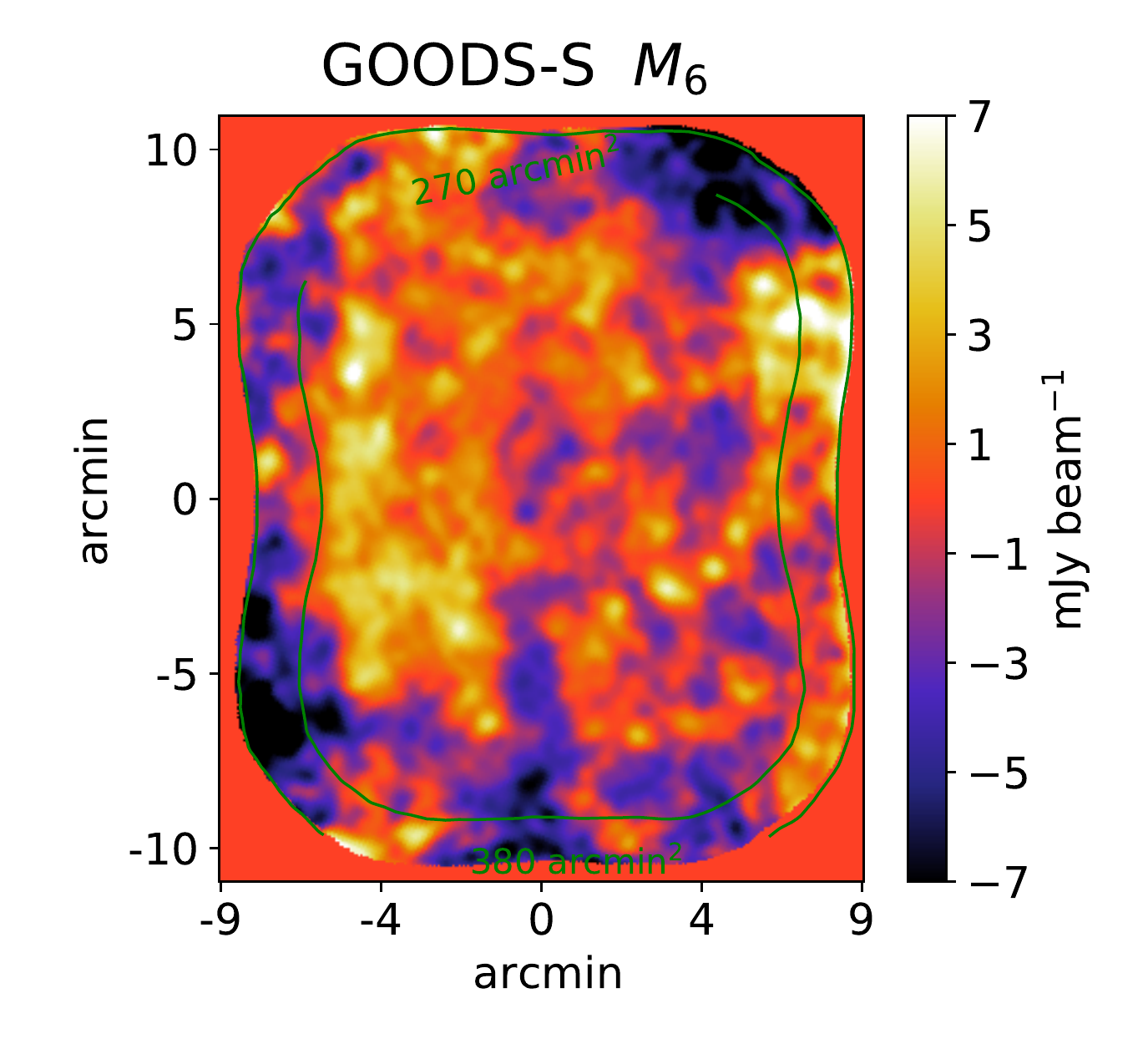} \\
    \vspace{1cm} \\
    \includegraphics[width=0.33\textwidth]{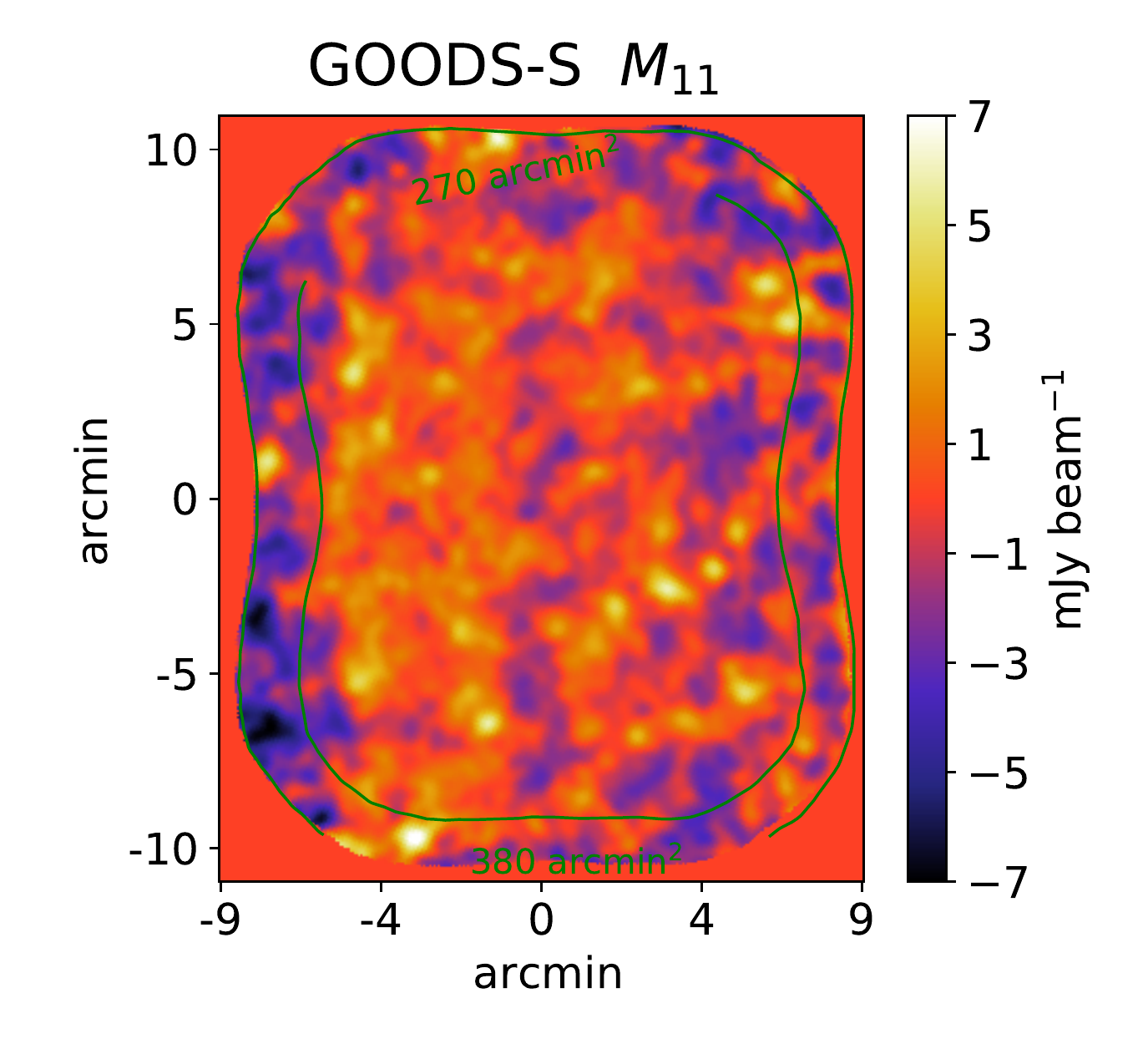} &
    \includegraphics[width=0.33\textwidth]{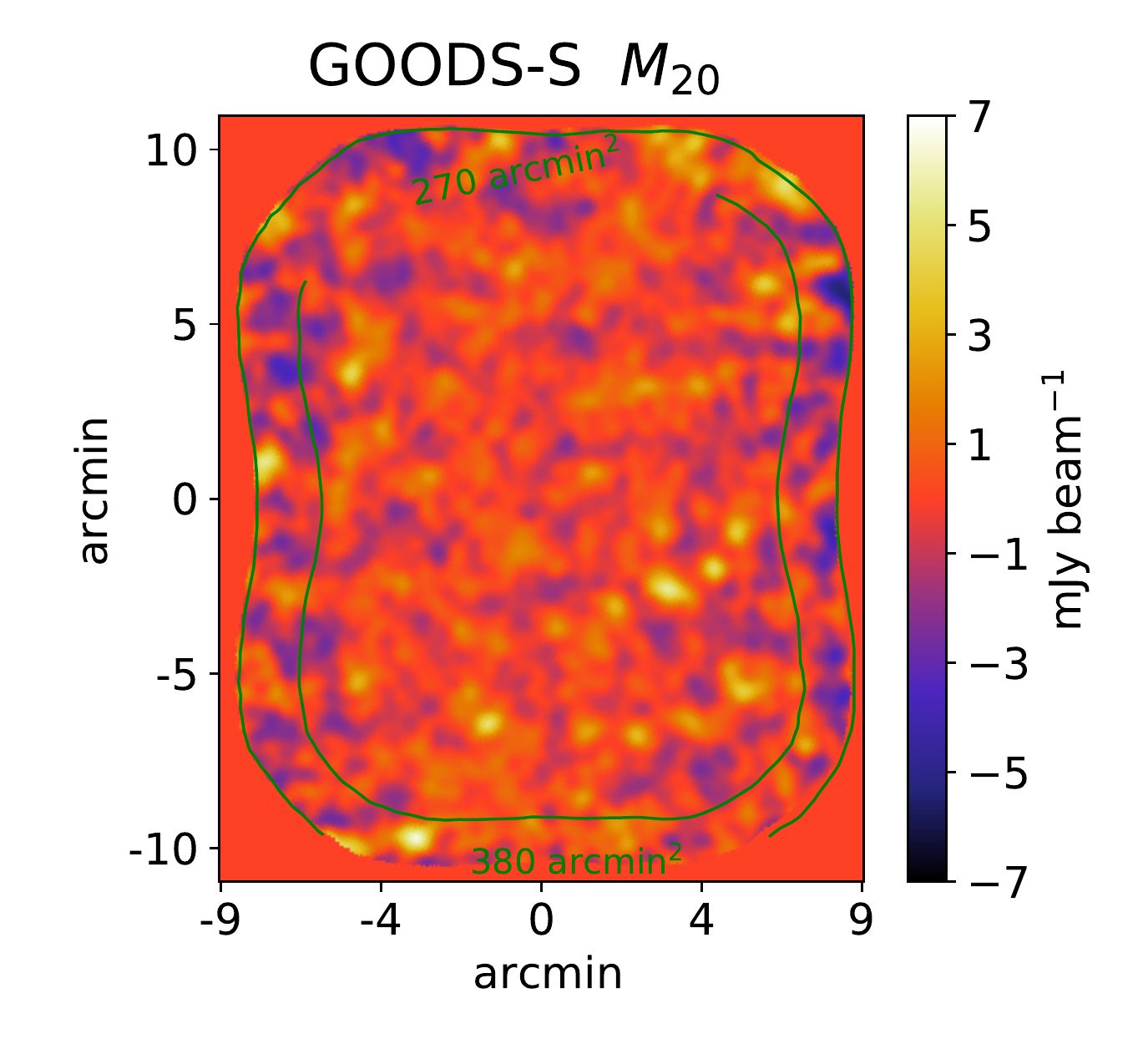} &
    \includegraphics[width=0.33\textwidth]{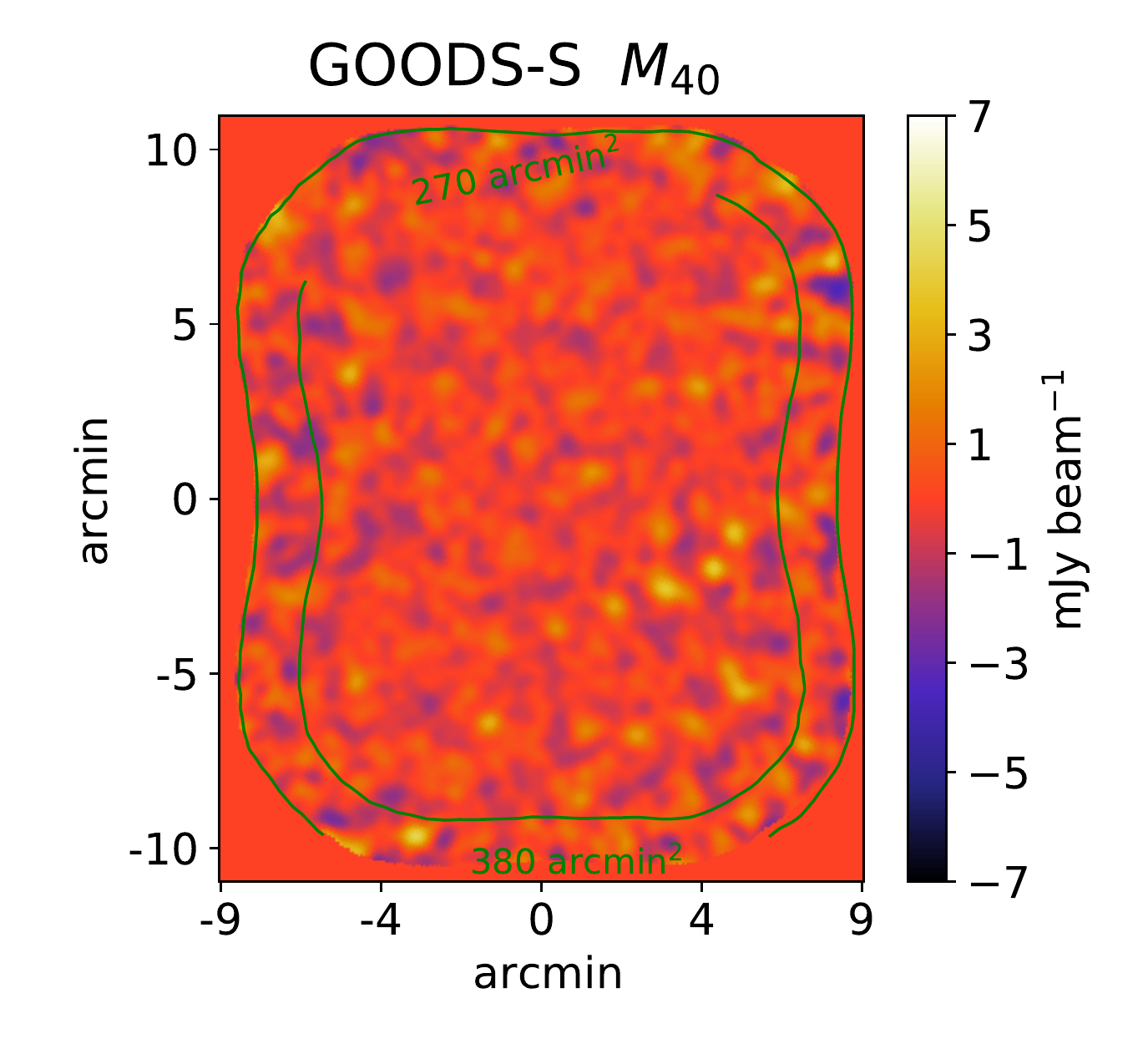} 
  \end{tabular}     }
  \caption{ Redundant maps of the GOODS-S field
            made with the AzTEC pipeline,
			removing $i$ principal components, respectively.}
     \label{fig_redundant_maps}
\end{figure*}

Following the same methodology, we compute the redundant maps 
$\left\lbrace M_i \right\rbrace$ for GOODS-S.
We show their statistical moments in figure \ref{fig_stats}, and
a few representative redundant maps in figure \ref{fig_redundant_maps}.
Notice the mixture of small and large structures in every map.
Next we compute the reference $M_0$ map as shown in figure \ref{fig_m0},
using the \textsc{pca}2.5$\sigma$ procedure (see \S\ref{subsec_macana}).

\begin{figure}
  \centering
    \includegraphics[width=0.47\textwidth]{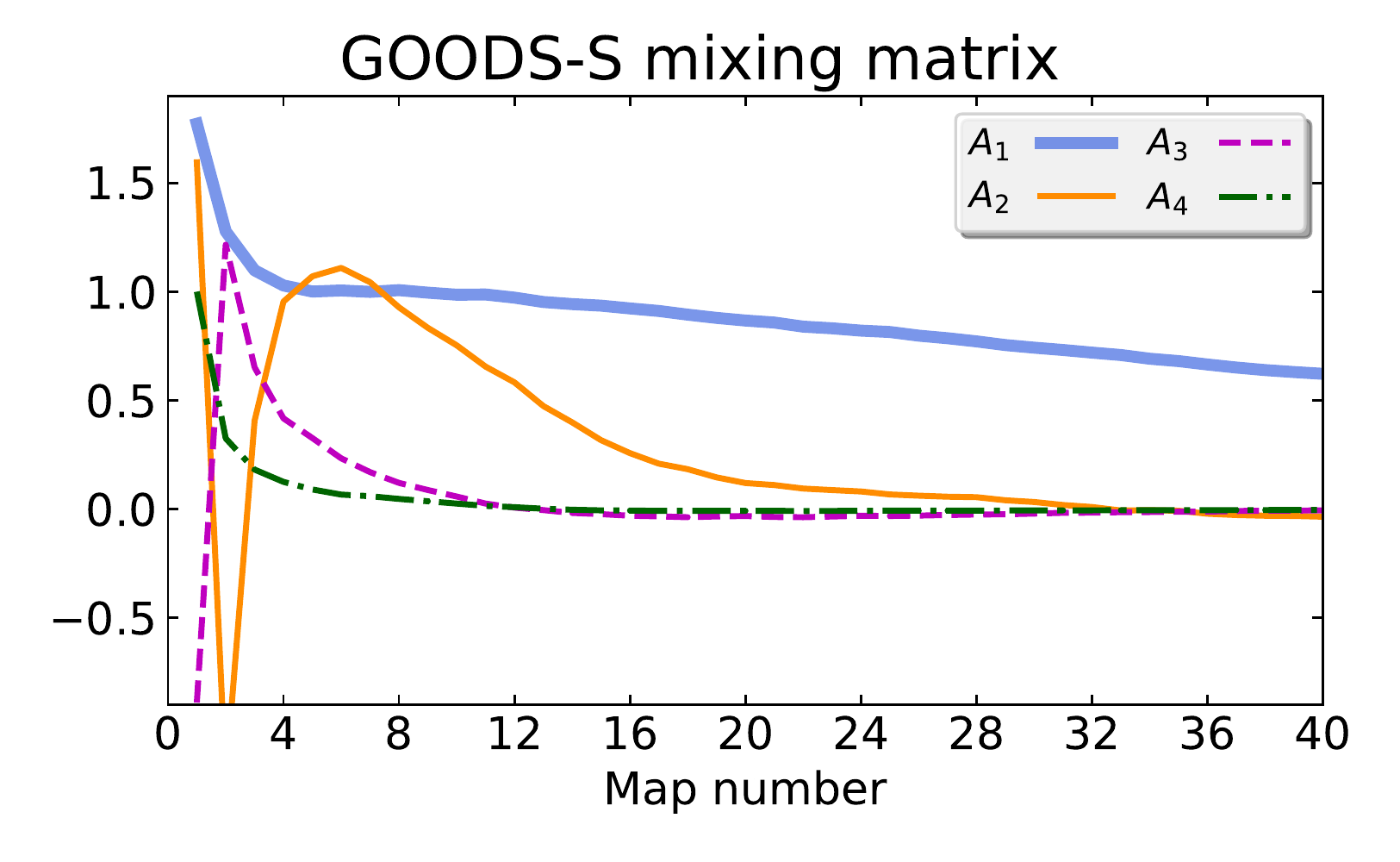} 
  \caption{Mixing matrix coefficients of the GOODS-S redundant maps, 
  		    decomposed with equation (\ref{eq_ica_maps}).}
  \label{fig_mix}
\end{figure}

\begin{figure*}
 \noindent\makebox[\textwidth]{
 \centering
  \begin{tabular}{cc}
    \includegraphics[width=0.47\textwidth]{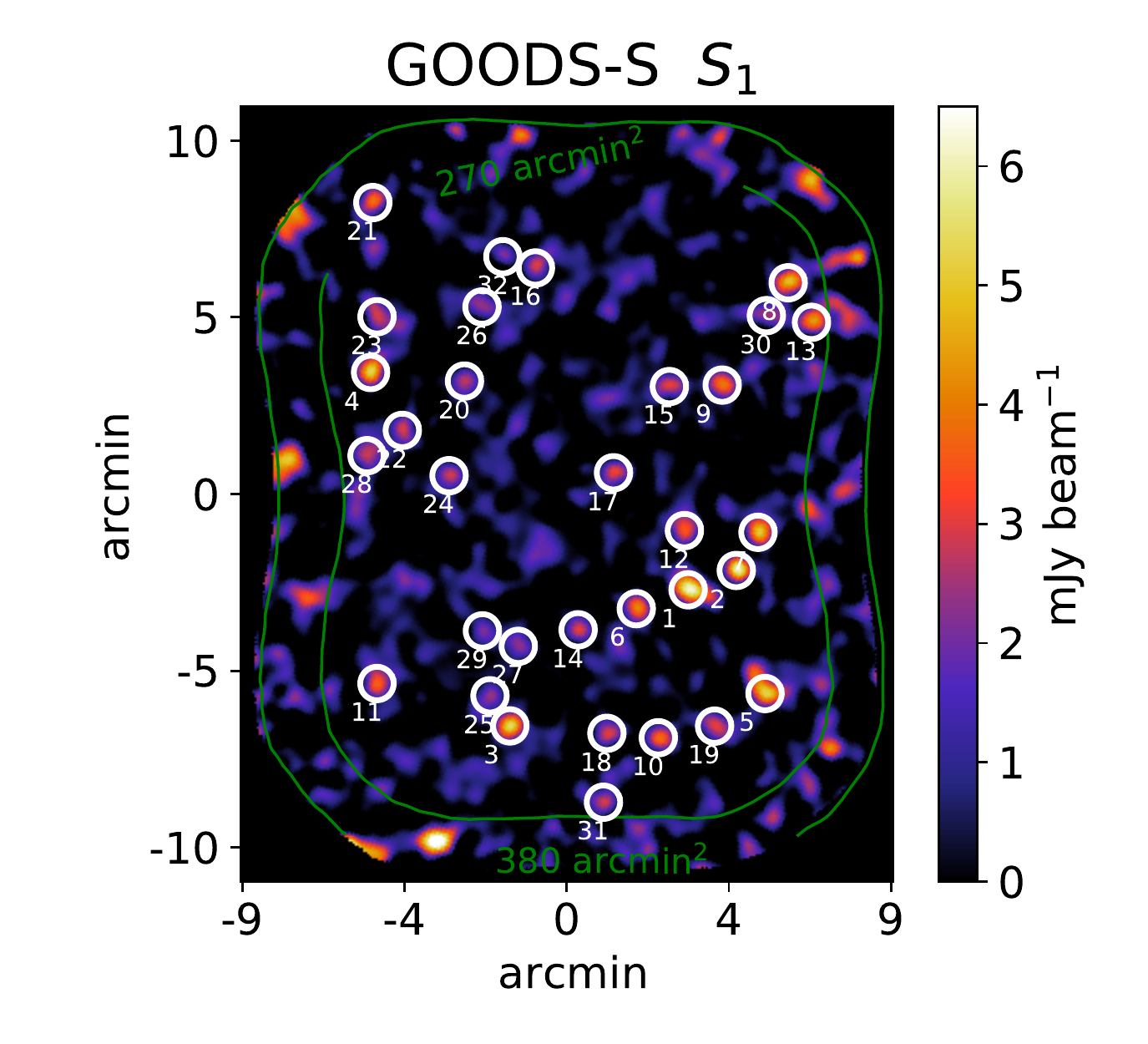} &
    \includegraphics[width=0.47\textwidth]{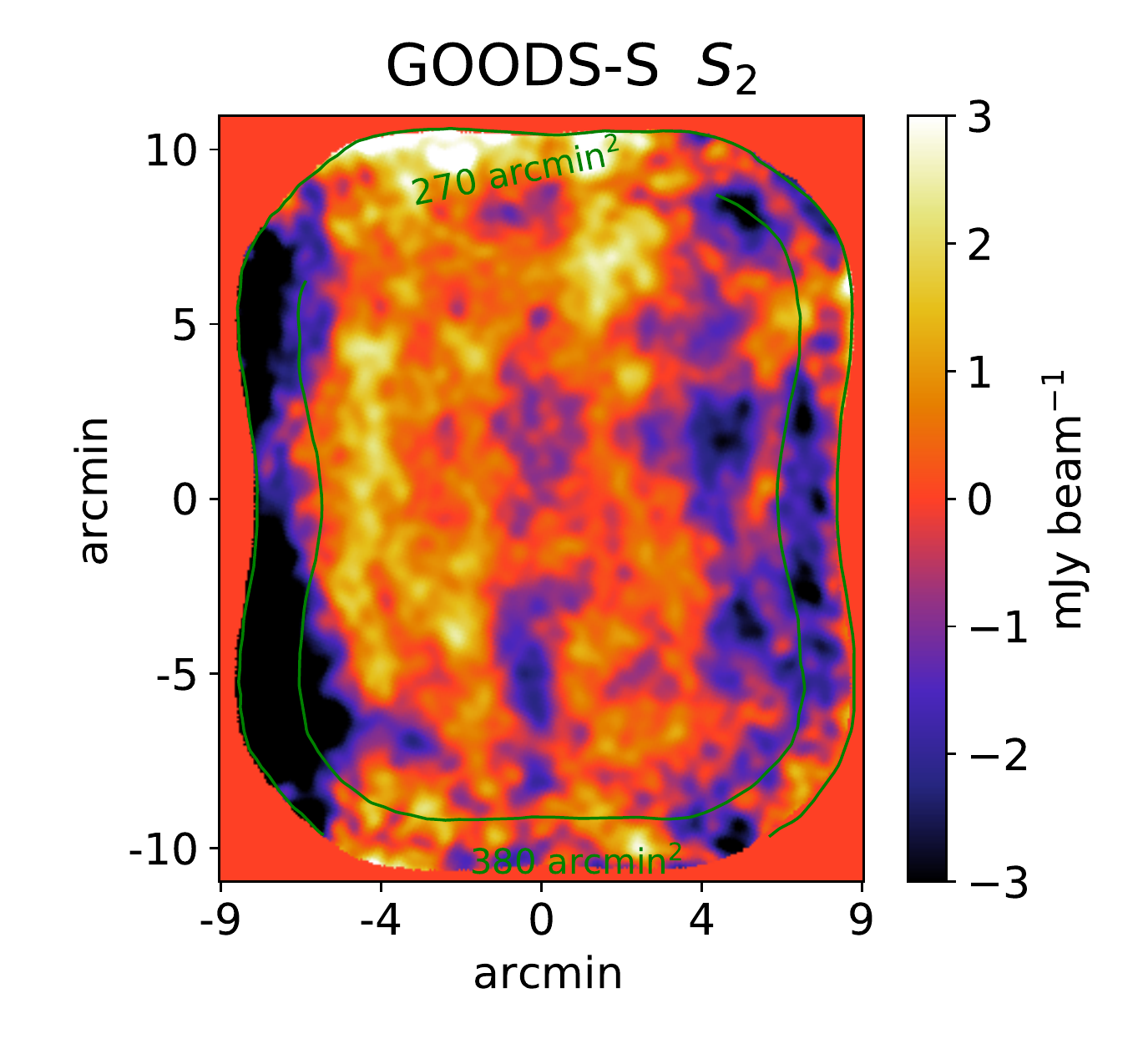} \\
    \includegraphics[width=0.47\textwidth]{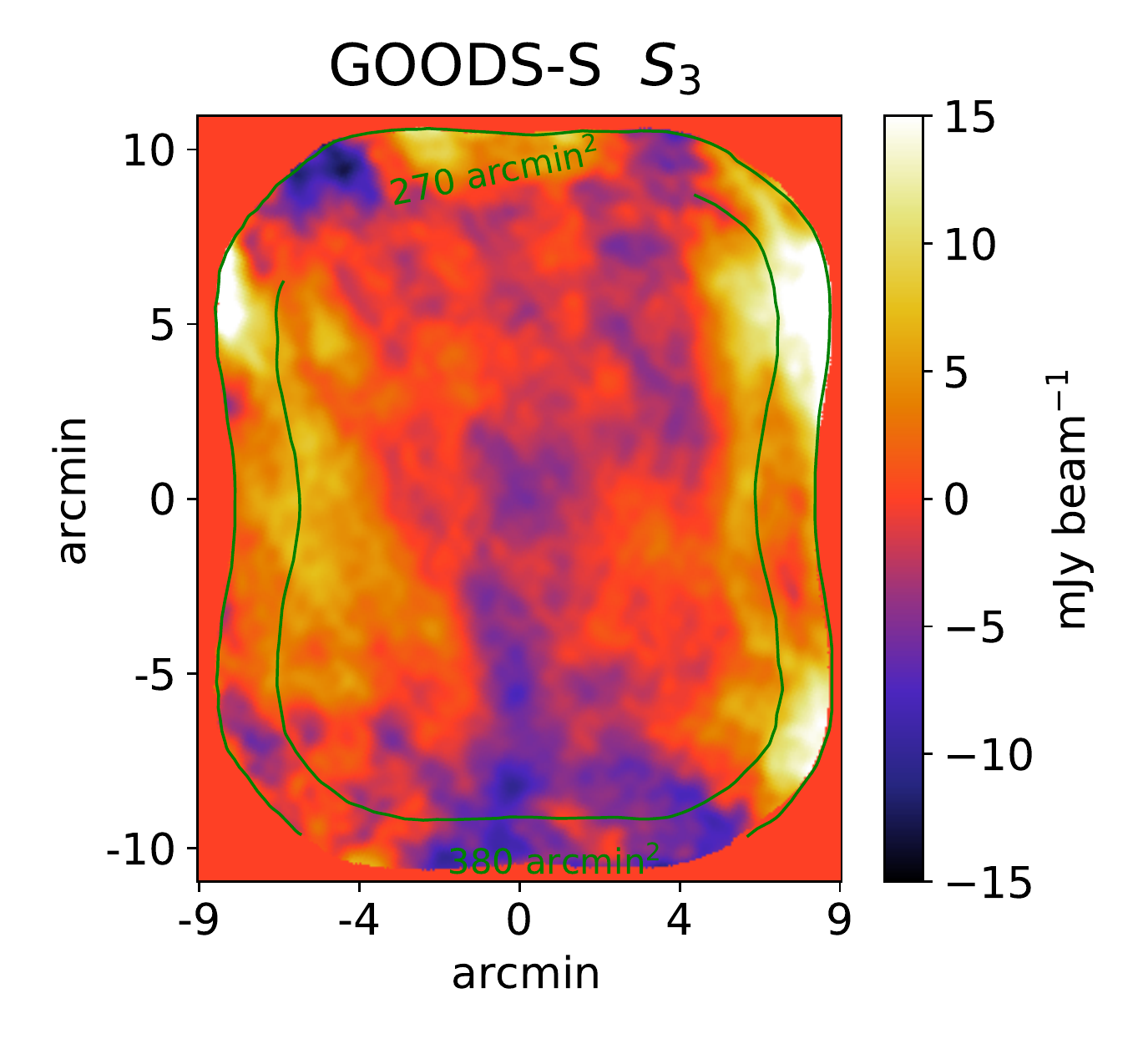} &
    \includegraphics[width=0.47\textwidth]{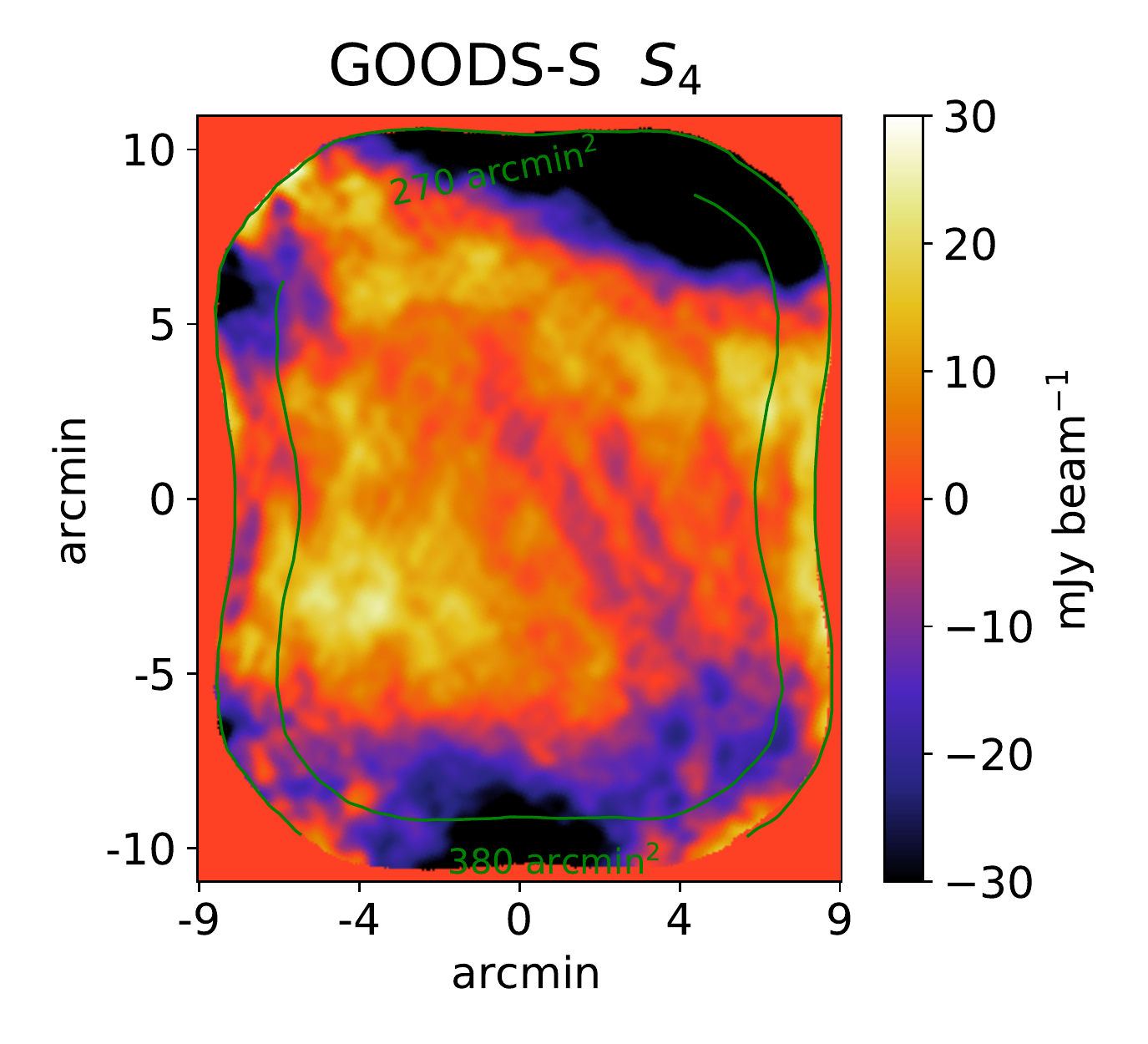} 
  \end{tabular}     }
  \caption{ Independent components decomposed
    from GOODS-S redundant maps
	and calibrated as explained in \S\ref{subsec_calibration}.
	$S_1$ is a point-like component of the GOODS-S field.
    White circles enclose 32 bright sources found with S/N$>$4.
	$S_2$ is an extended component, suspected of astrophysical nature,
	possibly the confusion background.
	$S_3$ and $S_4$ are interpreted as atmospheric foregrounds,
	and the effect of the Lissajous scanning pattern.}
     \label{fig_icagoodss}
\end{figure*}

We proceed with the same decomposition parameters used previously.
We show the mixing matrix in figure \ref{fig_mix} and 
the independent components in figure \ref{fig_icagoodss}.
As we discussed with simulations,
the independent components can be identified either by \textit{eye},
or from the behavior of the mixing coefficients.
$S_1$ is a point-source component,
whose mixing coefficients extend along all the redundant maps.
$S_2$ is an extended emission suspected for an astrophysical origin,
whose mixing coefficients survive to half the redundant maps.
The negative peak in $A_2$ is an effect due
to the positive/negative borders
between the $M_2$ and $M_{14}$ redundant maps.
$S_3$ is made of smooth bright fluctuations,
while $A_3$ survives only for the first dozen maps,
so it can be identified as an atmospheric foreground.
Finally, because of its symmetric stripes,
$S_4$ can be identified at least partially
with a systematic effect due to the Lissajous scan.
Given that $A_4$ mixes only $M_{1<i<3}$, we can assert that
the Lissajous systematic harms mostly the largest angular scales.

We continue with calibration
(see \S\ref{subsec_calibration} for details).
$S_1$ is scaled as $a_1 S_1$, with $a_1=1.22 \pm 0.05$
found employing the witness-based calibration;
its companion weight map $W_1$ is computed as before,
with the scale uncertainty added in quadrature as a systematic error.
For $S_2$, we proceed with a pixel-by-pixel fit to $M_{5<i<9}$,
which are the most correlated with $S_2$,
finding $a_2=1.57 \pm 0.41$.
(For comparison, from the same fit,
the $S_1$ scale is found to be $a_1=1.22\pm0.45$)
We finally scale $S_3$ and $S_4$ also with the pixel-by-pixel fit to 
their most akin redundant maps, $M_{2,3}$ and $M_1$ respectively.

%sig=30./2.35*(1./3600)
%beam=np.pi*sig**2
%confusion=1./(30*beam)

\begin{table}[h]
\centering
\caption{Statistical properties of GOODS-S maps}
\begin{tabular}{cccc}
	Map			&	Std [mJy]	&	Skewness	&	Negentropy	\\
	\hline 
	$M_0$		&	1.03		&	 0.84		&	   1.07	\\
	$S_1$	    &	1.07		&	 0.77		&	   1.18	\\
	$S_2$		&	1.04 		&  -0.67		&	   0.64	\\
	\hline
\end{tabular}
\tablecomments{	Statistical moments quantified within 
				270 \textsl{arcmin}$^2$ GOODS-S field.
				Negentropy is computed with equation (\ref{eq_negentropy_mine})}
\label{tabla_real_comp}
\end{table}

\begin{figure}
  \centering
    \includegraphics[width=0.47\textwidth]{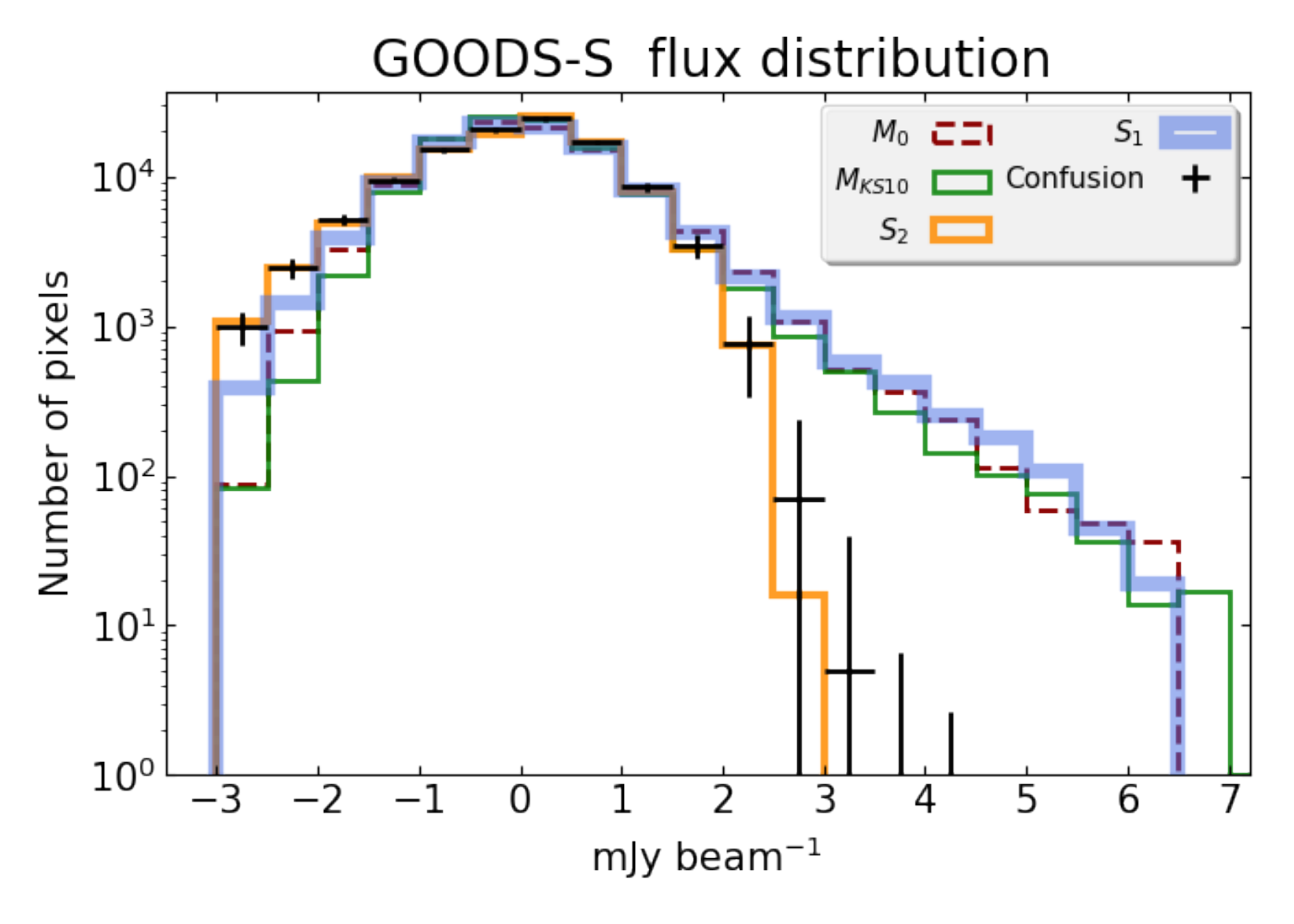} 
  \caption{ Flux distribution within the inner
    270 arcmin$^2$ region of the GOODS-S field,
    for the following maps:
	$M_0$ reference map computed with the
        \textsc{pca}2.5$\sigma$ procedure,
	the $M_{\text{KS10}}$ map computed with
        \textsc{pca}2.5$\sigma$ and Wiener filtering \citep{Scott2010},
    the independent components $S_1$ and $S_2$
    decomposed with our PCA-ICA technique.
	Black error bars represent the expectation from our
    simulations of the confusion background (details in text).}
  \label{fig_histos}
\end{figure}

\begin{figure}
	\centering
    \includegraphics[width=0.47\textwidth]{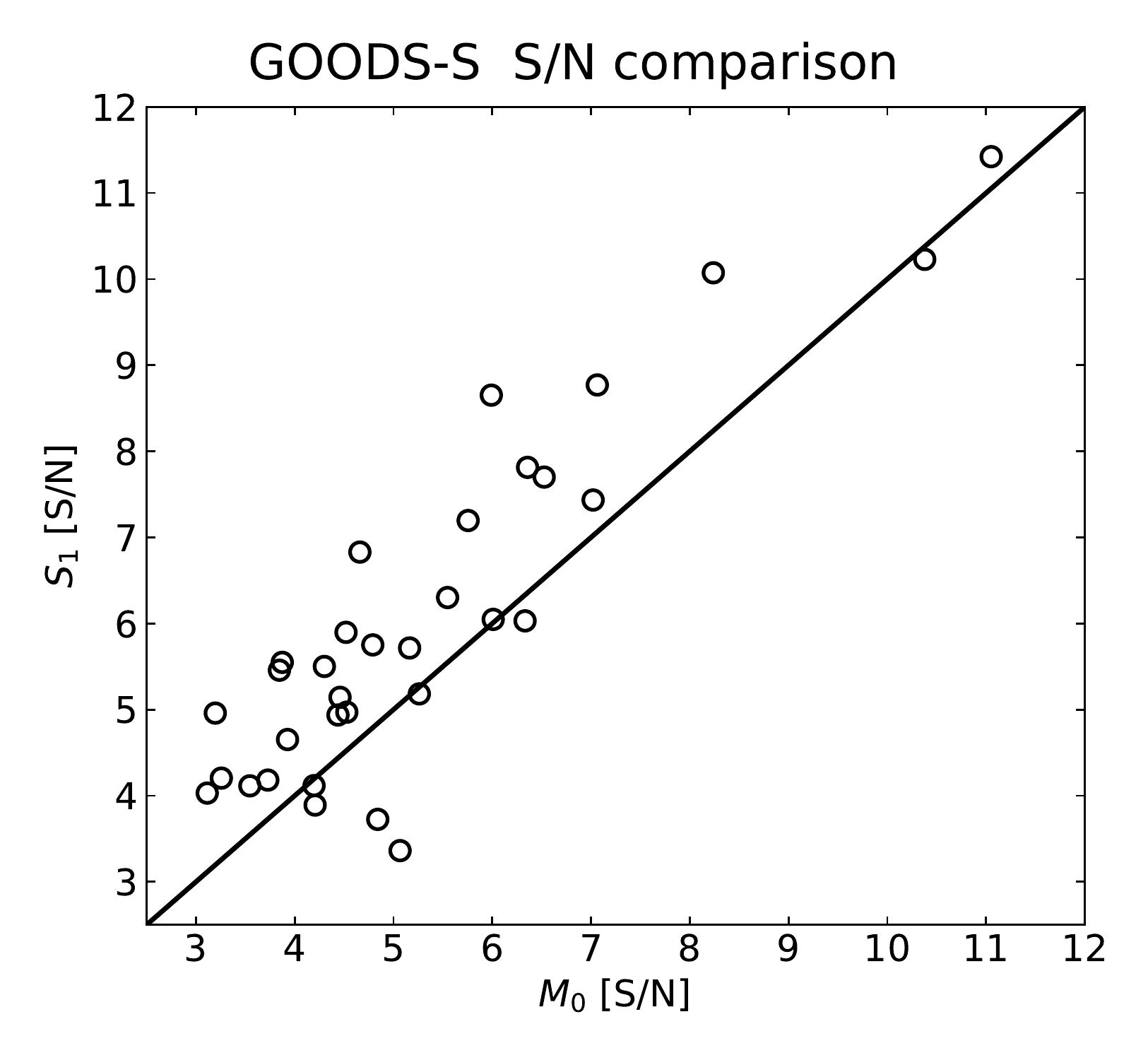}
  \caption{S/N comparison between point sources detected in
  			the GOODS-S field maps $S_1$ or $M_0$.}
  \label{fig_sncomp}
\end{figure}

\begin{figure*}
 \noindent\makebox[\textwidth]{
 \centering
  \begin{tabular}{c}
    \includegraphics[width=0.98\textwidth]{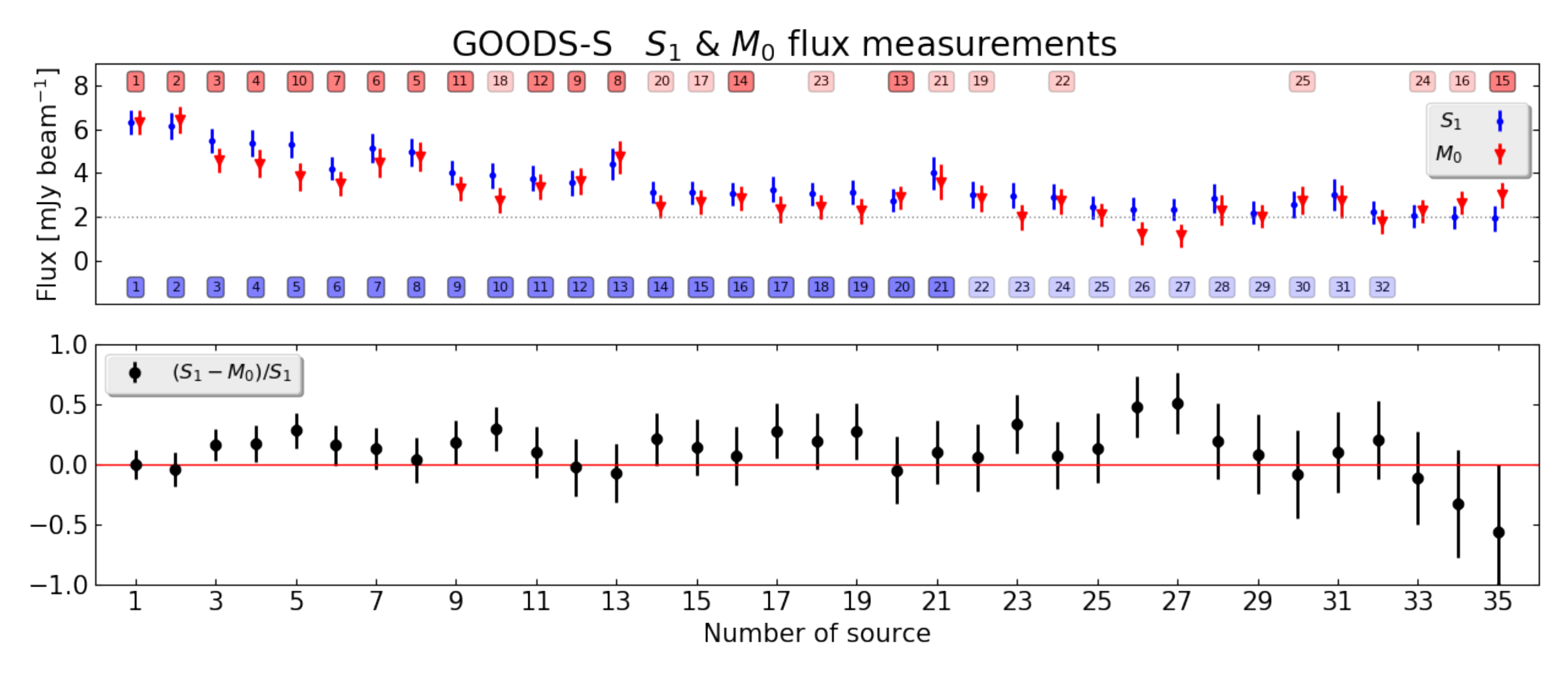} \\
    \includegraphics[width=0.98\textwidth]{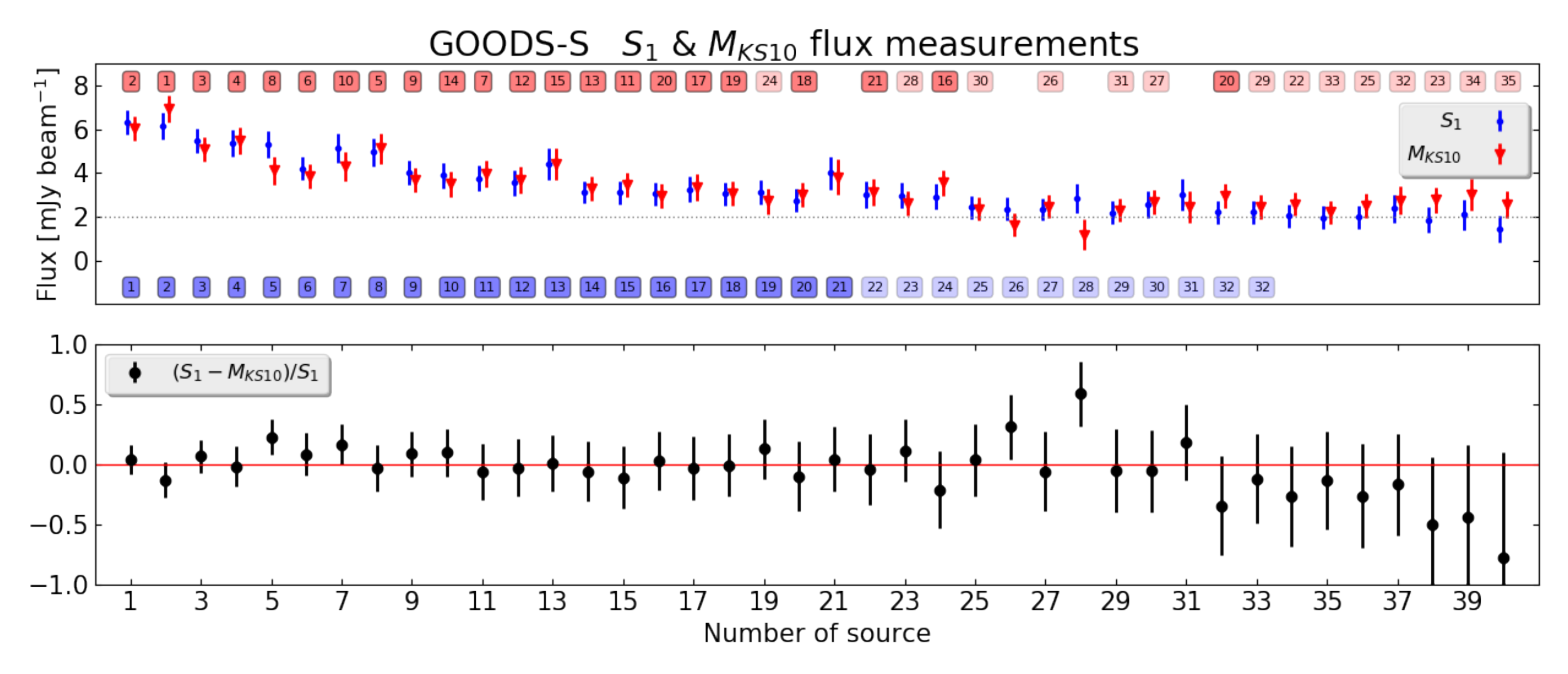} 
  \end{tabular}     }
  \caption{ GOODS-S flux measurements.
	Upper panel: bright-sources detected in $S_1$ and $M_0$.
	Lower panel: bright-sources detected
                 in $S_1$ and $M_{\text{KS10}}$.
	Square labels indicate the detection number on each map.
	Dark-color labels indicate detections with S/N$>$5,
	light-color labels indicate detections between 4$<$S/N$<$5.
	The confusion limit for this survey is 2 mJy \citep{Scott2010}.
	The flux residuals between each pair of maps are also shown.}
     \label{fig_flux_ica_vs_other}
\end{figure*}

The statistical properties of $M_0$, $S_1$, and $S_2$
(within 270 arcmin$^2$) are listed in table \ref{tabla_real_comp},
and their flux distribution can be observed in figure \ref{fig_histos}.
In this case, the improvement of the statistical properties of
$S_1$ compared to $M_0$ can only be noticed in the degree of negentropy.
Yet, the improvement is significant enough to boost the S/N
of point sources, as shown in figure \ref{fig_sncomp}.
In $M_0$, we count 25 bright sources with S/N$>$4.
In $S_1$, we count 32 bright sources with S/N$>$4.
We cannot check for bias as we did with mock data,
but we can compare the fluxes measured in $M_0$ and $S_1$
(see figure \ref{fig_flux_ica_vs_other}).

\begin{figure}
	\centering
    \includegraphics[width=0.47\textwidth]{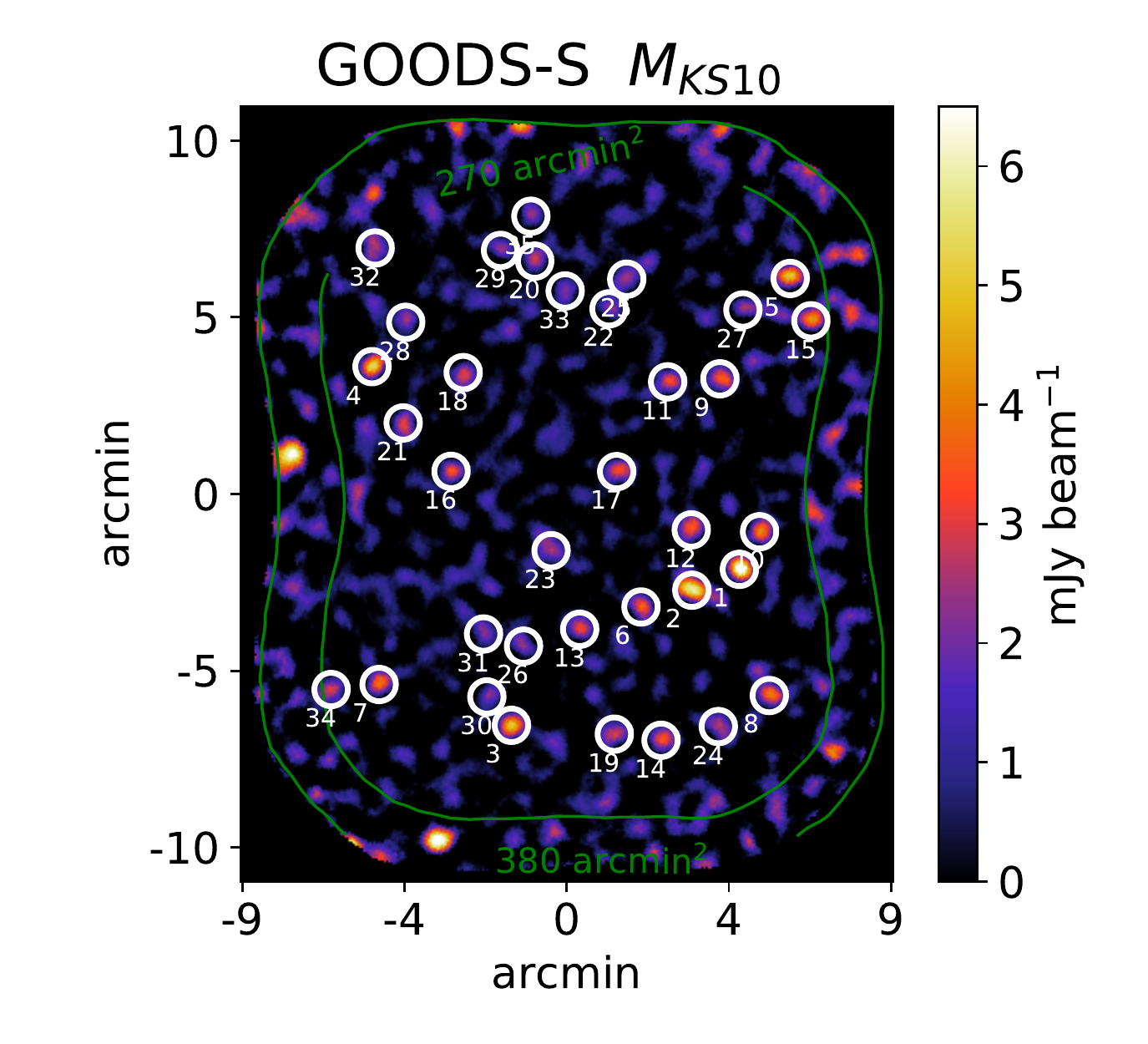}
  \caption{ GOODS-S $M_{\text{KS10}}$ map \citep{Scott2010}.
	Besides the \textsc{pca}2.5$\sigma$ procedure,
    an optimized Wiener filter was used to enhance point sources.
	Sources with S/N$>$4 are accounted as detections and circled
	with solid-white lines. Contours as in figure \ref{fig_model}.}
  \label{fig_scott}
\end{figure}

To enrich our discussion, we test the ability of our technique
to reproduce previously reported results from the same GOODS-S survey.
In $KS10$, the authors also applied
the \textsc{pca}2.5$\sigma$ procedure (with a different code),
but the main difference is that they applied a Wiener filter
with a specialized point-like kernel as a prior.
The assumptions taken in that approach and ours are
conceptually different and not mutually exclusive;
so, a direct comparison must be moderately assessed.
The $KS10$ map is shown in figure \ref{fig_scott},
in which there are 35 bright sources with S/N$>$4.

In figure \ref{fig_flux_ica_vs_other}
we also compare the flux measured on
$S_1$ and $M_{\scalebox{0.6}{KS10}}$.
For S/N$>$5, most of the fluxes are consistent,
only below S/N$\sim$5 there are some non-coincident detections.
These differences arise most likely due to the combined effect
of persistent foreground residuals after
the \textsc{pca}2.5$\sigma$ procedure
and their subsequent enhancement by the Wiener filter at unresolved scales.
This flux comparison is interesting for the purpose to probe
the recovery of high S/N detections using our PCA-ICA approach,
but lack of coincidences at low S/N should not be overstated.

As we mentioned, 
$S_2$ is suspected to have an astrophysical origin,
and an intuitive prospect is the confusion background.
To explore this possibility,
we simulate the fainter dusty star-forming galaxy population,
following the number counts measured by \citet{Fujimoto2016},
which includes the deep ALMA census of
faint sources ($\gtrsim$ 0.02 mJy) at 1.2 mm and
the bright-end AzTEC 1.1 mm counts of \citet{Scott2012}.
Sources are randomly distributed in space
within the simulated maps (i.e. no clustering).
As every realistic observation contains
some degree of instrumental noise,
we also add and convolve a 1.1 mJy white noise to our simulations,
in order to approximately match the negative flux tail of $S_2$.
We generate 200 random realizations of these confusion maps.
Using the weight map $W_1$,
we subtract bright sources (S/N$>$3.5),
and quantify the average flux distribution.
As seen in figure \ref{fig_histos},
$S_2$ is consistent with the flux distribution expected
from our simulated confusion background,
and reflects the 2 mJy confusion limit.
This interpretation of $S_2$ could help to explain
the detection differences of $S_1$ compared to
$M_0$ or $M_{\scalebox{0.6}{KS10}}$,
especially for point sources below the confusion limit
(see figure \ref{fig_flux_ica_vs_other}).

\begin{figure}
	\centering
    \includegraphics[width=0.47\textwidth]{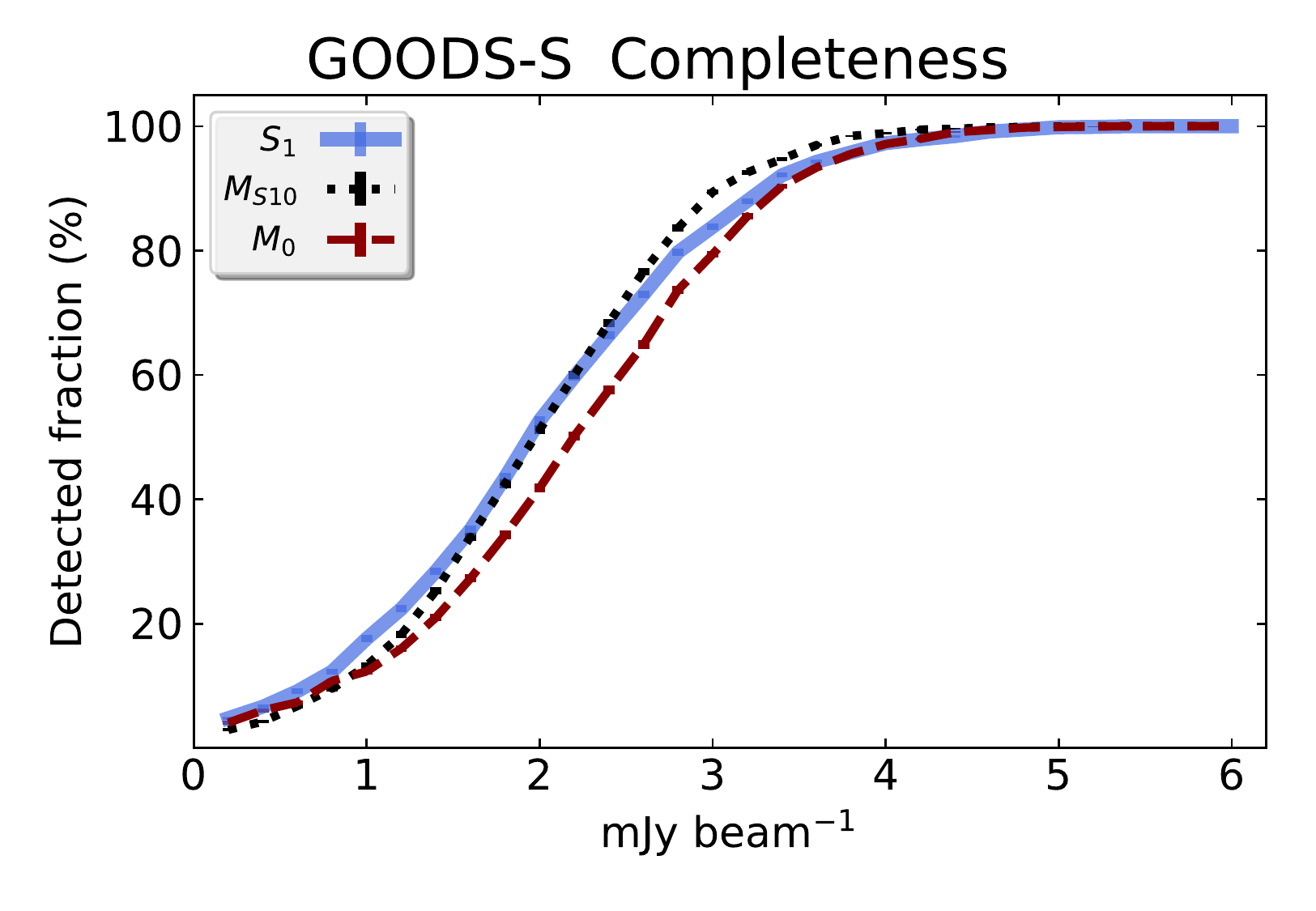}
  \caption{ GOODS-S completeness of $M_0$,
            $M_{\text{KS10}}$, and $S_1$ maps.}
  \label{fig_completeness}
\end{figure}

We also compute in figure \ref{fig_completeness} the completeness
for $M_0$, $M_{\scalebox{0.6}{KS10}}$, and $S_1$.
As expected from simulation results,
the completeness of $S_1$ is better compared to $M_0$.
Given that $M_{\scalebox{0.6}{KS10}}$ comes from a process
that suppresses extended objects in favor of point ones,
we would have expected a much larger completeness for
$M_{\scalebox{0.6}{KS10}}$ compared to $S_1$,
however, the results are comparable.

\section{Conclusions}
\label{sec_conclusions}

In this paper we are presenting a PCA-ICA algorithm capable to
separate atmospheric fluctuations, extended astrophysical foregrounds,
and point-like sources from single-wavelength millimeter surveys.
In order to probe the consistency of our results,
we have tested our methodologies on both mock and real data.

We confirm that our PCA-generation of redundant maps
allows a successful application of an ICA decomposition,
in defiance of the single-channel limitation.
We find a good agreement between simulation inputs
and the resulting independent components,
along with a good degree of isolation
(see table \ref{tabla_sim_comp}).

We have proposed and tested different strategies to calibrate
independent components, getting rid of the permutation and
scale ambiguities, inherent to ICA.
We find that our approach can be useful to remove
both atmospheric and astrophysical foregrounds,
minimizing information loss.
Consequently, our decomposition can help to prevent bias
in flux measurements and boost the signal-to-noise.

We also applied our techniques to
the AzTEC/ASTE survey of the GOODS-S field.
We find that a PCA-ICA decomposition $S_1$ is preferred over
the simplest \textsc{pca}2.5$\sigma$ procedure $M_0$,
as expected from the analogous result in simulations.
We confirm agreement with S/N$>$5 detected sources in $KS10$,
showing consistency to recover previously reported measurements.
An unexpected finding of this work was the measurement of
a feeble extended emission on the GOODS-S field
($S_2$ in figure \ref{fig_icagoodss}),
which according to simulations is consistent with the flux-distribution
of the faintest SMGs' confusion background.
We conclude that our PCA-ICA implementation is a viable and promising approach
to separate atmospheric and astrophysical (extended and compact) sources.

One route to extend our work is to improve redundancy
with algorithms other than PCA in time-domain.
One can also try different decomposition algorithms,
possibly more powerful than the simplest version of FastICA.
Besides, it should be possible to adapt our technique to multi-wavelength data,
increasing redundancy and decomposing signals in each channel,
before a multi-wavelength analysis.
The complementary maps decomposed by ICA are interesting by themselves.
Certainly, further investigations of this kind of techniques,
not only can improve atmospheric and instrumental models,
but also make new astrophysical emissions available.
The GOODS-S $S_2$ component is just an interesting example:
our astrophysical simulations indicate consistency with the confusion background,
yet, in future analyses we shall confirm this result through
exhaustive simulations of every systematic effect possibly sourcing $S_2$.
If confirmed, the characterization of the confusion background is 
of utmost importance in millimeter astronomy and cosmology,
allowing us to study the clustering properties of SMGs
and the distribution of matter in the Universe.

We finally stress that this kind of analysis is
particularly interesting at the advent of
the next generation of continuum cameras.
For example, MUSCAT
\footnote{http://gtr.rcuk.ac.uk/projects?ref=ST\%2FP002803\%2F1},
TolTEC \footnote{http://toltec.astro.umass.edu/},
SCUBA-2 \citep{SCUBA2013},
NIKA2 \citep{NIKA2016}.
Both MUSCAT and TolTEC are currently in construction
and appointed to work on the Large Millimeter Telescope \citep{LMT2010}.
Indeed, this paper is our first step towards
the ultimate goal of developing an efficient
multi-component decomposition pipeline for both instruments.
MUSCAT is a single-channel large-format camera,
comprising $\simeq1,200\,$ detectors at the 1.1 mm wavelength-band.
MUSCAT will record ten times more timestreams than AzTEC.
We will be able to generate many more levels of redundancy,
along broader ranges of angular scales;
thus the application of our PCA-ICA algorithms
will be very similar to this paper,
but much more promising in quality of the decomposition.
TolTEC will include 6,300 detectors distributed in three channels,
at 1.1, 1.4, and 2.1 mm,
each of them sensitive to linear polarization.
Thus, each single TolTEC observation will result in nine maps,
that can be further decomposed by our PCA (or alternative) technique
to generate higher levels of redundancy.
Altogether, TolTEC surveys and our PCA-ICA technique
bring out exciting expectations about the final data quality
and possible new astrophysical fields to be uncovered.

\acknowledgments 
This project was possible due to partial support from 
CONACyT research grants:
CB-2011/167291,
CB-2015/256961,
\textit{Fronteras de la Ciencia} 2016/1848,
and FONCICYT 2016/69.
We thank Emmaly Aguilar and Ivânio Puerari for useful discussions
and an anonymous referee for a critical review
that has improved the paper.

\bibliography{references}
\addcontentsline{toc}{section}{References}

\end{document}